\documentclass[12pt]{article}

\usepackage[margin=1.0in]{geometry}

\usepackage[round]{natbib}

\usepackage{amsthm, amsmath, amssymb}
\usepackage{mathtools, bbm}
\usepackage{setspace}
\usepackage{titlesec}
\titlespacing*{\section}{0pt}{2.5ex plus 1ex minus .2ex}{1.5ex plus .2ex}
\titlespacing*{\subsection}{0pt}{1.75ex plus 1ex minus .2ex}{1ex plus .2ex}
\usepackage{hyperref}
\hypersetup{colorlinks, linkcolor=blue, citecolor=blue, pdftitle={Semiparametric Identification of the Discount Factor and Payoff Function in Dynamic Discrete Choice Models}, pdfauthor={Yu Hao, Hiroyuki Kasahara, Katsumi Shimotsu}}
\usepackage{graphics}
\usepackage{multirow}
\usepackage{threeparttable}

\usepackage{xcolor}
\usepackage{booktabs}

\usepackage[normalem]{ulem}
 \usepackage{soul}

\newtheorem{proposition}{Proposition}

\newtheorem{remark}{Remark}

\newtheorem{lemma}{Lemma}
\newtheorem{corollary}{Corollary}
\newtheorem{assumption}{Assumption}
\newtheorem{example}{Example}

\DeclareMathOperator{\Ker}{Ker}
\DeclareMathOperator{\adj}{adj}
\DeclareMathOperator*{\argmax}{arg\,max}

\def\iid{\mathrm{i.i.d. }}
\def\bs{\boldsymbol}

\def\t{^{\top}}

\def\ve{\varepsilon}

\begin{document} 

{\singlespace

\title{\textbf{Semiparametric Identification of the Discount Factor and Payoff Function in Dynamic Discrete Choice Models}} 
\author{Yu Hao \\
Faculty of Business and Economics\\
University of Hong Kong \\
haoyu@hku.hk
\and 
Hiroyuki Kasahara \\
Vancouver School of Economics\\
University of British Columbia \\
hkasahar@mail.ubc.ca
\and  Katsumi Shimotsu\thanks{Corresponding author: Katsumi Shimotsu, Faculty of Economics, University of Tokyo, 7-3-1 Hongo, Bunkyo-ku, Tokyo 113-0033, Japan. Tel: +81-3-5841-5518; email: shimotsu@e.u-tokyo.ac.jp} \\
Faculty of Economics \\
University of Tokyo\\
shimotsu@e.u-tokyo.ac.jp}
\date{November 7, 2026}
\maketitle
}

\begin{abstract}
This paper studies the identification of the discount factor and payoff function in stationary infinite-horizon dynamic discrete choice models. In single-agent models, we show that nonparametric assumptions on per-period payoffs implied by economic theory, such as homogeneity, monotonicity, concavity, zero cross-differences, and complementarity, yield polynomial equalities and inequalities in the discount factor. These restrictions can identify the discount factor and the payoff function without normalizing the payoff of one action. Even when the payoff function is not point identified, they may still identify the discount factor. In dynamic games, we identify firm-specific discount factors by combining the irrelevance of other firms' lagged actions with restrictions such as exchangeability, even when the scrap value is unknown. We illustrate the approach with numerical examples and an empirical application to farmers' crop choices.
\end{abstract}



\newpage

\section{Introduction}

In empirical work on dynamic discrete choice, the discount factor is rarely estimated. Because it cannot be identified from conditional choice probabilities alone \citep{rust94hoe,magnac02em}, most studies fix it between 0.9 and 0.99 \citep[e.g.,][]{hendel2006, ryan2012, collardwexler2013demand,igami2020mergers, Miller2021}. Yet the studies that do estimate it often report values below that range \citep[e.g.,][]{yao2012determining,chung2014bonuses, GayleTomlin2018,DeGroote2019,kong2024nonparametric}, suggesting that individuals and firms may be more impatient than typically assumed.\footnote{Reviewing past studies, \citet{frederick2002time} find that discount factors vary considerably across contexts and populations.} Allowing for greater impatience can substantially change counterfactual policy conclusions \citep{ChingOsborne2019}.\footnote{\citet{fowlie2016market} and \citet{igami2017estimating} re-estimate their models over a range of discount factors around their baseline value. \citet{lau2024} proposes a framework for assessing the sensitivity of target parameters to the discount factor in dynamic discrete choice models.} This paper shows that, when the payoff function is left nonparametric, the discount factor can be identified by shape restrictions implied by economic theory, such as monotonicity and concavity.

The identification of the discount factor in dynamic structural models has received growing attention. \citet{abbring20qe} demonstrate that an exclusion restriction, which requires the payoff function to take the same value at two different action-state pairs, can identify the discount factor in single-agent dynamic discrete choice models up to a countable set, the zeros of a convergent power series in $\beta$. \citet{abbring20wp} sharpen this result by showing that the cardinality of the identified set is no greater than that of the state space. Other studies achieve identification by imposing linearity in parameters of the payoff function or by requiring a terminal action \citep{Bajari16qme, komarova18qe,chou24wp}. Neither assumption is generally implied by economic theory. The restrictions we study are, by contrast, shape restrictions that economic theory implies and that applied work already invokes for other purposes.

This paper makes three contributions. First, we show that nonparametric assumptions on per-period payoffs common in economic models generate equality and inequality restrictions with substantial identifying power. Examples include homogeneity, monotonicity, concavity, and zero cross-differences or complementarity across distinct state variables \citep[e.g.,][]{matzkin92em,matzkin94hoe}. Equality restrictions can point identify the discount factor or reduce its identified set to a few points, and inequality restrictions bound it. These restrictions can also identify the payoff function itself. Under $\rho$-finite dependence \citep{AltugMiller98restud, am11em} and suitable restrictions, the identified set for the discount factor has at most $\rho$ points, as in \citet{abbring20qe}.

Second, we show that these restrictions can identify the discount factor and the payoff function without normalizing the payoff of one action. \citet{aguirregabiriasuzuki14qme} show that fixed costs, entry costs, and scrap values in a single-agent dynamic entry-exit model are not separately identified without a normalization such as a zero scrap value, and that this normalization can affect counterfactual predictions \citep[see also][]{Kalouptsidi21qe}. Without this normalization, shape restrictions can identify the discount factor even when the payoff function is not point identified, and additional restrictions such as cost homogeneity can point identify the costs as well.

Third, we study dynamic games, for which identification of the discount factor has not been formally analyzed. Besides extending our single-agent results, we show that the irrelevance of other firms' lagged actions and the exchangeability of other firms' actions in per-period payoffs together provide equality restrictions that identify the discount factor. \citet{pesendorfer08restud} use the first assumption to identify payoffs when the discount factor is known. These results extend to an unknown scrap value. We also establish identification when agents have heterogeneous discount factors.

We illustrate our approach with farmers' crop choices in Story County, Iowa, over 2012--2023.
Because fields differ in their rotation behavior, we allow for latent types, each with its own payoff function and discount factor, and apply the identification results type by type.
For partial rotators, which plant corn after corn about 40 percent of the time, monotonicity of the payoff in market profitability gives the identified set $[0.681,0.848]$ for the discount factor.

\subsection{Related literature}

Our analysis builds on several strands of literature. Structural estimation of dynamic discrete choice models goes back to \citet{Wolpin84jpe}, \citet{Pakes86em}, and \citet{rust87em}, who estimate such models by maximum likelihood after solving the underlying dynamic program. The conditional choice probability approach of \citet{hotzmiller93restud} avoids solving the dynamic programming problem.

\citet{rust94hoe} shows that the discount factor is not identified in these models without additional restrictions. \citet{magnac02em} show that, without restrictions on preferences, the discount factor cannot be separately identified from current payoffs using conditional choice probabilities alone, and highlight the role of normalization. As \citet{abbring20qe} note, they identify $\beta$ with an exclusion restriction on a difference in expected discounted utilities between two choice sequences rather than on primitive payoffs. In the context of dynamic games, \citet{am07em}, \citet{bajari07ecta}, \citet{pakes07rand}, and \citet{pesendorfer08restud} develop tractable estimation methods. \citet{aguirregabiriasuzuki14qme}, \citet{noretstang14restud}, \citet{am20joe}, and \citet{Kalouptsidi21qe} analyze the identification of counterfactuals, highlighting how normalizations affect what can be learned about preferences. These studies typically treat the discount factor as known. We complement them with identification results for the discount factor. Closest to our approach, \citet{Kalouptsidi26res} use equality and inequality restrictions on the payoff function to partially identify counterfactual outcomes even when the payoff function is not point identified. We use analogous shape restrictions for a different purpose: to identify the discount factor, which their analysis takes as given.\footnote{In an extension, \citet{Kalouptsidi26res} allow $\beta$ to be unknown, but require it to lie in a prespecified compact interval rather than identifying it.}

Shape restrictions derived from economic theory have a long history in \textit{static} models. \citet{matzkin92em} shows that binary threshold crossing and binary choice models can be identified without parametric restrictions on either the utility function or the distribution of the unobservable term. \citet{matzkin91em} and \citet{matzkin93joe} develop related results for semiparametric estimation under monotonicity and concavity and for polychotomous choice, and \citet{matzkin94hoe} reviews this literature. \citet{matzkin03em} shows that homogeneity of degree one identifies models with nonadditive unobservable heterogeneity, and \citet{allen19em} use a variant of Slutsky symmetry to identify latent utility models with additively separable heterogeneity. In two-player binary choice games of complete information, a known sign of the interaction effect, together with exclusion and support conditions, identifies the payoff function or the distribution of its random coefficients \citep{berrytamer06book,foxlazzati17qe,dunker18joe}. Monotonicity restrictions are used widely in nonlinear and nonseparable models; see \citet{matzkin13are} and \citet{chetverikov18are} for surveys, and \citet{imbens94em}, \citet{chesher03em}, \citet{matzkin08em}, \citet{imbens09em}, \citet{shi18em}, \citet{pakes24qe}, and the references therein. To the best of our knowledge, shape restrictions on the payoff function have not been used to identify the discount factor in dynamic games with incomplete information.

The remainder of the paper is organized as follows. Section \ref{sec:model_single} presents the model and identification results for single-agent models, Section \ref{sec:fd} discusses finite dependence, and Section \ref{sec:examples} gives examples of payoff restrictions. Section \ref{sec:entry_example} presents a numerical example based on a dynamic entry model. Section \ref{sec:dynamic_game} extends the analysis to dynamic games, and Section \ref{sec:game_example} gives a numerical example for them. Section \ref{sec:empirical} applies our approach to agricultural land use, and Section \ref{sec:conclusion} concludes. The Online Appendix contains the proofs, an identification result based on log differences in payoffs, an additional numerical example, and the details of the empirical application.

\section{Identification in Single-Agent Models}\label{sec:model_single}

We use the following notation throughout the paper. Let $\bs{I}_k$ denote the $k \times k$ identity matrix, $\bs{1}_k$ the $k \times 1$ vector of ones, and $\bs{O}_{m,n}$ the $m \times n$ zero matrix. We suppress subscripts when no confusion arises. We use ``$ \coloneqq$'' to denote ``equals by definition.'' Let $\mathbbm{1}\{A\}$ denote the indicator function that takes the value one when $A$ is true and zero otherwise. For a set $\mathcal{S}$, let $|\mathcal{S}|$ denote its cardinality. We follow the convention that bold lowercase and uppercase letters typically denote vectors and matrices, respectively; stacked vectors such as $\bs{U}$ are also bold uppercase. For a matrix $\bs{A}$, let $\mathrm{rk}(\bs{A})$ denote its rank, and let $\Ker(\bs{A})$ denote its null space. For a square matrix $\bs{A}$, let $\adj(\bs{A})$ denote its adjoint, the transpose of its cofactor matrix, so that $\bs{A}^{-1}=\adj(\bs{A})/\det(\bs{A})$ when $\bs{A}$ is invertible \citep[][p.\ 47]{magnus19book}. With a slight abuse of notation, we write $(\bs{a}_1,\ldots,\bs{a}_k)$ to denote the vector $(\bs{a}_1\t,\ldots,\bs{a}_k\t)\t$ when no confusion arises.

\subsection{Framework and basic assumptions} 

We consider a stationary discrete-time infinite-horizon dynamic discrete choice model. Our presentation of the model follows \citet{abbring20qe}. In each period, an agent observes state variables $(\bs{x},\bs{\ve})$, where $\bs{x} \in \mathcal{X} = \{\bs{x}^1,\ldots,\bs{x}^J\}$ is observed by both the agent and the researcher, and $\bs{\ve} = (\ve_1,\ldots,\ve_K)\t$ is observed only by the agent. The distribution of $\bs{\ve}$ has a density with full support on $\mathbb{R}^K$ and a finite mean. The agent then chooses the action $a$ from the set of alternatives $\mathcal{A} = \{1,2,\ldots, K\}$ to maximize the expected present discounted value of current and future payoffs. Let $\beta \in [0,1)$ denote the time discount factor.

Let $u_k(\bs{x},\bs{\ve})$ be the per-period payoff function when choosing action $k \in \mathcal{A}$. Let $f_k(\bs{x}',\bs{\ve}'|\bs{x},\bs{\ve})$ denote the transition probability of $(\bs{x},\bs{\ve})$ given action $k \in \mathcal{A}$. The agent's value function $V(\bs{x},\bs{\ve})$ solves the Bellman equation
\[
V(\bs{x},\bs{\ve}) = \max_{k \in \mathcal{A}} \left[ u_k(\bs{x},\bs{\ve}) + \beta \int V(\bs{x}',\bs{\ve}') f_k(d\bs{x}',d\bs{\ve}'|\bs{x},\bs{\ve})\right] .
\]
We assume the per-period payoff function is additively separable as $u_k(\bs{x},\bs{\ve}) = u_k(\bs{x}) + \ve_k$ and the transition probability factors as $f_k(\bs{x}',\bs{\ve}'|\bs{x},\bs{\ve}) = g(\bs{\ve}'|\bs{x}') Q_k(\bs{x}'|\bs{x})$. Define the integrated value function (ex ante value function) as $V(\bs{x})  \coloneqq \int V(\bs{x},\bs{\ve}) g(d\bs{\ve}|\bs{x})$. Define the choice-specific value function for each $k \in \mathcal{A}$ as
\begin{equation}\label{v_k_x}
v_k(\bs{x})  \coloneqq u_k(\bs{x}) + \beta \sum_{\bs{x}' \in \mathcal{X}} V(\bs{x}') Q_k(\bs{x}'|\bs{x}).
\end{equation}

The conditional choice probability (CCP) $p_k(\bs{x})$ is the probability that alternative $k$ is the optimal choice given $\bs{x}$,
\[
p_k(\bs{x})  \coloneqq \int \mathbbm{1} \left\{ k = \argmax_{\ell \in \mathcal{A}} \left[ v_\ell (\bs{x}) + \ve_\ell\right]  \right\} g(d\bs{\ve}|\bs{x}).
\]
Define the CCP vector $\bs{p}(\bs{x}) \coloneqq (p_1(\bs{x}),\ldots,p_K(\bs{x}))\t$. \citet[][Lemma 1]{am11em} show that for every $k \in \mathcal{A}$, there exists a function $\psi_k(\cdot\,;\cdot)$ derived only from $g$ such that\footnote{\citet{am11em} assume $g(\bs{\ve}|\bs{x}) = g(\bs{\ve})$, but their proof of Lemma 1 remains valid even if $g$ depends on $\bs{x}$. Proposition 1 of \citet{hotzmiller93restud} also allows $g$ to depend on $\bs{x}$.}
\begin{equation}\label{HotzMiller}
\psi_k\left(\bs{p}(\bs{x});\bs{x}\right) = V(\bs{x}) - v_k(\bs{x}).
\end{equation}
We suppress the second argument in $\psi_k\left(\bs{p}(\bs{x});\bs{x}\right)$ and write $\psi_k(\bs{p}(\bs{x}))$. From Lemma 3 of \citet{am11em}, if $\ve_1,\ldots,\ve_K$ are independently drawn from the type-I extreme value distribution, then $\psi_k(\bs{p}(\bs{x})) = \gamma - \log (p_k(\bs{x}))$, where $\gamma$ is Euler's constant.\footnote{\citet{am11em} also consider other distributions from the generalized extreme value (GEV) family; in the nested logit case, $\psi_k(\bs{p}) = \gamma - \sigma\log(p_k) - (1-\sigma)\log\left(\sum_{k' \in \mathcal{K}} p_{k'}\right)$, where $\mathcal{K}$ is the nest containing $k$, and $\sigma$ captures the degree of correlation among alternatives within $\mathcal{K}$.} Substituting \eqref{HotzMiller} into \eqref{v_k_x} gives
\begin{align*}
v_k(\bs{x}) &= u_k(\bs{x}) + \beta \sum_{\bs{x}' \in \mathcal{X}} \left[ v_k(\bs{x}')+\psi_k (\bs{p}(\bs{x}')) \right] Q_k(\bs{x}'|\bs{x}).
\end{align*}

Let $\bs{v}_k$, $\bs{u}_k$, $\bs{p}_k$, $\bs{\psi}_k$, and $\bs{V}$ be $J\times 1$ vectors with the $j$-th elements $v_k(\bs{x}^j)$, $u_k(\bs{x}^j)$, $p_k(\bs{x}^j)$, $\psi_k(\bs{p}(\bs{x}^j))$, and $V(\bs{x}^j)$, respectively. Let $\bs{Q}_k$ be the $J \times J$ matrix with $(\ell,m)$-th entry $Q_k(\bs{x}^m|\bs{x}^\ell)$. The data identify $\bs{p}_k$ and $\bs{Q}_k$, and a distributional assumption on $g(\bs{\ve}|\bs x)$ identifies $(\bs{\psi}_1,\ldots,\bs{\psi}_K)$ from $(\bs{p}_1,\ldots,\bs{p}_K)$. We therefore treat $\{\bs{p}_k, \bs{Q}_k, \bs{\psi}_k; k \in \mathcal{A}\}$ as known and study the identification of $(\{\bs{v}_k,\bs{u}_k\}_{k=1}^K, \bs{V}, \beta)$.

Stacking \eqref{v_k_x} and \eqref{HotzMiller} over $\bs{x} \in \mathcal{X}$ gives
\begin{align}
\bs{v}_k &= \bs{u}_k + \beta \bs{Q}_k \bs{V}, \quad k=1,\ldots,K, \label{v_k}\\
\bs{\psi}_k &= \bs{V} - \bs{v}_k, \quad k=1,\ldots,K. \label{AM2}
\end{align}
These $2JK$ equations contain the $2JK+J+1$ unknowns $(\{\bs{v}_k,\bs{u}_k\}_{k=1}^K, \bs{V}, \beta )$, so identification requires at least $J+1$ additional restrictions.

Eliminating $\bs{v}_k$ from \eqref{v_k} and \eqref{AM2} gives $\bs{u}_k + \bs{\psi}_k = (\bs{I}- \beta \bs{Q}_k) \bs{V}$. It follows that, for $k=1,\ldots,K-1$, 
\begin{equation}\label{qudiff}
(\bs{I}- \beta \bs{Q}_k)^{-1}\bs{u}_k - (\bs{I}- \beta \bs{Q}_K)^{-1}\bs{u}_K  = - (\bs{I}- \beta \bs{Q}_k)^{-1}\bs{\psi}_k + (\bs{I}- \beta \bs{Q}_K)^{-1}\bs{\psi}_K.
\end{equation}		
Define a $J(K-1) \times JK$ matrix
\begin{equation}\label{eq:Pbeta}
\bs{P}(\beta) \coloneqq
\begin{bmatrix}
(\bs{I}- \beta \bs{Q}_1)^{-1} & \cdots &\bs{O} & -(\bs{I}- \beta \bs{Q}_{K})^{-1}\\
\vdots & \ddots & \vdots & \vdots \\
\bs{O} & \cdots & (\bs{I}- \beta \bs{Q}_{K-1})^{-1} & - (\bs{I}- \beta \bs{Q}_{K})^{-1}
\end{bmatrix}.
\end{equation}
Stack $\bs{u}_k$ and $\bs{\psi}_k$ for $k=1,\ldots,K$ into $JK \times 1$ vectors as
\[
\bs{U} \coloneqq
\begin{bmatrix}
\bs{u}_1 \\
\vdots \\
\bs{u}_{K}
\end{bmatrix}, \quad
\bs{\Psi} \coloneqq
\begin{bmatrix}
\bs{\psi}_1 \\
\vdots \\
\bs{\psi}_{K}
\end{bmatrix}
.
\]
Then, collecting \eqref{qudiff} for $k=1,\ldots, K-1$ gives
\begin{equation} \label{model_rest}
\bs{P}(\beta) \bs{U} = - \bs{P}(\beta) \bs{\Psi} .
\end{equation}
This system of $J(K-1)$ equations in the $JK \times 1$ unknown vector $\bs{U}$ contains all restrictions that the model places on $\beta$. All actions are treated symmetrically, so the results below do not depend on which action is used to construct $\bs{P}(\beta)$ in \eqref{eq:Pbeta}. Section \ref{sec: util_diff} adds linear restrictions on $\bs{U}$ that identify $\beta$.

\subsection{Identification of $\beta$ by economic restrictions on per-period payoff}\label{sec: util_diff}

Economic theory often restricts per-period payoffs, for example through homogeneity or monotonicity, and these restrictions are linear in $\bs{U}$. Section \ref{sec:examples} shows that homogeneity gives equalities $\bs{r}\t\bs{U}=0$ and that monotonicity and concavity give inequalities $\bs{r}\t\bs{U} \geq 0$. We start with equalities.
\begin{assumption}\label{assn_eq}
The payoff function $\bs{U}$ satisfies $\bs{R}\bs{U} =\bs{c}$ for a known $r\times JK$ full row rank matrix $\bs{R}$ and a known $r \times 1$ vector $\bs{c}$ with $r \geq J$.
\end{assumption}
As is standard, we normalize the scale of $\bs{\ve}$ to one; the identified set of $\beta$ does not depend on this normalization when $\bs{c}=\bs{0}$.

Define the $[J(K-1) +r] \times JK$ matrix and the $[J(K-1) +r] \times 1$ vector
\begin{equation}\label{R_beta_defn}
\bs{R}(\beta) \coloneqq 
\begin{bmatrix}
\bs{P}(\beta) \\
\bs{R}
\end{bmatrix}, \quad
\bs{c}(\beta) \coloneqq 
\begin{bmatrix}
- \bs{P}(\beta) \bs{\Psi} \\
\bs{c}
\end{bmatrix} .
\end{equation}
Combining the model restriction \eqref{model_rest} with Assumption \ref{assn_eq} yields 
\begin{equation}\label{eq:uplus_rest}
\bs{R}(\beta) \bs{U} = \bs{c}(\beta).
\end{equation} 
Partition $\bs{R}$ and $\bs{c}$ and define $\bs{R}_1(\beta)$ and $\bs{c}_1(\beta)$ as
\begin{equation}\label{RC_partition}
\bs{R} = 
\begin{bmatrix}
\bs{R}_{1} \\
\bs{R}_{2} 
\end{bmatrix}, \quad
\bs{c} = 
\begin{bmatrix}
\bs{c}_{1} \\
\bs{c}_{2}
\end{bmatrix} , \quad
\bs{R}_1(\beta) \coloneqq 
\begin{bmatrix} 
\bs{P}(\beta)\\
\bs{R}_{1} 
\end{bmatrix} , \quad
\bs{c}_1(\beta) \coloneqq 
\begin{bmatrix}
- \bs{P}(\beta) \bs{\Psi} \\
\bs{c}_1
\end{bmatrix} .
\end{equation}
Here, $\bs{R}_{1}$ is $J \times JK$, $\bs{R}_{2}$ is $(r-J)\times JK$, $\bs{c}_{1}$ is $J \times 1$, $\bs{c}_2$ is $(r-J) \times 1$, and $\bs{R}_{1}(\beta)$ is $JK \times JK$. With this notation, we can rewrite \eqref{eq:uplus_rest} as
\begin{equation}\label{R1R2}
\bs{R}_1(\beta) \bs{U}=\bs{c}_1(\beta), \quad \bs{R}_2 \bs{U} = \bs{c}_2.
\end{equation}
The following rank condition can be verified because $\bs{R}(\beta)$ is identified from the data.
\begin{assumption}\label{assn_eq_new}
For all $\beta \in [0,1)$, $\bs{R}(\beta)$ defined in \eqref{R_beta_defn} has full column rank.
\end{assumption}
Because $\bs{R}(\beta)$ has full column rank and $\bs{P}(\beta)$ has full row rank $J(K-1)$, for each $\beta$ we can select $J$ rows of $\bs{R}$ so that $\bs{R}_1(\beta)$ is nonsingular. Without loss of generality, we assume the rows in $\bs{R}_1$ are so chosen. If no single selection works for every $\beta$, we can partition $[0,1)$ into intervals such that the selection of $\bs{R}_1$ remains constant within each interval and compute the identifying equation for each interval.
Partition $\bs{R}_1$ as $\bs{R}_{1} = [\bs{R}_{11}, \ldots, \bs{R}_{1K}]$. Define
\begin{equation}\label{eq:def_MQ}
\bs{M}(\beta) \coloneqq \sum_{k=1}^{K} \bs{R}_{1k} (\bs{I} - \beta \bs{Q}_{k}), \quad \bs{Q}(\beta)
\coloneqq
\begin{bmatrix}
\bs{I} - \beta \bs{Q}_1\\
\vdots \\
\bs{I} - \beta \bs{Q}_K
\end{bmatrix} .
\end{equation}

\begin{proposition} \label{prop_equality_new}
Suppose Assumptions \ref{assn_eq} and \ref{assn_eq_new} hold. (a) If $\beta$ is known, then $\bs{U}$ is identified as $\bs{U} = \bs{R}_1(\beta)^{-1}\bs{c}_1(\beta)$. (b) If $r >J$, the identified set of $\beta$ is the intersection of $[0,1)$ with the roots of the following system of at most $(r-J)$ polynomials, each of degree at most $J$
\begin{equation} \label{eq:prop1_new}
\det(\bs{M}(\beta)) \bs{c}_2 + \det(\bs{M}(\beta)) \bs{R}_2 \bs{\Psi} - \bs{R}_2 \bs{Q}(\beta) \adj(\bs{M}(\beta)) (\bs{R}_1 \bs{\Psi} + \bs{c}_1) = \bs{0},
\end{equation}
provided that the left-hand side is not identically zero. Further, the per-period payoff corresponding to $\beta$ in the identified set is identified as $\bs{U} = \bs{R}_1(\beta)^{-1}\bs{c}_1(\beta)$.
\end{proposition}
\textit{Proof sketch.} Solving the first block of \eqref{R1R2} and using \eqref{RC_partition} yields $\bs{U}=\bs{R}_1(\beta)^{-1}\bs{c}_1(\beta)=-\bs{\Psi} + \bs{Q}(\beta)\bs{M}(\beta)^{-1}(\bs{R}_1\bs{\Psi}+\bs{c}_1)$. Substituting this into $\bs{R}_2\bs{U}=\bs{c}_2$, writing $\bs{M}(\beta)^{-1}=\adj(\bs{M}(\beta))/\det(\bs{M}(\beta))$, and multiplying by $\det(\bs{M}(\beta))$ gives \eqref{eq:prop1_new}. Online Appendix \ref{sec:proofs} gives the details.

Later results that identify $\beta$ also require that the left-hand side not be identically zero, and we do not repeat this condition.

The identified set of $\beta$ is also characterized as the intersection of $[0,1)$ and the roots of the polynomial greatest common divisor (GCD) of the $r-J$ polynomials in \eqref{eq:prop1_new}. Hence, when $r-J \geq 2$, the degree of the GCD is smaller than $J$ unless the $r-J$ polynomials have all their roots in common. The number of distinct real roots of the identifying polynomial in $[0,1)$ can be determined using Sturm's theorem \citep[][Theorem 2.7, p.\ 454]{lang02book}, which, for a polynomial $h(\beta)$, constructs a sequence of polynomials from $h(\beta)$ and $h'(\beta)$ through repeated polynomial division and counts the sign changes at the endpoints of the interval.

We rule out $\beta=1$ because $\bs{P}(\beta)$ is not defined there, as $\bs{I} - \bs{Q}_k$ is singular. The polynomials in \eqref{eq:prop1_new} are nevertheless defined at $\beta=1$, and they vanish there.
\begin{proposition} \label{prop_beta_1}
At $\beta=1$, we have $\det\left( \bs{M}(\beta)\right)=0$ and $\bs{Q}(\beta) \adj(\bs{M}(\beta))=\bs{O}$. Hence, all terms on the left-hand side of \eqref{eq:prop1_new} vanish at $\beta=1$. 
\end{proposition}

Because $\beta=1$ always solves the degree-$J$ system in Proposition \ref{prop_equality_new}, the identified set of $\beta$ in $[0,1)$ contains at most $J-1$ elements.

When $r \ge J$ linear restrictions on $\bs{U}$ are available, inequality restrictions on $\bs{U}$ can further tighten the identified set of $\beta$.

\begin{assumption}\label{assn_ineq}
The payoff function $\bs{U}$ satisfies
\[
\bs{R}_{\text{ineq}} \bs{U} \ge \bs{c}_{\text{ineq}},
\]
where $\bs{R}_{\text{ineq}}$ is a known $r_{\text{ineq}} \times JK$ full row rank matrix and $\bs{c}_{\text{ineq}}$ is a known $r_{\text{ineq}} \times 1$ vector.
\end{assumption}

\begin{proposition}\label{prop_ineq}
Suppose Assumptions~\ref{assn_eq}--\ref{assn_ineq} hold. Define the $r_{\text{ineq}} \times 1$ vector of polynomials
\begin{equation}\label{eq:prop2}
\bs{g}_{\text{ineq}}(\beta) \coloneqq \det(\bs{M}(\beta)) \bs{c}_{\text{ineq}}
+ \det(\bs{M}(\beta)) \bs{R}_{\text{ineq}} \bs{\Psi}
- \bs{R}_{\text{ineq}} \bs{Q}(\beta)
\adj(\bs{M}(\beta))
(\bs{R}_1 \bs{\Psi} + \bs{c}_1),
\end{equation}
whose entries have degree at most $J$ in $\beta$. Then the identified set of $\beta$ equals the intersection of $[0,1)$, the set of $\beta$ satisfying
\[
\bs{g}_{\text{ineq}}(\beta) \le \bs{0} \ \text{ if } \det(\bs{M}(\beta)) > 0,
\qquad
\bs{g}_{\text{ineq}}(\beta) \ge \bs{0} \ \text{ if } \det(\bs{M}(\beta)) < 0 ,
\]
and, when $r > J$, the set of roots of \eqref{eq:prop1_new}.
\end{proposition}

When $r=J$, the identified set of $\beta$ is determined solely by these sign conditions and is a union of intervals. When $r>J$, the sign conditions eliminate roots of \eqref{eq:prop1_new} that violate them.

\subsection{Relationship to the framework with $\bs{u}_K=\bs{0}$}\label{sec:standard}

This section discusses the identification of $\beta$ under the normalization $\bs{u}_K=\bs{0}$ and relates existing results to our framework. The normalization $\bs{u}_K=\bs{0}$ is common in empirical applications because the per-period payoff cannot be identified from the conditional choice probabilities alone without additional restrictions \citep[][Proposition 2]{magnac02em}, and because data on $\bs{u}_K$ or $\bs{V}$ are rarely available.\footnote{\citet{Kalouptsidi14aer, Kalouptsidi18res} utilize external data on entry costs and scrap values to estimate the value function $\bs{V}$, thereby avoiding the need to assume $\bs{u}_K=\bs{0}$.}
This assumption is not innocuous, however, as it can affect counterfactual predictions and other parameter estimates \citep{aguirregabiriasuzuki14qme, noretstang14restud, Kalouptsidi21qe}.

When $\bs{u}_K$ is known, the identifying equations \eqref{eq:prop1_new} and \eqref{eq:prop2} simplify as follows.
\begin{remark}[Known $\bs{u}_K$]\label{rem:standard}
Suppose $\bs{u}_K = \bs{c}_1$ for a known $\bs{c}_1$, which includes $\bs{u}_K = \bs{0}$ as a special case. This yields $J$ restrictions $\bs{R}_1 \bs{U} = \bs{c}_1$ with $\bs{R}_1 = [ \bs{O}_{J, J(K-1)}, \bs{I}_J]$. Then $\bs{R}_1(\beta)$ has full rank and $\bs{M}(\beta) = \bs{I}- \beta \bs{Q}_K$ has a positive determinant. Consequently, \eqref{eq:prop1_new} and \eqref{eq:prop2} become
\begin{align} 
&\det(\bs{I}-\beta\bs{Q}_K) \bs{c}_2 + \det(\bs{I}- \beta \bs{Q}_K) \bs{R}_2 \bs{\Psi}- \bs{R}_2 \bs{Q}(\beta) \adj (\bs{I}-\beta\bs{Q}_K) ( \bs{\psi}_K + \bs{c}_1) = \bs{0}, \label{ident_eq_known} \\
&\det(\bs{I}-\beta\bs{Q}_K) \bs{c}_{\text{ineq}} + \det(\bs{I}-\beta\bs{Q}_K) \bs{R}_{\text{ineq}} \bs{\Psi} - \bs{R}_{\text{ineq}} \bs{Q}(\beta) \adj(\bs{I}-\beta\bs{Q}_K)( \bs{\psi}_K + \bs{c}_1) \leq \bs{0}, \label{ident_ineq_known}
\end{align}
and $\beta$ is identified by restrictions on $\bs{u}_1, \ldots,\bs{u}_{K-1}$ implied by $\bs{R}_2\bs{U} = \bs{c}_2$ and $\bs{R}_{\text{ineq}} \bs{U} \ge \bs{c}_{\text{ineq}}$.
\end{remark}

Assuming $\bs{u}_K = \bs{0}$, it follows from \eqref{qudiff} that, for $k=1,\ldots,K-1$,\footnote{This equation corresponds to (3) in \citet{Kalouptsidi21qe} except that we impose $\bs{u}_K=\bs{0}$.}
\begin{equation}\label{ident_eqn1}
\bs{u}_k = -\bs{\psi}_k + (\bs{I} - \beta \bs{Q}_k) (\bs{I} - \beta \bs{Q}_K)^{-1}\bs{\psi}_K.
\end{equation}
\citet[][Result 5]{berrytamer06book} show that, in a monopoly exit model where exit is terminal, $\beta$ is identified from \eqref{ident_eqn1} if the payoff is known at one state.

\citet{abbring20qe} derive the identified set of $\beta$ under $\bs{u}_K = \bs{0}$ using an exclusion restriction of the form $u_k(\bs{x}_a) = u_\ell(\bs{x}_b)$ for some known choices $k \in \mathcal{A}\setminus \{K\}, \ell \in \mathcal{A}$ and known states $\bs{x}_a,\bs{x}_b \in \mathcal{X}$, where either $k \neq \ell$, $\bs{x}_a \neq \bs{x}_b$, or both. Under this exclusion restriction, \eqref{ident_eqn1} implies
\begin{equation} \label{abbring_thm1}
\psi_k(\bs{x}_a) - \psi_\ell(\bs{x}_b) = [(\bs{I}  -   \beta \bs{Q}_k)(\bs{x}_a) - (\bs{I} - \beta \bs{Q}_\ell)(\bs{x}_b) ] (\bs{I} - \beta\bs{Q}_K)^{-1} \bs{\psi}_K,
\end{equation}
where $(\bs{I} - \beta \bs{Q}_k)(\bs{x}_a)$ denotes the row of $\bs{I}-\beta\bs{Q}_k$ corresponding to $\bs{x}_a$, and similarly for $(\bs{I} - \beta \bs{Q}_\ell)(\bs{x}_b)$.\footnote{Equation \eqref{abbring_thm1} is equivalent to equation (12) in \citet{abbring20qe}; the equivalence follows from $\psi_K(\bs{x}_a)=(\bs{I} - \beta \bs{Q}_K)(\bs{x}_a)(\bs{I} - \beta\bs{Q}_K)^{-1}\bs{\psi}_K$.} The solution set to \eqref{abbring_thm1}, whose right-hand side is a convergent power series in $\beta$, is a closed discrete subset of $[0, 1)$ \citep{abbring20qe} with at most $J$ elements \citep[][Theorem 6]{abbring20wp}. In our framework, this exclusion restriction is a row of $\bs{R}_2$ with entries $1$ and $-1$ at the two equalized elements, and $\bs{c}_2=0$.

\subsection{Identification of $\beta$ without identifying $\bs{U}$}\label{sec:ident_rank_defi}

In some models, shape restrictions leave some elements of $\bs{U}$ unidentified no matter how many are imposed. This occurs, for example, in the entry-exit model with unrestricted scrap value in Section \ref{sec:scrap_value_example}. This section shows that $\beta$ can nonetheless be identified in such models, without identifying all the elements of $\bs{U}$.

Suppose we have $r$ linearly independent restrictions on $\bs{U}$ of the form $\bs{R}\bs{U} = \bs{c}$. We collect these restrictions as
\begin{equation} \label{RU_rank_deficient}
\bs{R}(\beta) \bs{U} = \bs{c}(\beta), \
 \text{ where }\ \bs{R}(\beta) = 
\begin{bmatrix}
\bs{P}(\beta) \\
\bs{R}
\end{bmatrix}, \quad
\bs{c}(\beta) =
\begin{bmatrix}
- \bs{P}(\beta) \bs{\Psi} \\
\bs{c}
\end{bmatrix} .
\end{equation}
Here, $\bs{R}(\beta)$ is $\overline{r} \times JK$, where $\overline{r}\coloneqq J(K-1)+r$. Let $q \coloneqq \mathrm{rk}(\bs{R}(\beta))$ denote its rank; $q \geq J(K-1)$ because the rows of $\bs{P}(\beta)$ are linearly independent. When $q< JK$, some elements of $\bs{U}$ are not identified even when $\beta$ is known. Normalizing some of these elements yields a system of full column rank. We then apply Proposition \ref{prop_equality_new} to this system and show that the identifying polynomials do not depend on the normalized values. Proposition \ref{prop_ident_partial} below formalizes this argument.
\begin{assumption} \label{assn:partial}
(a) The payoff function $\bs{U}$ satisfies $\bs{R}\bs{U} =\bs{c}$ for a known $r\times JK$ full row rank matrix $\bs{R}$ and a known $r \times 1$ vector $\bs{c}$. (b) $J(K-1) < q < JK$ for all $\beta \in [0,1)$. (c) $\overline{r} > q$. 
\end{assumption}
In Assumption \ref{assn:partial}, $J(K-1) < q$ requires that some row of $\bs{R}$ is not in the row space of $\bs{P}(\beta)$, and (c) requires that at least one restriction is available beyond those that determine the identified components of $\bs{U}$, as $r>J$ does in Proposition \ref{prop_equality_new}(b). 

Let $s \coloneqq q - J(K-1)>0$. Because $\bs{P}(\beta)$ has full row rank, we can select $s$ rows of $\bs{R}$ that, stacked below $\bs{P}(\beta)$, give a matrix of rank $q$. Collect them in $\bs{R}_1$, partition $\bs{R}$ and $\bs{c}$, and define $\bs{R}_1(\beta)$ and $\bs{c}_1(\beta)$ as in \eqref{RC_partition} but with $s$ rows in $\bs{R}_1$ instead of $J$,
\begin{equation}\label{eq:R_partition}
\bs{R} = \begin{bmatrix} \bs{R}_{1} \\ \bs{R}_{2} \end{bmatrix},
\quad
\bs{c} = 
\begin{bmatrix} \bs{c}_1 \\ \bs{c}_2 \end{bmatrix},  \quad
\bs{R}_1(\beta) \coloneqq 
\begin{bmatrix} 
\bs{P}(\beta)\\
\bs{R}_{1} 
\end{bmatrix} , \quad
\bs{c}_1(\beta) \coloneqq 
\begin{bmatrix}
- \bs{P}(\beta) \bs{\Psi} \\
\bs{c}_1
\end{bmatrix} ,
\end{equation}
where $\bs{R}_1$ is $s \times JK$, $\bs{R}_2$ is $(r-s) \times JK$, $\bs{c}_1$ is $s \times 1$, and $\bs{c}_2$ is $(r-s) \times 1$. Under Assumption \ref{assn:partial}, there exists a reordering of the columns of $\bs{R}(\beta)$, equivalently of the elements of $\bs{u}_K$, such that $\bs{R}(\beta)$ and $\bs{c}(\beta)$ are partitioned as
\begin{equation}\label{eq:Rbeta_partition}
\bs{R}(\beta) = 
\begin{bmatrix} \bs{R}_{1}(\beta) \\ \bs{R}_{2} \end{bmatrix}
=
\begin{bmatrix} \bs{R}_{1L}(\beta) & \bs{R}_{1R}(\beta) \\ \bs{R}_{2L} & \bs{R}_{2R} \end{bmatrix}
, \quad \bs{c}(\beta) = \begin{bmatrix} \bs{c}_1(\beta) \\ \bs{c}_2 \end{bmatrix},
\end{equation}
where $\bs{R}_{1L}(\beta)$ is $q \times q$ and nonsingular, $\bs{R}_{1R}(\beta)$ is $q \times (JK-q)$, $\bs{R}_{2L}$ is $(\overline{r}-q) \times q$, $\bs{R}_{2R}$ is $(\overline{r}-q) \times (JK-q)$, $\bs{c}_1(\beta)$ is $q \times 1$, and $\bs{c}_2$ is $(\overline{r}-q) \times 1$. This partition generalizes \eqref{R1R2}, which is the case $q=JK$ and $s=J$. Note that no reordering of the first $J(K-1)$ rows or columns of $\bs{R}(\beta)$ is necessary because $\bs{P}(\beta)$ has full row rank and its top-left $J(K-1) \times J(K-1)$ block is block diagonal. As in Section \ref{sec: util_diff}, if $q$ and the selection of $\bs{R}_{1L}(\beta)$ change with $\beta$, we can partition $[0,1)$ into intervals on which they remain constant and compute the identifying equation for each interval.

The columns of $\bs{R}_{1L}(\beta)$ are the first $J(K-1)$ columns of $\bs{R}(\beta)$ and $s$ columns of its last $J$ columns, which correspond to $\bs{u}_1,\ldots,\bs{u}_{K-1}$ and the elements of $\bs{u}_K$ at $s$ states. Let $\mathcal{X}_L \subset \mathcal{X}$ denote the set of these states. The columns of $\bs{R}_{1R}(\beta)$ correspond to the elements of $\bs{u}_K$ at $\bs{x}\in\mathcal{X}_R\coloneqq\mathcal{X}\setminus\mathcal{X}_L$. Let $\bs{E}_{\mathcal{X}_R}$ be the $(J-s)\times J$ matrix that selects the states in $\mathcal{X}_R$ in the order of the columns of $\bs{R}_{1R}(\beta)$. For a $(J-s)\times 1$ vector $\bs{v}$, we normalize the elements of $\bs{u}_K$ at $\bs{x}\in\mathcal{X}_R$ to $\bs{v}$ by adding $\bs{E}_{\mathcal{X}_R}\bs{u}_K=\bs{v}$ to $\bs{R}_1\bs{U}=\bs{c}_1$ and define
\begin{equation}\label{eq:Rtilde}
\widetilde{\bs{R}}_1 \coloneqq \begin{bmatrix} \multicolumn{2}{c}{\bs{R}_1} \\ \bs{O}_{J-s,J(K-1)} & \bs{E}_{\mathcal{X}_R} \end{bmatrix}, \quad \widetilde{\bs{c}}_1 \coloneqq \begin{bmatrix}\bs{c}_1\\ \bs{v}\end{bmatrix}, \quad \widetilde{\bs{M}}(\beta)\coloneqq \widetilde{\bs{R}}_1\bs{Q}(\beta),
\end{equation}
where $\bs{Q}(\beta)$ is defined in \eqref{eq:def_MQ}. Applying Proposition \ref{prop_equality_new} to this augmented system yields the following proposition, which has the same degree bound as Proposition \ref{prop_equality_new}.
\begin{proposition}\label{prop_ident_partial}
Suppose Assumption \ref{assn:partial} holds. Then, for any $\bs{v}$, the identified set of $\beta$ is the intersection of $[0,1)$ with the roots of the following system of at most $(\overline{r}-q)$ polynomials, each of degree at most $J$,
\begin{equation} \label{eq:prop_partial}
\det(\widetilde{\bs{M}}(\beta)) \bs{c}_2 + \det(\widetilde{\bs{M}}(\beta)) \bs{R}_{2} \bs{\Psi}
- \bs{R}_{2}\bs{Q}(\beta)\adj(\widetilde{\bs{M}}(\beta))
(\widetilde{\bs{R}}_{1}\bs{\Psi} + \widetilde{\bs{c}}_1)
= \bs{0},
\end{equation}
and these polynomials do not depend on $\bs{v}$.
\end{proposition}
\textit{Proof sketch.} Given $\beta$, the restrictions $\bs{R}(\beta)\bs{U}=\bs{c}(\beta)$ determine $\bs{U}$ only up to the values of $\bs{u}_K$ at the states in $\mathcal{X}_R$, and these values can be changed freely without violating the restrictions. Fixing them at $\bs{v}$ therefore leaves unchanged the set of $\beta$ for which the restrictions have a solution, and the augmented system has full column rank. Proposition \ref{prop_equality_new} applied to it then gives \eqref{eq:prop_partial}. Because the values at $\mathcal{X}_R$ do not affect whether the restrictions can be satisfied, the polynomials do not depend on $\bs{v}$. Online Appendix \ref{sec:proofs} gives the details.

\begin{proposition} \label{prop_beta_1_partial}
At $\beta=1$, every polynomial in \eqref{eq:prop_partial} vanishes. Hence, the identified set of $\beta$ contains at most $J-1$ elements.
\end{proposition}

Inequality restrictions $\bs{R}_{\text{ineq}}\bs{U}\ge\bs{c}_{\text{ineq}}$ can also be used here. A root of \eqref{eq:prop_partial} stays in the identified set only if some $\bs{U}$ satisfies $\bs{R}(\beta)\bs{U}=\bs{c}(\beta)$ together with these inequalities, which is a linear feasibility problem. Unlike the identifying polynomials, whether the inequalities hold depends on $\bs{v}$ unless the rows of $\bs{R}_{\text{ineq}}$ lie in the row space of $\bs{R}(\beta)$.

\citet{komarova18qe} consider the parametric payoff specification $\bs{U} = \bs{H} \bs{\theta}$, where $\bs{\theta}$ is a $d_\theta$-dimensional parameter vector and $\bs{H}$ is a known $JK \times d_\theta$ design matrix. They identify $(\beta,\bs{\theta})$ as a solution to
\[
\bs{a}(\beta) - \bs{B}(\beta)\bs{\theta} =\bs{0},
\]
where $\bs{a}(\beta)$ and $\bs{B}(\beta)$ are a $JK\times 1$ vector and a $JK\times d_\theta$ matrix, respectively, of nonlinear functions of $\beta$ that depend on $\bs{H}$ and the transition matrices. When $\bs{B}(\beta)$ is rank-deficient for all $\beta$, some components of $\bs{\theta}$ are not identified even when $\beta$ is known. In \citet{komarova18qe}, ``normalization'' refers to fixing the values of such unidentified parameters. Proposition~1 of \citet{komarova18qe} shows that this choice does not affect the identification of $\beta$. In our Proposition \ref{prop_ident_partial}, the identifying polynomials do not depend on the values of the normalized components of $\bs{U}$, which generalizes their Proposition~1 to the nonparametric case. Counterfactual predictions that depend on these components, however, depend on the normalization.

\section{Finite Dependence}\label{sec:fd}

\citet{AltugMiller98restud} and \citet{am11em} introduce finite dependence and show that it reduces the computational cost of estimating dynamic discrete choice models. A model exhibits finite dependence if two action sequences with different initial actions lead to the same state distribution after a finite number of periods. This section shows that finite dependence can reduce the degree of the identifying polynomial.

We use three versions of finite dependence from the literature, which differ in the initial action-state pairs.
\begin{enumerate}
\item \citep[][p.\ 1836]{am11em} Starting from state $\bs{x}_a$, taking two different actions, $k_a$ and $k_b$, from $\mathcal{A}\setminus \{K\}$, followed by $K$ for the next $\rho$ periods results in the same state distribution.
\[
\bs{Q}_{k_a}(\bs{x}_a) \bs{Q}_K^\rho = \bs{Q}_{k_b}(\bs{x}_a)\bs{Q}_K^\rho.
\]

\item \citep{abbring20qe} Starting from state $\bs{x}_a$, taking actions $k_a \in \mathcal{A} \setminus \{K\}$ or $K$, followed by $K$ for the next $\rho$ periods results in the same state distribution. Furthermore, the same applies to another pair $(\bs{x}_b,k_b)$ with $k_b \in \mathcal{A} \setminus \{K\}$.
\[
\bs{Q}_{k_a}(\bs{x}_a) \bs{Q}_K^\rho = \bs{Q}_{K}(\bs{x}_a)\bs{Q}_K^\rho \text{ and } \bs{Q}_{k_b}(\bs{x}_b) \bs{Q}_K^\rho = \bs{Q}_{K}(\bs{x}_b)\bs{Q}_K^\rho.
\]
This condition allows $\bs{x}_a=\bs{x}_b$ or $k_a=k_b$. \citet[][Theorem 2]{abbring20qe} show that this finite dependence can reduce the degree of the identifying polynomial \eqref{abbring_thm1} to $\rho$.
 
\item \citep[][p.\ 483]{abbring20qe} Starting from two different states $\bs{x}_a$ and $\bs{x}_b$, taking a common action $k_a \in \mathcal{A} \setminus \{K\}$, followed by $K$ for the next $\rho$ periods results in the same state distribution. Furthermore, the same holds when $k_a$ is replaced with $K$.
\[
\bs{Q}_{k_a}(\bs{x}_a) \bs{Q}_K^\rho = \bs{Q}_{k_a}(\bs{x}_b)\bs{Q}_K^\rho \text{ and } \bs{Q}_{K}(\bs{x}_a) \bs{Q}_K^\rho = \bs{Q}_{K}(\bs{x}_b)\bs{Q}_K^\rho.
\]
\end{enumerate}
Each of the three conditions implies the following assumption. For example, setting $\bs{x}_a=\bs{x}_b$ in \eqref{assn_finite_eq} gives the first specification.
\begin{assumption}\label{assn_finite}
$\{\bs{Q}_k: k \in \mathcal{A}\}$ satisfies, for two action-state pairs $(k_a,\bs{x}_a)$ and $(k_b,\bs{x}_b)$ with $k_a,k_b \in \mathcal{A}\setminus \{K\}$ and some finite positive integer $\rho$,
\begin{equation}\label{assn_finite_eq}
\left[\bs{Q}_{k_a}(\bs{x}_a)- \bs{Q}_{k_b}(\bs{x}_b)\right]\bs{Q}_K^\rho = \left[\bs{Q}_{K}(\bs{x}_a)- \bs{Q}_{K}(\bs{x}_b)\right]\bs{Q}_K^\rho.
\end{equation}
\end{assumption}
Dividing both sides of the identifying equation \eqref{eq:prop1_new} by $\det(\bs{M}(\beta))$ gives
\begin{equation}\label{eq:prop_fd}
\bs{c}_2 + \bs{R}_2 \bs{\Psi} - \bs{R}_2 \bs{Q}(\beta) \bs{M}(\beta)^{-1} (\bs{R}_1 \bs{\Psi} + \bs{c}_1) = \bs{0}.
\end{equation}
Recall that $(\bs{I}-\beta\bs{Q}_k)^{-1} = \bs{I}+\beta\bs{Q}_k +\beta^2\bs{Q}_k^2+ \cdots$ for $\beta\in[0,1)$, so the entries of $\bs{R}_2 \bs{Q}(\beta) \bs{M}(\beta)^{-1}$ are in general power series in $\beta$. The following lemma shows that finite dependence, together with structure in $\bs{R}_1$ and $\bs{R}_2$, reduces the degree of $\bs{R}_2 \bs{Q}(\beta) \bs{M}(\beta)^{-1}$ to at most $\rho$. 
Restriction \eqref{restriction_homo3} in Example \ref{example_homo3} satisfies the conditions of the lemma.

\begin{lemma}\label{lem_fd_suff}
Suppose $\bs{R}_1 = [\bs{O}_{J,J(K-1)},\, \bs{I}_J]$, so that $\bs{u}_K$ is normalized and $\bs{M}(\beta) = \bs{I} - \beta \bs{Q}_K$. Suppose further that each restriction in $\bs{R}_2$ is a linear combination of utility differences $u_{k_{a}}(\bs{x}_{a}) - u_{k_{b}}(\bs{x}_{b})$ whose action-state pairs $\{(k_{a}, \bs{x}_{a}),(k_{b}, \bs{x}_{b})\}$ satisfy Assumption \ref{assn_finite}. Then each entry of $\bs{R}_2 \bs{Q}(\beta) \bs{M}(\beta)^{-1}$ is a polynomial in $\beta$ of degree at most $\rho$.
\end{lemma}

This degree bound carries over to the identifying system \eqref{eq:prop_fd}.
\begin{proposition}\label{prop_fd}
Suppose that the assumptions of Proposition \ref{prop_equality_new} hold and that each entry of $\bs{R}_2 \bs{Q}(\beta) \bs{M}(\beta)^{-1}$ is a polynomial in $\beta$ of degree at most $\rho$. Then the identified set of $\beta$ equals the intersection of $[0,1)$ with the roots of \eqref{eq:prop_fd}, a system of at most $(r-J)$ polynomials each of degree at most $\rho$.
\end{proposition}

When $\bs{R}(\beta)$ does not have full column rank, Proposition \ref{prop_fd} applies to the augmented system of Proposition \ref{prop_ident_partial}: if each entry of $\bs{R}_2\bs{Q}(\beta)\widetilde{\bs{M}}(\beta)^{-1}$ is a polynomial in $\beta$ of degree at most $\rho$, then the identified set of $\beta$ equals the intersection of $[0,1)$ with the roots of the system of polynomials of degree at most $\rho$
\begin{equation}\label{eq:prop_fd_partial}
\bs{c}_2 + \bs{R}_2 \bs{\Psi} - \bs{R}_2 \bs{Q}(\beta) \widetilde{\bs{M}}(\beta)^{-1} (\widetilde{\bs{R}}_1 \bs{\Psi} + \widetilde{\bs{c}}_1) = \bs{0}.
\end{equation}

\section{Examples of Economic Restrictions on Per-Period Payoff Functions}\label{sec:examples}

This section gives examples of payoff restrictions that take the forms in Section \ref{sec:model_single}. In many applied economic models, payoff functions satisfy nonparametric restrictions such as monotonicity, concavity, and homogeneity of degree $\nu$; see \citet[Section 5]{matzkin92em} for examples. They give equalities $\bs{r}_1\t \bs{U}=c_1$ or inequalities $\bs{r}_2\t \bs{U}\geq c_2$. 
These restrictions have economic content: the payoff $u_k(\bs{x})$ is typically an indirect function, such as a maximized profit or a minimized cost given the state and action, and inherits properties of technology and preferences.\footnote{For example, maximizing $z w^{\alpha} l^{\gamma}-l$ over labor $l$, with $\alpha>0$ and $0<\gamma<1$, gives a profit proportional to $z^{1/(1-\gamma)}w^{\alpha/(1-\gamma)}$, which is increasing, supermodular, convex in $z$, concave in $w$ if and only if $\alpha+\gamma\le 1$, and homogeneous in $w$ of unknown degree (Example \ref{example_homo_unknown} in Online Appendix \ref{sec:log_diff}).}

We partition the state variable $\bs{x}$ as $\bs{x}=(\bs{w},\bs{z})$, where $\bs{z}$ may be empty, and write the payoff function as $u_k(\bs{w},\bs{z})$. Assume $\mathcal{X}=\mathcal{W}\times \mathcal{Z}$, where $\mathcal{W} = \{\bs{w}^1, \bs{w}^2, \dots, \bs{w}^{m_w}\}$ and $\mathcal{Z} = \{\bs{z}^1, \bs{z}^2, \dots, \bs{z}^{m_z}\}$. When $\bs{z}$ is empty, we let $m_z =1$. For simplicity, the domain of $\bs{w}$ does not depend on $\bs{z}$, but the results remain valid if it does. The following examples restrict the payoff of a single action $k$. A restriction that holds for several actions applies to each of them. 

\subsection{Equality restrictions based on homogeneity}\label{subsec:equality}

We first consider payoffs that are homogeneous in $\bs{w}$. We assume that $u_k(\bs{w},\bs{z})$ is defined at the $L \le m_w$ points $\widetilde{\bs{w}}, \lambda_2 \widetilde{\bs{w}}, \ldots, \lambda_L \widetilde{\bs{w}} \in \mathcal{W}$ for some $\lambda_2,\ldots,\lambda_L \in \mathbb{R}^+\setminus\{1\}$. 

\begin{example}[Homogeneous function of known degree $\nu$]\label{example_homo}

Homogeneous functions model returns to scale in consumer and production theory. For example, cost and profit functions are homogeneous of degree one in prices for any technology \citep[Propositions~5.C.1 and~5.C.2]{MasColell1995}.

Suppose the payoff function $u_k(\bs{w},\bs{z})$ is known to be homogeneous of degree $\nu$ in $\bs{w}$, i.e., for all $(\bs{w},\bs{z}) \in \mathcal{X}$ and all $\lambda > 0$,
\[
u_k(\lambda \bs{w}, \bs{z}) = \lambda^\nu u_k(\bs{w}, \bs{z}).
\]
This gives $m_z (L-1)$ linear equality restrictions on $\bs{U}$ of the form
\[
u_k(\lambda_\ell \widetilde{\bs{w}}, \bs{z}) - \lambda_\ell^\nu u_k( \widetilde{\bs{w}},\bs{z}) = 0, \quad \bs{z} \in \mathcal{Z},\ \ell=2,\ldots,L.
\]
\end{example}

\begin{example}[Log transformation of homogeneous function of degree $\nu$]\label{example_homo2}
Suppose $u_k(\bs{w}, \bs{z}) = \log f_k(\bs{w}, \bs{z})$, where $f_k(\bs{w},\bs{z})$ is homogeneous of degree $\nu$ in $\bs{w}$ with possibly unknown $\nu$. Then, for all $(\bs{w},\bs{z}) \in \mathcal{X}$ and all $\lambda > 0$,
\[
u_k(\lambda \bs{w}, \bs{z}) = \log f_k(\lambda \bs{w}, \bs{z}) = \nu \log(\lambda) + \log f_k(\bs{w},\bs{z}) = \nu\log \lambda + u_k(\bs{w},\bs{z}).
\] 
It follows that, for $\ell = 2,\ldots,L$,
\[
u_k(\lambda_\ell \widetilde{\bs{w}}, \bs{z}) - u_k(\widetilde{\bs{w}},\bs{z}) = \nu\log \lambda_\ell.
\]
Because $\nu$ is the same for all $\bs{z}$, this gives $m_z (L-1)-1$ restrictions on $\bs{U}$ of the form
\begin{equation}\label{restriction_homo2}
\frac{u_k(\lambda_\ell \widetilde{\bs{w}}, \bs{z}) - u_k(\widetilde{\bs{w}},\bs{z})}{\log \lambda_\ell} - \frac{u_k(\lambda_2 \widetilde{\bs{w}},\bs{z}^1) - u_k(\widetilde{\bs{w}},\bs{z}^1)}{\log \lambda_2} = 0
\end{equation}
for $\bs{z} \in \mathcal{Z}$ and $\ell = 2,\ldots,L$ with $(\bs{z},\ell)\neq(\bs{z}^1,2)$. 
\end{example}

\begin{example}[Additively separable functions with a homogeneous component of known degree $\nu$]\label{example_homo3}

Suppose $u_k(\bs{w},\bs{z})$ is additively separable in $\bs{w}$ and $\bs{z}$ as
\[
u_k(\bs{w},\bs{z})  = u_k^w(\bs{w},\bs{z}) + u_k^z(\bs{z}),
\]
and $u_k^w(\bs{w},\bs{z})$ is known to be homogeneous of degree $\nu \neq 0$ in $\bs{w}$, i.e., for all $(\bs{w},\bs{z}) \in \mathcal{X}$ and all $\lambda > 0$,
\begin{equation}\label{u_w}
u_k^w(\lambda \bs{w},\bs{z}) = \lambda^\nu  u_k^w(\bs{w},\bs{z}).
\end{equation}
Then $u_k(\lambda_\ell \widetilde{\bs{w}},\bs{z}) - u_k(\widetilde{\bs{w}}, \bs{z}) = (\lambda_\ell^\nu-1)u_k^w(\widetilde{\bs{w}},\bs{z})$, which gives $m_z (L-2)$ restrictions of the form
\begin{equation}\label{restriction_homo3}
\frac{u_k(\lambda_\ell \widetilde{\bs{w}},\bs{z}) - u_k(\widetilde{\bs{w}}, \bs{z})}{ \lambda_\ell^\nu-1} - \frac{u_k(\lambda_2 \widetilde{\bs{w}},\bs{z}) - u_k(\widetilde{\bs{w}}, \bs{z})}{ \lambda_2^\nu-1} = 0
\end{equation}
for $\bs{z} \in \mathcal{Z}$ and $\ell = 3,\ldots,L$.
For example, variable profit proportional to market size $w$, net of a fixed cost, gives $u_k(w,z)=w\,v_k(z)-F_k(z)$ with $\nu=1$ \citep[equation~(8)]{BresnahanReiss1991}.

Alternatively, suppose $u_k^w(\bs{w}, \bs{z})$ is the logarithm of a function that is homogeneous of degree $\nu$, with possibly unknown $\nu$, so that $u_k(\lambda \bs{w}, \bs{z})  = u_k^w(\lambda \bs{w}, \bs{z}) + u_k^z(\bs{z}) = \nu \log \lambda + u_k^w(\bs{w}, \bs{z})+ u_k^z(\bs{z})$. Then $u_k(\lambda_\ell \widetilde{\bs{w}}, \bs{z}) - u_k(\widetilde{\bs{w}},\bs{z}) = \nu \log \lambda_\ell$, which leads to $m_z (L-1)-1$ restrictions of the form \eqref{restriction_homo2}. 

\end{example}

\subsection{Equality restrictions based on zero cross-differences}\label{subsec:equality_cross}

\begin{example}[Zero cross-difference with respect to two state variables]\label{example_zero_cross} 
In some payoff functions, the difference in $u_k(\bs{w},\bs{z})$ with respect to $\bs{w}$ is constant across certain values of $\bs{z}$. Suppose there exist $k \in \mathcal{A}$, $(\bs{w}_1,\bs{z}_1)$, and $(\bs{w}_2,\bs{z}_2)$ such that 
\begin{equation} \label{eq:cross_diff}
u_k(\bs{w}_2,\bs{z}_1) - u_k(\bs{w}_1,\bs{z}_1)  =  u_k(\bs{w}_2,\bs{z}_2) - u_k(\bs{w}_1,\bs{z}_2) .
\end{equation}
Let $\mathcal{W}_a \subset \mathcal{W}$ and $\mathcal{Z}_a \subset \mathcal{Z}$ be the sets of $\bs{w}$ and $\bs{z}$ for which condition \eqref{eq:cross_diff} holds. Then, \eqref{eq:cross_diff} provides $(|\mathcal{W}_a|-1)\cdot (|\mathcal{Z}_a|-1)$ restrictions.

For example, in a dynamic entry model with state $\bs{x}=(w,\bs{z})$, where $w=a_{t-1}$ and $\bs{z}$ is exogenous, \eqref{eq:cross_diff} holds if the entry cost $u_1(1,\bs{z})-u_1(2,\bs{z})$ does not vary with $\bs{z}$. Section \ref{sec:ex_rest} gives a concrete example.
 
Many empirical studies assume additive separability of payoff functions, $u_k(\bs{w},\bs{z}) = u_k^w(\bs{w}) + u_k^z(\bs{z})$, in which case \eqref{eq:cross_diff} holds. However, \eqref{eq:cross_diff} can hold even when $u_k(\bs{w},\bs{z})$ is not additively separable. For instance, if $w$ and $z$ are scalars, then $u_k(w,z) = \exp(w) \cdot \mathbbm{1}\{z \geq 0\}$ satisfies \eqref{eq:cross_diff} whenever $z_1$ and $z_2$ share the same sign.
\end{example}

\citet[][Section 5]{abbring20qe} observe that excluding a variable from current utility yields multiple exclusion restrictions. Suppose $\bs{z}$ does not affect utilities for some $k \in \mathcal{A}\setminus \{K\}$: $u_k(\bs{w},\bs{z})= u_k(\bs{w})$  for all $(\bs{w},\bs{z}) \in \mathcal{W} \times \mathcal{Z}$. This condition gives $|\mathcal{W}| (|\mathcal{Z}|-1)$ restrictions. If the restriction holds for multiple values of $k$, it may point identify $\beta$. The zero cross-difference restriction generalizes this condition. It requires only that the difference $u_k(\bs{w}_2,\bs{z}) - u_k(\bs{w}_1,\bs{z})$ be invariant to $\bs{z}$.

\subsection{Inequality restrictions based on monotonicity, concavity, and complementarity/substitutability}\label{subsec:inequality}

\begin{example}[Monotonicity]
Suppose $w$ is scalar with $w^{\ell}<w^{\ell+1}$ for all $\ell$, and $u_k(w,\bs{z})$ is weakly increasing in $w$, i.e., $u_k(w, \bs{z})\geq u_k(w',\bs{z})$ for any $w \geq w'$ and all $\bs{z}\in\mathcal{Z}$. This assumption gives $(m_w-1)m_{z}$ inequality restrictions of the form $u_k(w^{\ell+1},\bs{z}) -u_k(w^{\ell},\bs{z})\geq 0$ for $\ell = 1,\ldots, m_w-1$ and $\bs{z}\in \mathcal{Z}$.
\end{example}

\begin{example}[Concavity]  

Suppose that \( u_k(\bs{x}) \) is weakly concave:  for any \( \lambda \in [0,1] \) and \( \bs{x}, \bs{x}' \in \mathcal{X} \) with \( \lambda \bs{x} + (1 - \lambda) \bs{x}' \in \mathcal{X} \), we have
$u_k(\lambda \bs{x} + (1 - \lambda) \bs{x}') \geq \lambda u_k(\bs{x}) + (1 - \lambda) u_k(\bs{x}')$. Each such triple $(\bs{x},\bs{x}',\lambda)$ yields one inequality restriction.
    
Concavity also implies non-positive discrete second differences in each variable. Suppose \( \bs{x} = (w, z) \), $w$ and $z$ are scalar with $m_w,m_z \geq 3$, and $u_k(w,z)$ is weakly concave. If we order the elements of $\mathcal{X}$ so that $w^{\ell} < w^{\ell+1}$ and $z^{\ell} < z^{\ell+1}$ hold for all $\ell$, then concavity implies 
 \begin{align*}
&\frac{u_k(w^{\ell+1}, z) - u_k(w^\ell, z)}{w^{\ell+1} - w^\ell} - \frac{u_k(w^{\ell+2}, z) - u_k(w^{\ell+1}, z)}{w^{\ell+2} - w^{\ell+1}} \geq 0 \text{ and}\\
&\frac{u_k(w, z^{\ell+1}) - u_k(w, z^\ell)}{z^{\ell+1} - z^\ell} - \frac{u_k(w, z^{\ell+2}) - u_k(w, z^{\ell+1})}{z^{\ell+2} - z^{\ell+1}} \geq 0 
\end{align*}
for all relevant $w \in \mathcal{W}$, $z \in \mathcal{Z}$, and $\ell = 1, \ldots, m_w - 2$ or $\ell = 1, \ldots, m_z - 2$, respectively. 
These conditions give $m_w (m_z - 2) + (m_w - 2) m_z$ inequality restrictions.

\end{example}

\begin{example}[Complementarity and substitutability]
Inputs or goods are complements when the marginal product or marginal utility of one increases with the quantity of the other.
 
Suppose \( \bs{x} = (w, z) \), where $w$ and $z$ are scalar. We order the elements of $\mathcal{X}$ so that $w^{\ell} \leq w^{\ell+1}$ and $z^{\ell} \leq z^{\ell+1}$ for all $\ell$. Complementarity between $w$ and $z$ implies nonnegative cross-differences, so that for any $(\ell,m)\in \{1,\ldots,m_w-1\}\times \{1,\ldots,m_z-1\}$,
\begin{equation}\label{eq:comple}
u_k(w^{\ell+1}, z^{m+1}) - u_k(w^{\ell+1}, z^{m}) - u_k(w^{\ell}, z^{m+1}) + u_k(w^{\ell}, z^{m}) \geq 0.
\end{equation}
This condition gives $(m_w - 1) (m_z - 1)$ inequality restrictions. Conversely, substitutability between $w$ and $z$ leads to the same number of restrictions with the direction of the inequality in \eqref{eq:comple} reversed.

\end{example}

\subsection{Linear-in-parameters payoff function}\label{sec:linear}

When the payoff is linear in parameters, the identifying restrictions are easy to derive.

Suppose that $\bs{U} = \bs{H} \bs{\theta}$ for a parameter vector $\bs{\theta}$ and a known $JK \times \dim(\bs{\theta})$ matrix $\bs{H}$ of full column rank.
Let $d_H =JK - \dim(\bs{\theta})$ be the dimension of $\Ker(\bs{H}\t)$, and let $\bs{R}$ be a $d_H \times JK$ matrix whose rows form a basis for $\Ker(\bs{H}\t)$. Then, we obtain $d_H$ identifying restrictions of the form
\[
\bs{R}\bs{U} = \bs{R}\bs{H} \bs{\theta} =\bs{0}.
\]

These restrictions can provide strong identifying power when $\dim(\bs{\theta})$ is small. A fully nonparametric payoff, $u_k(\bs{x})= \sum_{j=1}^J \theta_{kj} \mathbbm{1}\{\bs{x} =\bs{x}^j\}$, is also linear in parameters, but then $\dim(\bs{\theta})=JK$, so $d_H=0$ and there are no restrictions.

Online Appendix \ref{sec:log_diff} shows that restrictions on log differences in payoffs, such as homogeneity of unknown degree, also identify $\beta$.

\section{Numerical Example 1: Dynamic Entry Model}\label{sec:entry_example}

Using a firm entry-exit model, this section illustrates the identifiability of $\beta$ and $\bs{U}$ without the normalization $\bs{u}_K=\bs{0}$.
\label{sec:scrap_value_example}

We consider a class of firm entry-exit models used, for example, in \citet{aguirregabiriasuzuki14qme, igami2017estimating, Kalouptsidi21qe}. In each period, a firm chooses to be active ($a_t = 1$) or inactive ($a_t = 2$) in a market, so $\mathcal{A} = \{1,2\}$. The state variable is $(\bs{x}_t, \ve_t) = (\bs{w}_{t}, y_{t},\ve_t)= (w_{vt}, w_{ct}, y_{t},\ve_t)$, where $w_{vt} \in \mathcal{W}_v$ is a profit shifter, $w_{ct} \in \mathcal{W}_c$ is a cost shifter, $y_t=a_{t-1}$, and $\ve_t$ follows a type-I extreme value distribution. Let $m_v = |\mathcal{W}_v|$, $m_c = |\mathcal{W}_c|$, and $m_w = m_v m_c$. Because $y_t$ takes two values, there are $J = 2 m_w$ states.

We specify the firm's period profit (net of $\ve_t$) as
\[
u_1(\bs{w},1) =  \pi(w_v) -f\!c(w_c), \quad u_1(\bs{w},2) = \pi(w_v) - f\!c(w_c) -ec(w_c), \quad u_2(\bs{w},1) = sv(w_c), \quad u_2(\bs{w},2) = 0,
\]
where $\pi(w_v)$ is the variable operating profit, $f\!c(w_c)$ is the fixed cost, $ec(w_c)$ is the entry cost, and $sv(w_c)$ is the scrap value. As in \citet{aguirregabiriasuzuki14qme} and \citet{Kalouptsidi21qe}, we assume that the econometrician knows $\pi(w_v)$ but not $(f\!c(\cdot),ec(\cdot),sv(\cdot))$, which do not depend on the profit shifter $w_v$. The restriction $u_2(\bs{w},2) = 0$ normalizes the per-period payoff of a firm that is inactive in both the current and previous periods. This normalization imposes $m_w$ restrictions on $\bs{u}_2$ but does not imply zero scrap value.

We set $\pi(w_v) = \exp(w_v)$, $f\!c(w_c)=3.0 w_c$, $ec(w_c)=3.0 w_c$, and $sv(w_c) = 2.0 w_c$. The profit and cost shifters $(w_{vt},w_{ct})$ are exogenous and follow the $\text{AR}(1)$ processes $w_{vt} = 0.5 w_{v,t-1} + e_{vt}$ and $w_{ct} = 1.0 + 0.5 w_{c,t-1} + e_{ct}$, where $e_{vt}$ and $e_{ct}$ are mutually independent and $\text{iid } N(0,1)$. We discretize these processes using the method of \citet{tauchen1986finite}. For $w_{vt}$, we use a fixed grid $\{0.5, 1.5\}$ when $m_v=2$ and $\{0.5, 1.0, 1.5\}$ when $m_v=3$. For $w_{ct}$, we set the grid endpoints at the $0.5/m_c$ and $1 - 0.5/m_c$ quantiles of its stationary distribution with equispaced points between them. The discount factor is set to $\beta=0.95$.

\subsection{Exclusion restrictions}\label{sec:ex_rest}

In this model, the cost functions and scrap value satisfy exclusion restrictions (ER) because they do not depend on $w_v$.

\noindent \textbf{ER on Scrap Value ($m_w -m_c$ restrictions):} 
The scrap value satisfies $u_2(w_v, w_c, 1) = u_2(w_v', w_c, 1)$ for all $w_v, w_v' \in \mathcal{W}_v$ and $w_c \in \mathcal{W}_c$, giving $m_c(m_v-1)=m_w-m_c$ restrictions.

\noindent \textbf{ER on Fixed and Entry Costs ($2 (m_w -m_c)$ restrictions):}
The exclusion restrictions on fixed and entry costs imply that, for all $w_c \in \mathcal{W}_c$, $w_v \neq w_v'$ in $\mathcal{W}_v$, and $y \in \{1,2\}$,
\[
u_1(w_v, w_c, y) - u_1(w_v', w_c, y) = \pi(w_v) - \pi(w_v').
\]
The two cases $y=1$ and $y=2$ say that the fixed cost and the sum of the fixed and entry costs, respectively, do not depend on $w_v$.
Since $\pi(w_v) - \pi(w_v')$ is assumed to be known, these yield $2m_c(m_v-1)=2 (m_w -m_c)$ independent restrictions.

\subsection{Additional assumptions on scrap value and cost functions}\label{sec:assn_homo}

\citet{aguirregabiriasuzuki14qme} show that $(f\!c(\cdot),ec(\cdot),sv(\cdot))$ are not separately identified if we use only exclusion restrictions. We consider three additional assumptions on the scrap value and cost functions.

\noindent \textbf{Scrap Value Homogeneity ($m_c-1$ restrictions):} Scrap values are homogeneous of degree 1 in $w_c$. This implies that $w_c \cdot u_2(w_v, w_c', 1) - w_c' \cdot u_2(w_v, w_c, 1) =0$ for all $w_c \neq w_c'$ and provides $m_c - 1$ restrictions in addition to ER.

\noindent \textbf{Known Scrap Value ($m_w$ restrictions):}
The scrap value $sv(w_c)$ is known. This includes zero scrap value as a special case and provides $m_w$ restrictions. From equation \eqref{ident_eq_known}, $\beta$ can be identified by shape restrictions on $(f\!c(w_c),ec(w_c))$ discussed in Section \ref{sec:examples}.

\noindent \textbf{Cost Homogeneity ($2(m_c-1)$ restrictions):} Fixed and entry costs are homogeneous of degree 1 in $w_c$. This implies that, for all $w_v \in \mathcal{W}_v$, $w_c \neq w_c'$ in $\mathcal{W}_c$, and $y \in \{1,2\}$,
\[
w_c \cdot u_1(w_v, w_c', y) - w_c' \cdot u_1(w_v, w_c, y) = \pi(w_v) \cdot (w_c - w_c').
\]
These yield $2(m_c-1)$ independent restrictions in addition to ER.

\subsection{Identification results}

Table \ref{tab:id_mw4} summarizes the identification results for two cases, $m_v=m_c=2$ and $m_v=m_c=3$. ER and HR denote the exclusion restrictions of Section \ref{sec:ex_rest} and the homogeneity restrictions of Section \ref{sec:assn_homo}. Here, $r$ is the number of restrictions imposed by $\bs{R}$, comprising those introduced in Sections \ref{sec:ex_rest} and \ref{sec:assn_homo} together with the $m_w$ restrictions imposed by $u_2(\bs{w},2)=0$. The quantity $\overline{r}$ denotes the number of restrictions in $\bs{R}(\beta)$, which has $JK$ columns. Define $\delta \coloneqq \overline{r}- \mathrm{rk}(\bs{R}(\beta))$, the number of overidentifying restrictions available to identify $\beta$. From Propositions \ref{prop_equality_new} and \ref{prop_ident_partial}, $\beta$ is identified if $\delta > 0$, and both $\beta$ and $\bs{U}$ are identified if $\delta>0$ and $\mathrm{rk}(\bs{R}(\beta)) = JK$. When only ER is imposed on the cost functions, $\bs{U}$ is not identified unless the scrap value is known, because $\bs{R}(\beta)$ does not have full column rank. The quantity $JK-\mathrm{rk}(\bs{R}(\beta))$ represents the degree of underidentification of $\bs{U}$. Combining ER and HR on the cost functions identifies both $\beta$ and $\bs{U}$. Increasing $m_v$ and $m_c$ from 2 to 3 does not resolve the rank deficiency of $\bs{R}(\beta)$.
\begin{table}[htbp]
\centering
\caption{Identification Results}
\label{tab:id_mw4}
\small
\begin{tabular}{lrrrrr@{\hspace{2em}}rrrrr}
\toprule
 & \multicolumn{5}{c}{\begin{tabular}{@{}c@{}}Fixed and Entry\\Costs: ER\end{tabular}} & \multicolumn{5}{c}{\begin{tabular}{@{}c@{}}Fixed and Entry\\Costs: ER+HR\end{tabular}} \\
\cmidrule(lr){2-6} \cmidrule(lr){7-11}
 & \multicolumn{1}{c}{$r$} & \multicolumn{1}{c}{$\overline{r}$} & \multicolumn{1}{c}{$JK$} & \multicolumn{1}{c}{$\mathrm{rk}(\bs{R}(\beta))$} & \multicolumn{1}{c}{$\delta$} & \multicolumn{1}{c}{$r$} & \multicolumn{1}{c}{$\overline{r}$} & \multicolumn{1}{c}{$JK$} & \multicolumn{1}{c}{$\mathrm{rk}(\bs{R}(\beta))$} & \multicolumn{1}{c}{$\delta$} \\
\midrule
\multicolumn{11}{l}{\textit{Panel A: $m_v=m_c=2$, $m_w=4$}} \\
\addlinespace[0.5ex]
SV: ER & 10 & 18 & 16 & 14\rlap{$^\dagger$} & 4 & 12 & 20 & 16 & 16 & 4 \\
SV: ER+HR & 11 & 19 & 16 & 15\rlap{$^\dagger$} & 4 & 13 & 21 & 16 & 16 & 5 \\
SV: Known & 12 & 20 & 16 & 16 & 4 & 14 & 22 & 16 & 16 & 6 \\
\addlinespace[0.5ex]
\midrule
\addlinespace[0.5ex]
\multicolumn{11}{l}{\textit{Panel B: $m_v=m_c=3$, $m_w=9$}} \\
\addlinespace[0.5ex]
SV: ER & 27 & 45 & 36 & 33\rlap{$^\dagger$} & 12 & 31 & 49 & 36 & 36 & 13 \\
SV: ER+HR & 29 & 47 & 36 & 35\rlap{$^\dagger$} & 12 & 33 & 51 & 36 & 36 & 15 \\
SV: Known & 30 & 48 & 36 & 36 & 12 & 34 & 52 & 36 & 36 & 16 \\
\bottomrule
\end{tabular}
\medskip

\raggedright
\small
\textit{Note:} $\beta$ is identified in every case. $^\dagger$\,$\mathrm{rk}(\bs{R}(\beta))<JK$, so $\bs{U}$ is not identified; in all other cases, $\bs{U}$ is identified.
\end{table}

\begin{figure}[!t]
	\centering
	\includegraphics[width=1\linewidth]{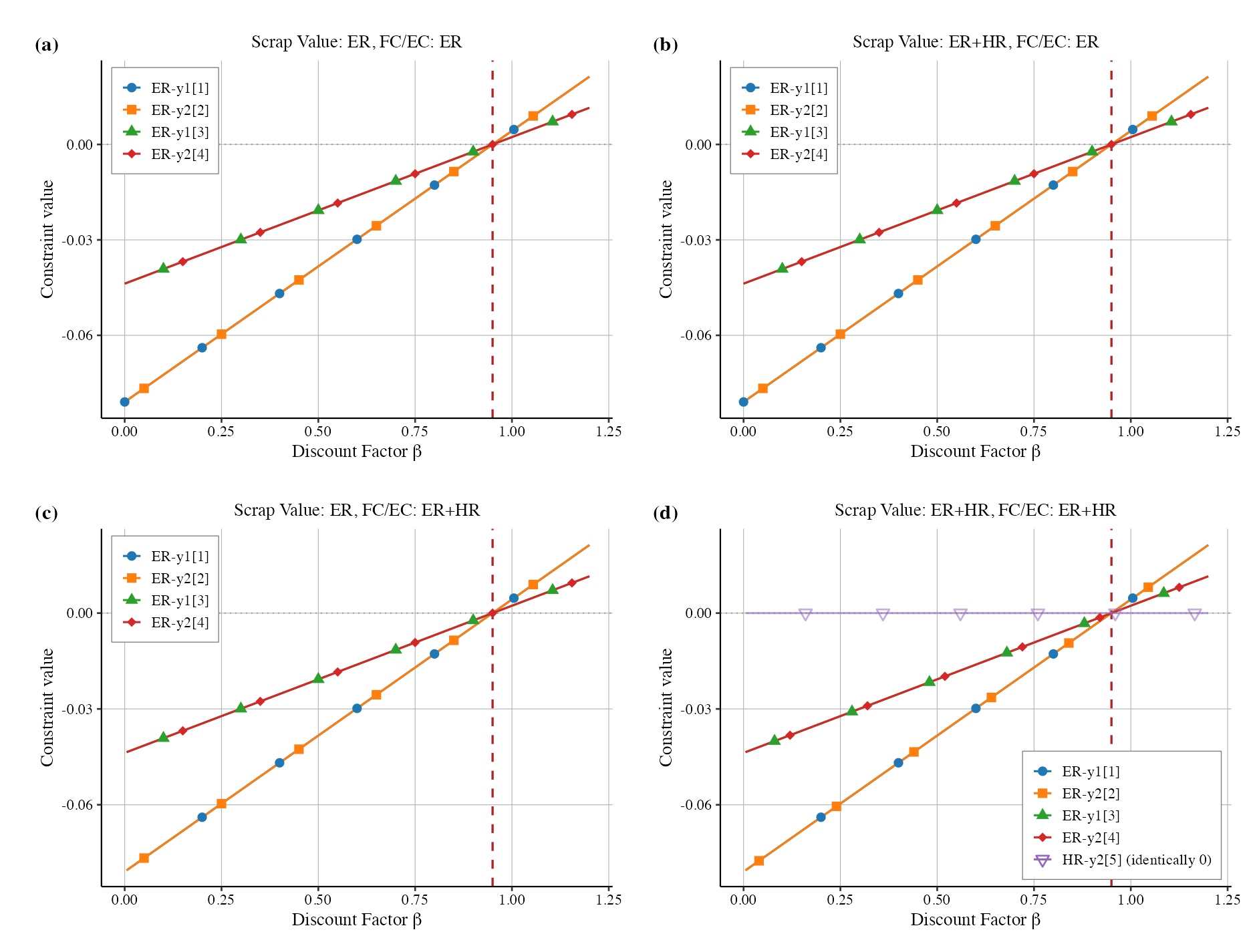}
	\caption[]{Identifying Polynomials for $\beta \in [0,1.2]$: Model without Zero Scrap Value. The red dashed vertical line marks the true value $\beta=0.95$.}
	\label{fig1}
\end{figure}

Figure \ref{fig1} plots the identifying polynomials implied by the restrictions described in the first two rows of Panel A ($m_v=m_c=2$) of Table \ref{tab:id_mw4}. This model exhibits finite dependence, and the identifying polynomials are linear in $\beta$. Panels (a) and (b) correspond to the left half of Panel A of Table \ref{tab:id_mw4}. Both impose ER on the cost functions. On the scrap value, (a) imposes ER and (b) imposes ER+HR. Here $\bs{U}$ is not identified, and the polynomials are those in \eqref{eq:prop_fd_partial}. The matrix $\bs{R}_2$ consists of the four cost-ER restrictions, one for each $y\in\{1,2\}$ and each of the two $w_c$ grid points. Adding HR on the scrap value raises $\mathrm{rk}(\bs{R}(\beta))$ by one but leaves $\delta$ unchanged, because the added restriction reduces the underidentification of $\bs{U}$ rather than restricting $\beta$. All plotted lines intersect the horizontal axis at $\beta=0.95$ and identify $\beta$ even though $\bs{U}$ is not identified. In these panels, some polynomials coincide even though the corresponding rows of $\bs{R}$ are linearly independent. Panels (c) and (d) correspond to the right half. Both impose ER+HR on the cost functions. On the scrap value, (c) imposes ER and (d) imposes ER+HR. The polynomials are those in \eqref{eq:prop_fd}. Adding HR to the cost functions increases $\mathrm{rk}(\bs{R}(\beta))$ to $JK$ and identifies $\bs{U}$ as well as $\beta$. In (c), the HR restrictions are used to solve for $\bs{U}$, and $\bs{R}_2$ consists of the same four cost-ER restrictions plotted in (a) and (b). In (d), $\bs{R}_2$ adds one HR restriction, but it is implied by the restrictions used to solve for $\bs{U}$, so its identifying polynomial is identically zero. All plotted lines again intersect the horizontal axis at $\beta=0.95$.

Multiplying a polynomial by a constant rescales its slope, so the slopes in these figures are not informative on their own. The strength of identification depends on a slope's magnitude relative to the sampling variation in the estimated curve.

Online Appendix \ref{sec:oa_zero_scrap} presents a second entry model that imposes $\bs{u}_K=\bs{0}$ and does not exhibit finite dependence, and applies the restrictions of Section \ref{sec:examples}. Homogeneity or zero cross-difference alone point identifies $\beta$. The inequality restrictions together give $\beta\in[0.939,0.972]$ around the true value $0.95$.

\section{Models of Dynamic Discrete Games}\label{sec:dynamic_game}

\subsection{Framework and basic assumptions} 

We consider the dynamic discrete game of \citet{am07em}. There are $N$ firms operating in the market. Each firm $i$ selects an action $a_{it}$ in period $t$ from the set $\mathcal{A}_i = \{1,\ldots, K\}$, where action $K$ is being inactive. The joint action space is $\mathcal{A} \coloneqq \times_{i=1}^N \mathcal{A}_i$. Define $\bs{a}_t \coloneqq (a_{1t}, \ldots, a_{Nt})$. Define $\bs{a}_{-it}$ as the action profile of all firms except $i$, with corresponding action space $\mathcal{A}_{-i} \coloneqq \times_{j \neq i} \mathcal{A}_j$. The observable state $\bs{x}_t$ takes values in $\mathcal{X} = \{\bs{x}^1,\ldots,\bs{x}^{m_x}\}$. Firm $i$'s private information shock $\bs{\ve}_{it} \coloneqq (\ve_{it}(1), \ldots, \ve_{it}(K)) \in \mathbb{R}^K$ has density $g_i(\bs{\ve}_{it})$ and is i.i.d.\ over time and independent across firms. Firm $i$'s per-period payoff is $\widetilde\pi_i(\bs{a}_{t}, \bs{x}_t,\bs{\ve}_{it})$. The joint private information vector is $\bs{\ve}_t \coloneqq(\bs{\ve}_{1t}, \ldots, \bs{\ve}_{Nt})$. The sequence $\{\bs{x}_t, \bs{\ve}_t\}$ follows a controlled Markov process with transition probability $p(\bs{x}_{t+1}, \bs{\ve}_{t+1}|\bs{x}_t, \bs{\ve}_t, \bs{a}_t)$.

Each firm maximizes the expected discounted sum of current and future payoffs,
\[
E \left\{ \sum_{s=t}^\infty \beta_i^{s-t} \, \widetilde\pi_i(\bs{a}_s,\bs{x}_s,\bs{\ve}_{is}) \middle|\bs{x}_t, \bs{\ve}_{it} \right\},
\]
where the discount factor $\beta_i \in [0,1)$ may differ across firms. Let $Q(\bs{x}_{t+1}|\bs{x}_t, \bs{a}_t)$ denote the transition probability of $\bs{x}_t$. We adopt the following standard assumptions.
\begin{assumption}\label{assn_multi}
(a) Additive separability: $\widetilde\pi_i(\bs{a}_t,\bs{x}_t,\bs{\ve}_{it}) = \pi_i(\bs{a}_t,\bs{x}_t) + \ve_{it}(a_{it})$, (b) Conditional independence: $p(\bs{x}_{t+1}, \bs{\ve}_{t+1}|\bs{x}_t, \bs{\ve}_t, \bs{a}_t) = p_\ve(\bs{\ve}_{t+1})Q(\bs{x}_{t+1}|\bs{x}_t, \bs{a}_t)$, and (c) Independent private information shocks: $p_\ve(\bs{\ve}_t)= \prod_{i=1}^N g_i(\bs{\ve}_{it})$. 
\end{assumption}

Firms follow stationary Markov strategies, so we drop the time subscript. Let $\bs{x}'$ and $\bs{\ve}'$ denote the next-period state variables. Let $\sigma = \{\sigma_i(\bs{x},\bs{\ve}_i)\}_{i=1}^N$ denote the strategy profile, where each firm $i$'s strategy function is $\sigma_i: \mathcal{X} \times \mathbb{R}^K \to \mathcal{A}_i$. \citet{am07em} establish the existence of Markov perfect equilibrium (MPE) strategies in this model. Let $\bs{P}^* \coloneqq \{[P_i^*(a|\bs{x})]_{a \in \mathcal{A}_i, \bs{x}\in\mathcal{X}}\}_{i=1}^N$ denote the equilibrium choice probabilities corresponding to an MPE strategy profile $\sigma^*$. The equilibrium probability that the firms other than $i$ choose $\bs{a}_{-i}$ is
\begin{equation} \label{eq:p_minus}
P_{-i}^*(\bs{a}_{-i}|\bs{x}) \coloneqq \prod_{j \neq i} P_j^* (a_{-i}[j]|\bs{x}),
\end{equation}
where $a_{-i}[j]$ is firm $j$'s action in $\bs{a}_{-i}$. Under equilibrium choice probabilities $\bs{P}^*$, firm $i$'s expected payoff and expected transition probability are
\begin{equation} \label{eq:expected_p}
\pi_i^*(a_i,\bs{x}) = \sum_{\bs{a}_{-i} \in \mathcal{A}_{-i}}P_{-i}^*(\bs{a}_{-i}|\bs{x}) \pi_i(a_i,\bs{a}_{-i},\bs{x})
\end{equation}
and
\[
Q_i^*(\bs{x}'|\bs{x},a_i) \coloneqq \sum_{\bs{a}_{-i} \in \mathcal{A}_{-i}} P_{-i}^*(\bs{a}_{-i}|\bs{x}) Q(\bs{x}'|\bs{x},a_i,\bs{a}_{-i}).
\]

As shown by \citet{am07em}, the equilibrium choice probability satisfies
\[
P^*_i(a_i|\bs x) = \int \mathbbm{1}\left\{a_i=\argmax_{k \in \mathcal{A}_i} \left\{ v_i^{*}(k,\bs{x}) + \ve_i(k) \right\}\right\} g_i(d\bs{\ve}_i),
\]
where the equilibrium choice-specific value function $v_i^{*}(a_i,\bs{x})$ is defined as
\begin{equation}\label{v_k_x_am}
v_i^*(a_i,\bs{x}) \coloneqq \pi_i^*(a_i,\bs{x}) + \beta_i \sum_{\bs{x}' \in \mathcal{X}} V_i^*(\bs{x}') Q_i^*(\bs{x}'|\bs{x},a_i),
\end{equation}
and the integrated value function $V_i^*$ satisfies the integrated Bellman equation
\begin{equation}\label{int_Bellman}
V_i^*(\bs{x}) = \int \max_{a_i \in \mathcal{A}_i} \left\{ v_i^*(a_i,\bs{x}) + \ve_i(a_i) \right\} g_i(d\bs{\ve}_i).
\end{equation}
Following \citet{am07em} and \citet{pesendorfer08restud}, we assume the data are generated by a single MPE, though the model may admit multiple equilibria.

Define $\bs{P}_{-i}^{*}(\bs{x}) \coloneqq[P_{-i}^*(\bs{a}_{-i}|\bs{x})]_{\bs{a}_{-i} \in \mathcal{A}_{-i}}$ as the $K^{N-1} \times 1$ vector of conditional choice probabilities of firms other than firm $i$. Similarly, define the $K^{N-1} \times 1$ vector $\bs{\pi}_i^{-i}(a_i,\bs{x}) \coloneqq [\pi_i(a_i,\bs{a}_{-i},\bs{x})]_{\bs{a}_{-i} \in \mathcal{A}_{-i}}$. Then, from \eqref{eq:expected_p}, 
\[
\pi_i^*(a_i,\bs{x}) = \bs{P}_{-i}^{*}(\bs{x})\t \bs{\pi}_i^{-i}(a_i,\bs{x}).
\]
Collect $v_i^*(k,\bs{x})$, $V_i^*(\bs{x})$, and $\bs{\pi}_{i}^{-i}(k,\bs{x})$ across all $\bs{x} \in \mathcal{X}$ into two $m_x \times 1$ vectors and one $(K^{N-1} \cdot m_x) \times 1$ vector, respectively, as 
\[
\bs{v}_{ik}^* \coloneqq
\begin{bmatrix}
v_i^*(k,\bs{x}^1)\\
\vdots \\
v_i^*(k,\bs{x}^{m_x})
\end{bmatrix},\quad
\bs{V}_{i}^* \coloneqq
\begin{bmatrix}
V_i^*(\bs{x}^1)\\
\vdots \\
V_i^*(\bs{x}^{m_x})
\end{bmatrix},\quad
\bs{\pi}_{ik}^{-i} \coloneqq
\begin{bmatrix}
\bs{\pi}_{i}^{-i}(k,\bs{x}^1)\\
\vdots \\
\bs{\pi}_{i}^{-i}(k,\bs{x}^{m_x})
\end{bmatrix}.
\]
Collect $\bs{P}_{-i}^{*}(\bs{x})\t$ across all $\bs{x} \in \mathcal{X}$ into an $m_x \times (K^{N-1}\cdot m_x)$ matrix as
\[
\bs{P}_{-i}^{*} \coloneqq 
\begin{bmatrix}
\bs{P}_{-i}^{*}(\bs{x}^1)\t & & \bs{O}\\
& \ddots & \\
\bs{O} & & \bs{P}_{-i}^{*}(\bs{x}^{m_x})\t
\end{bmatrix}.
\]
Let $\bs{Q}_{ik}^*$ be the $m_x \times m_x$ matrix with the $(\ell,m)$-th entry $Q_i^*(\bs{x}^m|\bs{x}^\ell,k)$. Stacking \eqref{v_k_x_am} at $a_i=k$ across $\bs{x} \in \mathcal{X}$ gives the multi-agent counterpart of \eqref{v_k},
\begin{equation}\label{Bellman4_am_star}
\bs{v}_{ik}^* = \bs{P}_{-i}^{*} \bs{\pi}_{ik}^{-i} + \beta_i \bs{Q}_{ik}^* \bs{V}_i^*, \quad k=1,\ldots,K.
\end{equation}

The following lemma is a multi-agent counterpart of Lemma 1 of \citet{am11em}. Its proof is in Online Appendix \ref{sec:proofs}. Collect $P_i^*(a_i|\bs{x})$ into the vector $\bs{P}_i^*(\bs{x}) \coloneqq (P_i^*(1|\bs{x}),\ldots,P_i^*(K|\bs{x}))\t$. 
\begin{lemma} \label{psi_lemma}
Define $\bs{p} \coloneqq (p(1),\ldots,p(K))\t$, where $\sum_{k=1}^K p(k) = 1$ and $p(k)>0$ for all $k$. Then there exists a real-valued function $\psi_{ik}(\bs{p})$ for every $i \in \{1,\ldots,N\}$ and $k \in \{1,\ldots,K\}$ such that
\begin{equation}\label{eq:psi_lemma}
\psi_{ik}(\bs{P}_i^*(\bs{x})) = V_i^{*}(\bs{x}) - v_i^*(k,\bs{x}),
\end{equation}
for all $\bs{x} \in \mathcal{X}$. Furthermore, if the elements of $\bs{\ve}_i$ follow mutually independent type-I extreme value distributions, then $\psi_{ik}(\bs{p}) = \gamma - \log (p(k))$, where $\gamma$ is Euler's constant.
\end{lemma}
Let $\bs{\psi}_{ik}^*$ be the $m_x \times 1$ vector whose $j$th element is $\psi_{ik}(\bs{P}_i^*(\bs{x}^j))$. Collecting \eqref{eq:psi_lemma} across all $\bs{x} \in \mathcal{X}$ gives the following multi-agent counterpart of \eqref{AM2}:
\begin{equation}\label{psi_star}
\bs{\psi}_{ik}^* = \bs{V}_i^* - \bs{v}_{ik}^*, \quad k=1,\ldots,K.
\end{equation}
Eliminating $\bs{v}_{ik}^*$ from \eqref{Bellman4_am_star} and \eqref{psi_star} gives $\bs{P}_{-i}^{*} \bs{\pi}_{ik}^{-i} + \bs{\psi}_{ik}^* = (\bs{I} - \beta_i\bs{Q}_{ik}^*)\bs{V}_i^*$. It follows that, for $k=1,\ldots,K-1$,
\begin{equation}\label{psi_diff_multi2}
(\bs{I} - \beta_i \bs{Q}_{ik}^*)^{-1}\bs{P}_{-i}^{*} \bs{\pi}_{ik}^{-i} - (\bs{I} - \beta_i \bs{Q}_{iK}^*)^{-1}\bs{P}_{-i}^{*} \bs{\pi}_{iK}^{-i} = -(\bs{I} - \beta_i \bs{Q}_{ik}^*)^{-1} \bs{\psi}_{ik}^* + ( \bs{I} - \beta_i\bs{Q}_{iK}^*)^{-1} \bs{\psi}_{iK}^*.
\end{equation}
Stack $\bs{\pi}_{ik}^{-i}$ and $\bs{\psi}_{ik}^*$ for $k=1,\ldots,K$ into vectors as
\[
\bs{\Pi}_i \coloneqq 
\begin{bmatrix}
\bs{\pi}_{i1}^{-i} \\
\vdots \\
\bs{\pi}_{iK}^{-i}
\end{bmatrix}, \quad
\bs{\Psi}_i^* \coloneqq
\begin{bmatrix}
\bs{\psi}_{i1}^* \\
\vdots \\
\bs{\psi}_{iK}^*
\end{bmatrix}.
\]
Let $m_{\Pi} \coloneqq \dim(\bs{\Pi}_i) = K^{N} m_x$.
Define the $(K-1)m_x \times K m_x$ matrix $\bs{P}_i^*(\beta_i)$ and the $K m_x \times m_{\Pi}$ block-diagonal matrix $\overline{\bs{P}}_i^*$ as
\begin{equation*}\bs{P}_i^*(\beta_i) \coloneqq
\begin{bmatrix}
(\bs{I}- \beta_i \bs{Q}_{i1}^*)^{-1} & \cdots &\bs{O} & -(\bs{I}- \beta_i \bs{Q}_{iK}^*)^{-1}\\
\vdots & \ddots & \vdots & \vdots \\
\bs{O} & \cdots & (\bs{I}- \beta_i \bs{Q}_{i,K-1}^*)^{-1} & - (\bs{I}- \beta_i \bs{Q}_{iK}^*)^{-1}
\end{bmatrix}, \quad
\overline{\bs{P}}_i^* \coloneqq 
\left[\smash{\underbrace{
\left.\begin{array}{ccc}
\bs{P}_{-i}^{*} & &\\
& \ddots & \\
& & \bs{P}_{-i}^{*}\end{array}\right.}_{K \text{ times}}}
\left.\begin{array}{c}
\\
\\
\\
\end{array}\right.\hspace{-1.1em}
\right].\end{equation*}
Collecting \eqref{psi_diff_multi2} for $k=1,\ldots, K-1$ gives the multi-agent counterpart of \eqref{model_rest},
\begin{equation} \label{model_rest_star}
\bs{P}_i^*(\beta_i) \overline{\bs{P}}_i^* \bs{\Pi}_i = - \bs{P}_i^*(\beta_i) \bs{\Psi}_i^* .
\end{equation}
This is a system of $(K-1) m_x$ equations in the $K^N m_x\times 1$ unknown payoff vector $\bs{\Pi}_i$. If $\beta_i$ were known, identifying $\bs{\Pi}_i$ would require $(K^{N} - K + 1) m_x$ additional linear restrictions \citep[][Proposition 2]{pesendorfer08restud}.

\subsection{Identification of $\beta_i$ by economic restrictions}\label{sec: util_diff_game}

To identify $\beta_i$, we augment \eqref{model_rest_star} with additional restrictions on $\bs{\Pi}_i$. The restrictions of Section \ref{sec:examples} apply here, but we first use restrictions specific to games. We set $\bs{x}_t = (S_t, \bs{a}_{t-1})$, where $S_t \in \mathcal{S}$ is an exogenous state (e.g., market size) with $m_s \coloneqq |\mathcal{S}|$ and $\bs{a}_{t-1}$ is the lagged action profile. Hence $m_x = |\mathcal{X}| = K^{N} m_s$.

Two restrictions are common in applications: (a) a known payoff for one action (typically non-entry) and (b) the irrelevance of other firms' lagged actions. Restriction (a), imposed in equation (17) of \citet{pesendorfer08restud}, is strong but supplies $K^{N-1} m_x$ restrictions. Section \ref{sec:beta_no_scrap} relaxes it.
\begin{assumption}[Known payoff for one action]\label{assn_prof_K}
Firm $i$'s payoff under action $a_i=K$ is known: for all $\bs{a}_{-i} \in \mathcal{A}_{-i}$ and $\bs{x} \in \mathcal{X}$, we have $\pi_i(a_{i}=K,\bs{a}_{-i},\bs{x}) = r_i(\bs{a}_{-i},\bs{x})$ for known $r_i(\cdot,\cdot)$.
\end{assumption}

Restriction (b) is equation (16) of \citet{pesendorfer08restud}.
\begin{assumption}[Irrelevance of other firms' lagged actions] \label{assn_prof_lagged}
Firm $i$'s payoff under action $a_{it}\neq K$ does not depend on the lagged actions of other firms: for all $a_{it} \in \mathcal{A}_i \setminus \{K\}$, $(\bs{a}_{-it}, S_t, a_{i,t-1}) \in \mathcal{A}_{-i}\times \mathcal{S} \times \mathcal{A}_i$, and $\widetilde{\bs{a}}_{-i,t-1} \in \mathcal{A}_{-i}$,
\[
\pi_i(a_{it},\bs{a}_{-it},S_t,a_{i,t-1},\bs{a}_{-i,t-1}=\bs{1}_{N-1}) = \pi_i(a_{it},\bs{a}_{-it},S_t,a_{i,t-1},\widetilde{\bs{a}}_{-i,t-1}).
\]
\end{assumption}

Assumption \ref{assn_prof_lagged} generates $(K-1)|\mathcal{A}_{-i}|\, m_s |\mathcal{A}_i|(|\mathcal{A}_{-i}|-1) = (K-1)(K^{N-1}-1)m_x$ restrictions. Henceforth, we impose Assumption \ref{assn_prof_lagged} and write the payoff as $\pi_i(a_{it},\bs{a}_{-it},S_t,a_{i,t-1})$. This notation also covers $a_{it}=K$ when we relax Assumption \ref{assn_prof_K} in Section \ref{sec:beta_no_scrap}, where Assumptions \ref{assn_KK} and \ref{assn_scrap} make the payoff independent of $\bs{a}_{-i,t-1}$. Because $(K-1)(K^{N-1}-1)m_x + K^{N-1}m_x = (K^N-K+1)m_x$, Assumptions \ref{assn_prof_K} and \ref{assn_prof_lagged} together just identify $\bs{\Pi}_i$ when $\beta_i$ is known, provided that a rank condition holds.\footnote{\citet{pesendorfer08restud} define the payoff function as $\pi_i(\bs{a},\bs{x})$, where $\bs{x} = (\bs{x}_{1}, \ldots, \bs{x}_{N})$ and $\bs{x}_{i} \in \mathcal{X}_i$ with $|\mathcal{X}_i|=L$. They show that the payoff function is identified if $\beta_i$ is known, assumptions similar to Assumptions \ref{assn_prof_lagged} and \ref{assn_prof_K} hold, $L \geq K$, and a rank condition is satisfied. They state this condition as ``$L \geq K+1$'' because they define $\mathcal{A}_i$ as $\{0,1,\ldots,K\}$. In our setting, the condition $L \geq K$ is automatically satisfied because the state variable $\bs{x}$ includes the lagged action profile $\bs{a}_{t-1}$.}

Our first additional restriction is the exchangeability of other firms' actions.
\begin{assumption}[Exchangeability (EX)]\label{assn_prof_ex}
Firm $i$'s per-period payoff under some action $a_{it}\neq K$ and some state $(S_t, a_{i,t-1})$ is exchangeable in the action profile of the other firms. Namely, there exists $(a_{it},S_t,a_{i,t-1}) \in (\mathcal{A}_i \setminus \{K\})\times \mathcal{S}\times \mathcal{A}_i$ such that, for all $\bs{a}_{-it}\in\mathcal{A}_{-i}$ and every permutation $\tau(\bs{a}_{-it})$ of $\bs{a}_{-it}$,
\[
\pi_i(a_{it},\bs{a}_{-it},S_t,a_{i,t-1}) = \pi_i(a_{it},\tau(\bs{a}_{-it}),S_t,a_{i,t-1}).
\]
\end{assumption}
Assumption \ref{assn_prof_ex}, or a variant of it, holds when firm decisions are based on aggregate industry states rather than on the individual states of other firms, a popular specification in the literature (see, e.g., \citet{EricsonPakes1995}, \citet{pakes07rand}, \citet{am07em}, \citet{ryan2012}, \citet{collardwexler2013demand}, \citet{Benkard2015}, \citet{igami2017estimating}, and \citet{igami2020mergers}).\footnote{
\citet{Benkard2015} assumes that a firm's profit depends on its own status and an aggregate industry state, measured by the number of firms at each quality level, while \citet{igami2017estimating} defines aggregate states as the counts of firms in each of the four technological categories: (1) ``old only,'' (2) ``both,'' (3) ``new only,'' and (4) ``potential entrant.''} For example, Assumption \ref{assn_prof_ex} holds if the payoff depends on $\bs{a}_{-it}$ only through the sum $\sum_{j \neq i} a_{jt}$. When $K=2$, Assumption \ref{assn_prof_ex} implies $2^{N-1}-N$ restrictions for each triple $(a_{it},S_t,a_{i,t-1})$ at which it holds, so that it yields $\kappa_{\text{ex}} \cdot (2^{N-1}-N)$ restrictions in total, where $\kappa_{\text{ex}}$ is the number of such triples. It has identifying content only if $N \geq 3$. 

Our second additional restriction is a zero cross-difference between the exogenous state $S_t$ and the other firms' current actions $\bs{a}_{-it}$.
\begin{assumption}[Zero cross-difference in $S_t$ and other firms' actions (CD)]\label{assn_prof_cross}
There exists $(a_{it},a_{i,t-1}) \in (\mathcal{A}_i \setminus \{K\}) \times \mathcal{A}_i$ such that
\[
\pi_i(a_{it},\bs{a}_{-it},S_t',a_{i,t-1}) - \pi_i(a_{it},\bs{a}_{-it},S_t,a_{i,t-1})
= \pi_i(a_{it},\widetilde{\bs{a}}_{-it},S_t',a_{i,t-1}) - \pi_i(a_{it},\widetilde{\bs{a}}_{-it},S_t,a_{i,t-1}),
\]
for all $S_t, S_t' \in \mathcal{S}$ and all $\bs{a}_{-it},\widetilde{\bs{a}}_{-it} \in \mathcal{A}_{-i}$.
\end{assumption}
Assumption \ref{assn_prof_cross} is the game-theoretic counterpart of Example \ref{example_zero_cross}, with the exogenous state in the role of $\bs{w}$ and the other firms' actions in the role of $\bs{z}$. It holds when the payoff is additively separable in $S_t$ and $\bs{a}_{-it}$, for example when market size shifts the payoff without interacting with the number of active rivals. As Example \ref{example_zero_cross} shows, separability is sufficient but not necessary. Assumption \ref{assn_prof_cross} implies $(m_s-1)(K^{N-1}-1)$ restrictions for each pair $(a_{it},a_{i,t-1})$ at which it holds, so that it yields $\kappa_{\text{cd}} \cdot (m_s-1)(K^{N-1}-1)$ restrictions in total, where $\kappa_{\text{cd}}$ is the number of such pairs. Unlike Assumption \ref{assn_prof_ex}, it does not require $N \geq 3$.

In our game-theoretic model, stacking restrictions on $\bs{\Pi}_i$ together with \eqref{model_rest_star} and proceeding as in the single-agent case leads to an identifying polynomial in $\beta_i$ of very high degree. Solving for $\bs{\Pi}_i$ requires inverting an $m_\Pi \times m_\Pi$ matrix whose entries, once the inverses $(\bs{I}- \beta_i \bs{Q}_{ik}^*)^{-1}$ in $\bs{P}_i^*(\beta_i)$ have been cleared, are polynomials of degree $m_x-1$ in $\beta_i$. Since $m_\Pi = K^N m_x$, the degree grows exponentially in the number of firms.

Instead, we use the fact that Assumption \ref{assn_prof_K} fixes the expected payoff $\bs{P}_{-i}^{*}\bs{\pi}_{iK}^{-i}$ directly. The expected payoffs under the other actions then follow from \eqref{psi_diff_multi2} row by row, without inverting any matrix of order $m_\Pi$.

Collecting the two inverses in \eqref{psi_diff_multi2} gives $(\bs{I} - \beta_i \bs{Q}_{ik}^*)^{-1}(\bs{P}_{-i}^{*}\bs{\pi}_{ik}^{-i} + \bs{\psi}_{ik}^*) = (\bs{I} - \beta_i \bs{Q}_{iK}^*)^{-1}(\bs{P}_{-i}^{*}\bs{\pi}_{iK}^{-i} + \bs{\psi}_{iK}^*)$, and premultiplying by $\bs{I} - \beta_i \bs{Q}_{ik}^*$ yields
\begin{equation}\label{U_row_game}
\bs{P}_{-i}^{*}\bs{\pi}_{ik}^{-i} = - \bs{\psi}_{ik}^* + (\bs{I} - \beta_i \bs{Q}_{ik}^*)(\bs{I} - \beta_i \bs{Q}_{iK}^*)^{-1} (\bs{P}_{-i}^{*}\bs{\pi}_{iK}^{-i} + \bs{\psi}_{iK}^*), \quad k=1,\ldots,K-1 .
\end{equation}
Equation \eqref{U_row_game} expresses the expected payoff under each action $k \neq K$ in terms of the expected payoff under action $K$, $\bs{P}_{-i}^{*}\bs{\pi}_{iK}^{-i}$. Stacking Assumption \ref{assn_prof_K} for all $\bs{a}_{-i}\in \mathcal{A}_{-i}$ and $\bs{x} \in \mathcal{X}$ as $\bs{\pi}_{iK}^{-i} = \bs{r}_i^{-i}$ and substituting into \eqref{U_row_game} gives
\begin{equation}\label{U_inv_game}
\bs{P}_{-i}^{*}\bs{\pi}_{ik}^{-i} = - \bs{\psi}_{ik}^* + (\bs{I} - \beta_i \bs{Q}_{ik}^*)(\bs{I} - \beta_i \bs{Q}_{iK}^*)^{-1} (\bs{\psi}_{iK}^* + \bs{P}_{-i}^{*}\bs{r}_i^{-i}), \quad k=1,\ldots,K-1 ,
\end{equation}
whose right-hand side is known up to $\beta_i$. Collect the unknown subvectors $\bs{\pi}_{i1}^{-i},\ldots,\bs{\pi}_{i,K-1}^{-i}$ into a vector as
\[
\bs{\Pi}_i^{-K} \coloneqq 
\begin{bmatrix}
\bs{\pi}_{i1}^{-i} \\
\vdots \\
\bs{\pi}_{i,K-1}^{-i}
\end{bmatrix},
\]
where $m_\Pi^{-K} \coloneqq \dim(\bs{\Pi}_i^{-K}) = (K-1)K^{N-1} m_x$. Let $\overline{\bs{P}}_i^{*-K}$ denote the $(K-1)m_x \times m_\Pi^{-K}$ block-diagonal matrix with $K-1$ diagonal blocks equal to $\bs{P}_{-i}^{*}$, so that $\overline{\bs{P}}_i^{*-K}\bs{\Pi}_i^{-K}$ collects the expected payoffs for $k=1,\ldots,K-1$.

Define $d_i(\beta_i) \coloneqq \det(\bs{I} - \beta_i\bs{Q}_{iK}^*)$. Multiplying \eqref{U_inv_game} by $d_i(\beta_i)$ to clear the inverse and stacking over $k=1,\ldots,K-1$ gives
\begin{equation}\label{pi_under_R1}
d_i(\beta_i)\, \overline{\bs{P}}_i^{*-K} \bs{\Pi}_i^{-K} = \bs{b}_i^*(\beta_i),
\end{equation}
where $\bs{b}_i^*(\beta_i) \coloneqq (\bs{b}_{i1}^*(\beta_i),\ldots,\bs{b}_{i,K-1}^*(\beta_i))$ is the $(K-1)m_x \times 1$ vector with block-rows
\[
\bs{b}_{ik}^*(\beta_i) \coloneqq - d_i(\beta_i)\bs{\psi}_{ik}^* + (\bs{I} - \beta_i \bs{Q}_{ik}^*) \adj(\bs{I} - \beta_i \bs{Q}_{iK}^*) (\bs{\psi}_{iK}^* + \bs{P}_{-i}^{*}\bs{r}_i^{-i}), \quad k=1,\ldots,K-1 .
\]
Both sides of \eqref{pi_under_R1} are polynomials of degree at most $m_x$ in $\beta_i$. Write the restriction from Assumption \ref{assn_prof_lagged} as $\bs{R}_2 \bs{\Pi}_i^{-K} = \bs{0}$. Since Assumption \ref{assn_prof_lagged} imposes $(K-1)(K^{N-1}-1)m_x$ restrictions, \eqref{pi_under_R1} and Assumption \ref{assn_prof_lagged} together supply $(K-1)m_x + (K-1)(K^{N-1}-1)m_x = m_\Pi^{-K}$ equations, which matches $\dim(\bs{\Pi}_i^{-K})$. The following assumption adds restrictions on $\bs{\Pi}_i^{-K}$ such as Assumption \ref{assn_prof_ex} or the shape restrictions of Section \ref{sec:examples}.

\begin{assumption}\label{assn_add}
In addition to Assumptions \ref{assn_prof_K} and \ref{assn_prof_lagged}, the payoff function satisfies $q_3$ restrictions of the form $\bs{R}_3\bs{\Pi}_i^{-K}=\bs{c}_3$, where $\bs{R}_3$ is a known $q_3 \times m_{\Pi}^{-K}$ full row rank matrix, and $\bs{c}_3$ is a known $q_3 \times 1$ vector. Combine equation \eqref{pi_under_R1}, Assumption \ref{assn_prof_lagged}, and this restriction as
\begin{equation}\label{XY_beta}
d_i(\beta_i) \bs{X}_i\bs{\Pi}_i^{-K} = \bs{Y}_i(\beta_i),
\end{equation}
where
\[
\underset{((m_{\Pi}^{-K}+q_3 )\times m_{\Pi}^{-K})}{\bs{X}_i} \coloneqq \begin{bmatrix} \ \overline{\bs{P}}_i^{*-K} \ \\ \bs{R}_2 \\ \bs{R}_3\end{bmatrix}, \quad
\underset{((m_{\Pi}^{-K}+q_3) \times 1)}{\bs{Y}_i(\beta_i)} \coloneqq
\begin{bmatrix}
\bs{b}_i^*(\beta_i) \\
\bs{0}\\
d_i(\beta_i) \bs{c}_3
\end{bmatrix}.
\]
Assume that $\bs{X}_i$ has full column rank.
\end{assumption}
We proceed to derive the identifying polynomial in $\beta_i$ under Assumption \ref{assn_add}. Because $\bs{X}_i$ has full column rank, we can, after rearranging rows if necessary, write
\[
\bs{X}_i = \begin{bmatrix}\bs{X}_{i1} \\ \bs{X}_{i2}\end{bmatrix},
\]
where $\bs{X}_{i1}$ is an $m_{\Pi}^{-K}\times m_{\Pi}^{-K}$ invertible matrix. Split $\bs{Y}_i(\beta_i)$ conformably as $\bs{Y}_{i1}(\beta_i)$ and $\bs{Y}_{i2}(\beta_i)$, so that \eqref{XY_beta} becomes $d_i(\beta_i) \bs{X}_{i1}\bs{\Pi}_i^{-K} = \bs{Y}_{i1}(\beta_i)$ and $d_i(\beta_i) \bs{X}_{i2}\bs{\Pi}_i^{-K} = \bs{Y}_{i2}(\beta_i)$. The first equation implies $\bs{\Pi}_i^{-K} = \bs{X}_{i1}^{-1}\bs{Y}_{i1}(\beta_i)/d_i(\beta_i)$. Substituting this into the second equation gives the following proposition.
\begin{proposition}\label{prop_equality_game}
Suppose Assumptions \ref{assn_prof_K}, \ref{assn_prof_lagged}, and \ref{assn_add} hold. Then the identified set of $\beta_i$ is the intersection of the interval $[0, 1)$ and the roots of the following system of at most $q_3$ polynomials, each of degree at most $m_x$
\begin{equation}\label{eq:ident_poly_game}
\bs{X}_{i2} \bs{X}_{i1}^{-1}\bs{Y}_{i1}(\beta_i) - \bs{Y}_{i2}(\beta_i) = \bs{0}.
\end{equation}
Moreover, for each $\beta_i$ in the identified set, the corresponding per-period payoff is identified as $\bs{\Pi}_i^{-K} = \bs{X}_{i1}^{-1}\bs{Y}_{i1}(\beta_i)/d_i(\beta_i)$, and $\bs{\pi}_{iK}^{-i}$ is known by Assumption \ref{assn_prof_K}.
\end{proposition}
The degree of each polynomial in this system is at most $m_x$ because the elements of $\bs{Y}_i(\beta_i)$ are linear functions of $d_i(\beta_i)$ and $\{(\bs{I} - \beta_i \bs{Q}_{ik}^*)\adj(\bs{I} - \beta_i\bs{Q}_{iK}^*)\}_{k=1}^{K-1}$.

\begin{remark}
We allow players to have different discount factors because the identification result in Proposition \ref{prop_equality_game} applies separately to each player. If all players share the same discount factor, the restrictions can be combined across players.
\end{remark}

Inequality restrictions from monotonicity, concavity, and complementarity (Section \ref{subsec:inequality}) can refine the identified set of $\beta_i$.
\begin{assumption}\label{assn_add_inequality}
The payoff function satisfies $\bs{R}_4 \bs{\Pi}_i^{-K} \geq \bs{c}_4$, where $\bs{R}_4$ is a known $q_4\times m_{\Pi}^{-K}$ full row rank matrix and $\bs{c}_4$ is a known $q_4 \times 1$ vector.
\end{assumption}
The following proposition follows immediately from substituting $\bs{\Pi}_i^{-K} = \bs{X}_{i1}^{-1}\bs{Y}_{i1}(\beta_i)/d_i(\beta_i)$ into $\bs{R}_4 \bs{\Pi}_i^{-K} \geq \bs{c}_4$ and multiplying by $d_i(\beta_i)>0$.
\begin{proposition} \label{prop_ineq_game2}
Suppose the assumptions of Proposition \ref{prop_equality_game} and Assumption \ref{assn_add_inequality} hold. Then the identified set of $\beta_i$ is the intersection of the interval $[0, 1)$, the set of the roots of \eqref{eq:ident_poly_game}, and the set of $\beta_i$ satisfying $\bs{R}_4\bs{X}_{i1}^{-1}\bs{Y}_{i1}(\beta_i) - d_i(\beta_i) \bs{c}_4 \geq \bs{0}$.
\end{proposition}

\subsection{Identification of $\beta_i$ without the zero scrap value assumption}\label{sec:beta_no_scrap}

Section \ref{sec: util_diff_game} assumes, in Assumption \ref{assn_prof_K}, that firm $i$'s payoff from being inactive ($a_{it}=K$) is known. We now replace Assumption \ref{assn_prof_K} with two assumptions on that payoff. The first normalizes firm $i$'s payoff to zero when it is inactive in both the current and previous periods ($a_{it}=a_{i,t-1}=K$). It does not imply zero scrap value.
\begin{assumption}[Zero payoff when inactive in both periods]\label{assn_KK}
For all $\bs{a}_{-it} \in \mathcal{A}_{-i}$, $S_t \in \mathcal{S}$, and $\bs{a}_{-i,t-1} \in \mathcal{A}_{-i}$,
\[
\pi_i(a_{it}=K,\bs{a}_{-it},S_t,a_{i,t-1}=K,\bs{a}_{-i,t-1}) = 0.
\]
\end{assumption}
Assumption \ref{assn_KK} imposes $|\mathcal{A}_{-i}|^2 m_s = K^{N-2} m_x$ restrictions on $\pi_i(\bs{a}_{t},S_t,\bs{a}_{t-1})$. The second assumption makes firm $i$'s scrap value independent of other firms' current and lagged actions.
\begin{assumption}[Independence of scrap value from other firms' actions]\label{assn_scrap}
For all $(S_t,a_{i,t-1}) \in \mathcal{S} \times (\mathcal{A}_i \setminus \{K\})$ and all $(\bs{a}_{-it},\bs{a}_{-i,t-1}) \in \mathcal{A}_{-i}^2$,
\[
\pi_i(a_{it}=K,\bs{a}_{-it}=\bs{1}_{N-1},S_t,a_{i,t-1},\bs{a}_{-i,t-1}=\bs{1}_{N-1})
= \pi_i(a_{it}=K,\bs{a}_{-it},S_t,a_{i,t-1},\bs{a}_{-i,t-1}).
\]
\end{assumption}
Assumption \ref{assn_scrap} imposes $(K-1)(|\mathcal{A}_{-i}|^2-1)m_s = (K-1)(K^{2N-2}-1)m_s$ restrictions. We collect Assumptions \ref{assn_KK} and \ref{assn_scrap}, together with other restrictions such as Assumptions \ref{assn_prof_lagged} and \ref{assn_prof_ex} and the shape restrictions of Section \ref{sec:examples}, into the following assumption.

\begin{assumption}\label{assn_reduction}
The payoff function satisfies $\bs{R}\bs{\Pi}_i=\bs{c}$ for a known $r \times m_\Pi$ matrix $\bs{R}$
of full row rank and a known $r \times 1$ vector $\bs{c}$.
\end{assumption}

The following lemma turns a restriction on $\bs{\Pi}_i$ into a restriction on the expected payoff $\overline{\bs{P}}_i^* \bs{\Pi}_i$, which can then be combined with the model restriction \eqref{model_rest_star} as in the single-agent case. Its proof is in Online Appendix \ref{sec:proofs}. Let $\bs{B}$ be an $m_\Pi \times (m_\Pi - r)$ matrix whose columns form a basis of $\Ker(\bs{R})$, and let $\bs{\pi}_i^0$ satisfy $\bs{R}\bs{\pi}_i^0=\bs{c}$. Such $\bs{\pi}_i^0$ exists because $\bs{R}$ has full row rank. Define $\bs{D}$ as a matrix whose rows form a basis of $\Ker\big((\overline{\bs{P}}_i^*\bs{B})\t\big)$, and define $\bs{c}_i^* \coloneqq \bs{D}\,\overline{\bs{P}}_i^*\bs{\pi}_i^0$.

\begin{lemma}[Reduction to the expected payoff]\label{lem_reduction}
$\bs{D}$ does not depend on $\beta_i$ and is identified from the data. Moreover, a vector $\bs{U}_i^*$ can be written as $\bs{U}_i^* = \overline{\bs{P}}_i^*\bs{\Pi}_i$ for some $\bs{\Pi}_i$ satisfying $\bs{R}\bs{\Pi}_i=\bs{c}$ if and only if it satisfies $\bs{D}\bs{U}_i^*=\bs{c}_i^*$.
Consequently, for every $\beta_i \in [0,1)$ there exists $\bs{\Pi}_i$ satisfying
$\bs{R}\bs{\Pi}_i=\bs{c}$ and \eqref{model_rest_star} if and only if there exists
$\bs{U}_i^*$ satisfying $\bs{D}\bs{U}_i^*=\bs{c}_i^*$ and
$\bs{P}_i^*(\beta_i)\bs{U}_i^*=-\bs{P}_i^*(\beta_i)\bs{\Psi}_i^*$.
\end{lemma}

Write $r_i^*$ for the number of rows of $\bs{D}$, so that $\bs{D}$ is $r_i^* \times K m_x$. The reduction discards part of the content of $\bs{R}\bs{\Pi}_i=\bs{c}$. Because the expected payoff averages $\bs{\Pi}_i$ over the current actions of the other firms, payoff vectors that differ by an element of $\Ker(\overline{\bs{P}}_i^*)$ have the same expected payoff. Restrictions that constrain only such differences are lost, so $r_i^*$ may be smaller than $r$. Moreover, the rows of $\bs{D}$ are combinations of the rows of $\bs{R}$ with coefficients that depend on the other firms' equilibrium choice probabilities. They need not have an economic interpretation even when the rows of $\bs{R}$ do. Neither feature impairs the identification of $\beta_i$: by Lemma \ref{lem_reduction} the two systems are satisfiable at exactly the same values of $\beta_i$, because $\beta_i$ enters \eqref{model_rest_star} only through $\bs{P}_i^*(\beta_i)$ acting on the expected payoff. What the reduction gives up is information about $\bs{\Pi}_i$ itself, which this section does not seek to identify.

Collecting the model restriction \eqref{model_rest_star} and $\bs{D} \overline{\bs{P}}_i^* \bs{\Pi}_i = \bs{c}_i^*$ gives the multi-agent counterpart of \eqref{eq:uplus_rest},
\begin{equation} \label{RU_star_rank_deficient}
\bs{R}_i^*(\beta_i) \overline{\bs{P}}_i^* \bs{\Pi}_i = \bs{c}_i^*(\beta_i), \
\text{ where }\
\bs{R}_i^*(\beta_i) \coloneqq
\begin{bmatrix}
\bs{P}_i^*(\beta_i) \\
\bs{D}
\end{bmatrix}, \quad
\bs{c}_i^*(\beta_i) \coloneqq
\begin{bmatrix}
  - \bs{P}_i^*(\beta_i) \bs{\Psi}_i^* \\
\bs{c}_i^*
\end{bmatrix},
\end{equation}
in which $(\bs{R}_i^*(\beta_i), \overline{\bs{P}}_i^* \bs{\Pi}_i,\bs{c}_i^*(\beta_i))$ plays the role of $(\bs{R}(\beta), \bs{U}, \bs{c}(\beta))$. The matrix $\bs{R}_i^*(\beta_i)$ is $\overline{r}_i^* \times K m_x$, where $\overline{r}_i^* \coloneqq (K-1) m_x+r_i^*$. As with $\bs{P}_i^*$, a star indicates an object built from the equilibrium choice probabilities; the counts $r_i^*$, $\overline{r}_i^*$, $q_i^*$, and $s_i^*$ refer to this system. Define $q_i^* \coloneqq \mathrm{rk}(\bs{R}_i^*(\beta_i))$. Because the rows of $\bs{P}_i^*(\beta_i)$ are linearly independent, $q_i^* \geq (K-1)m_x$. When $\bs{R}_i^*(\beta_i)$ does not have full column rank, some elements of $\overline{\bs{P}}_i^* \bs{\Pi}_i$ are not identified.

Combining Sections \ref{sec: util_diff} and \ref{sec:ident_rank_defi}, we treat the full-column-rank and rank-deficient cases of $\bs{R}_i^*(\beta_i)$ together under the following assumption.
\begin{assumption} \label{assn:rank_star}
(a) $(K-1) m_x < q_i^* \leq K m_x$ for all $\beta_i \in [0,1)$. (b) $\overline{r}_i^* > q_i^*$.
\end{assumption}

Under Assumption \ref{assn:rank_star}, we partition $\bs{D}$ and $\bs{c}_i^*$ into blocks $\bs{D}_1$, $\bs{D}_2$, $\bs{c}_{i1}^*$, and $\bs{c}_{i2}^*$ as in \eqref{eq:R_partition}, and $\bs{R}_i^*(\beta_i)$ and $\bs{c}_i^*(\beta_i)$ as in \eqref{eq:Rbeta_partition}. When $\bs{R}_i^*(\beta_i)$ is rank-deficient, we augment $\bs{D}_1$ and $\bs{c}_{i1}^*$ as in \eqref{eq:Rtilde}, which normalizes the elements of the expected payoff of action $K$ at the $m_x-s_i^*$ states in $\mathcal{X}_R$, where $s_i^*\coloneqq q_i^*-(K-1)m_x$. By Proposition \ref{prop_ident_partial}, this leaves the identified set and the identifying polynomials unchanged. Denote the results by $\widetilde{\bs{D}}_1$ and $\widetilde{\bs{c}}_{i1}^*$. When $\bs{R}_i^*(\beta_i)$ has full column rank, $\widetilde{\bs{D}}_1=\bs{D}_1$ and $\widetilde{\bs{c}}_{i1}^*=\bs{c}_{i1}^*$. Let $\bs{Q}_i^*(\beta_i)$ stack $\bs{I}-\beta_i\bs{Q}_{ik}^*$ over $k=1,\ldots,K$, and let $\widetilde{\bs{M}}_i^*(\beta_i)\coloneqq\widetilde{\bs{D}}_1\bs{Q}_i^*(\beta_i)$. Applying Propositions \ref{prop_equality_new} and \ref{prop_ident_partial} to the system \eqref{RU_star_rank_deficient} gives the following corollary.
\begin{corollary} \label{cor_equality_new}
Suppose Assumptions \ref{assn_reduction} and \ref{assn:rank_star} hold. Then the identified set of $\beta_i$ is the intersection of $[0,1)$ with the roots of the following system of at most $\overline{r}_i^*-q_i^*$ polynomials, each of degree at most $m_x$
\begin{equation} \label{eq:prop1_new_game}
\det(\widetilde{\bs{M}}_i^*(\beta_i)) \bs{c}_{i2}^* + \det(\widetilde{\bs{M}}_i^*(\beta_i)) \bs{D}_2 \bs{\Psi}_i^* - \bs{D}_2 \bs{Q}_{i}^*(\beta_i) \adj(\widetilde{\bs{M}}_i^*(\beta_i)) (\widetilde{\bs{D}}_1 \bs{\Psi}_i^* + \widetilde{\bs{c}}_{i1}^*) = \bs{0}.
\end{equation}
When $q_i^*<Km_x$, $\overline{\bs{P}}_i^* \bs{\Pi}_i$ is not identified.
\end{corollary}
Once $\beta_i$ is identified, \eqref{model_rest_star} and $\bs{R}\bs{\Pi}_i=\bs{c}$ identify $\bs{\Pi}_i$ if the matrix that stacks $\bs{P}_i^*(\beta_i)\overline{\bs{P}}_i^*$ and $\bs{R}$ has full column rank $m_\Pi$. Linear inequality restrictions on $\bs{\Pi}_i$ can also be used here. A root of \eqref{eq:prop1_new_game} stays in the identified set only if some $\bs{\Pi}_i$ satisfies \eqref{model_rest_star} and $\bs{R}\bs{\Pi}_i=\bs{c}$ together with these inequalities, which is a linear feasibility problem.

\section{Numerical Example 2: Dynamic Game Model}\label{sec:game_example}

This section illustrates the application of the identifying restrictions to the dynamic game of Section \ref{sec:dynamic_game}. We consider a market with $N = 3$ firms. In each period, firm $i$ decides whether to operate ($a_{it} = 1$) or not ($a_{it} = 2$). The current-period payoff for firm $i$, $\widetilde\pi_i(\bs{a}_{t}, \bs{x}_t,\bs{\ve}_{it})$, is
\[
\theta_{RS} \log S_{t} - \theta_{RN} \log \left(1+\sum_{j\neq i}(2-a_{jt})\right) -\theta_{FC,i} -\theta_{EC}(a_{i,t-1}-1)+\ve_{it}(1), \quad \text{if } a_{it} = 1,
\]
and $sv_i(S_t)\mathbbm{1}\{a_{i,t-1} = 1\}+\ve_{it}(2)$ if $a_{it} = 2$. The shocks $\ve_{it}(1)$ and $\ve_{it}(2)$ are independent type-I extreme value draws. The market size $S_{t}$ follows an exogenous first-order Markov process.\footnote{The market size takes values in $\{2,6,10\}$. Its transition matrix is
$\begin{bmatrix}
0.8&0.2&0.0\\
0.2&0.6&0.2\\
0.0&0.2&0.8
\end{bmatrix} .
$}
We set the firm-specific discount factors to $(\beta_1,\beta_2,\beta_3) = (0.8, 0.9, 0.95)$ and the remaining parameters to $\theta_{RS}=1.0$, $\theta_{RN}=2.0$, $\theta_{EC}=1.0$, $\theta_{FC,1}=1.0$, $\theta_{FC,2}=0.9$, $\theta_{FC,3}=0.8$, and $sv_i(S_t) = c_i S_t$ with $(c_1,c_2,c_3) = (0.05,0.04,0.06)$. Because $sv_i$ is proportional to $S_t$, the scrap value is homogeneous of degree one in $S_t$. The econometrician does not know $sv_i$.

Alongside Assumptions \ref{assn_prof_lagged}, \ref{assn_KK}, and \ref{assn_scrap}, we consider the restriction that the entry cost excludes $S_t$: for all $\bs{a}_{-it}$, $\pi_i(1,\bs{a}_{-it},S_t,1)-\pi_i(1,\bs{a}_{-it},S_t,2)$ does not depend on $S_t$. We maintain these four restrictions throughout and ask which additional assumptions identify $\beta_i$ and $\bs{\Pi}_i$.

Whether the system must be augmented depends on the rank of $\bs{R}_i^*(\beta_i)$. Imposing exchangeability (Assumption \ref{assn_prof_ex}, EX) or the zero cross-difference (Assumption \ref{assn_prof_cross}, CD), alone or together, leaves $\bs{R}_i^*(\beta_i)$ rank-deficient at every $\beta_i \in [0,1)$: its rank is $47$, against $K m_x = 48$. Both parts of Assumption \ref{assn:rank_star} then hold, and \eqref{eq:prop1_new_game} applies to the augmented system. The discount factor is identified, but the payoff function is not. Adding scrap-value homogeneity of degree one in $S_t$ gives $\bs{R}_i^*(\beta_i)$ full column rank, so that \eqref{eq:prop1_new_game} applies without augmentation and $\bs{\Pi}_i$ is identified. Without the entry-cost exclusion, exchangeability or the cross-difference combined with homogeneity reaches only rank $47$, so the entry-cost exclusion and scrap-value homogeneity are both needed for full rank.

Table \ref{tab:id_game_scrap} collects the results. We omit the case with neither EX nor CD, where $\beta_i$ is not identified regardless of scrap-value homogeneity. Here $r_i^*$ is the number of rows of $\bs{D}$, which is the number of restrictions on the expected payoff that remain after the reduction of Lemma \ref{lem_reduction}. These restrictions are generated by Assumptions \ref{assn_prof_lagged}, \ref{assn_KK}, and \ref{assn_scrap}, the exclusion of $S_t$ from the entry cost, and the restriction listed in each row. As explained in Section \ref{sec:beta_no_scrap}, $r_i^*$ is smaller than the number of restrictions these assumptions impose on $\bs{\Pi}_i$ itself. The columns $\overline{r}_i^*$ and $K m_x$ show the numbers of rows and columns of $\bs{R}_i^*(\beta_i)$. Define $\delta_i^* \coloneqq \overline{r}_i^*- \mathrm{rk}(\bs{R}_i^*(\beta_i))$, the number of overidentifying restrictions available to identify $\beta_i$.

\begin{table}[htbp]
\centering
\caption{Identification with an Unknown Scrap Value}
\label{tab:id_game_scrap}
\small
\begin{tabular}{lrrrrr@{\hspace{2em}}rrrrr}
\toprule
 & \multicolumn{5}{c}{Scrap value: unrestricted} & \multicolumn{5}{c}{Scrap value: homogeneous} \\
\cmidrule(lr){2-6} \cmidrule(lr){7-11}
Additional restriction & \multicolumn{1}{c}{$r_i^*$} & \multicolumn{1}{c}{$\overline{r}_i^*$} & \multicolumn{1}{c}{$K m_x$} & \multicolumn{1}{c}{$\mathrm{rk}(\bs{R}_i^*(\beta_i))$} & \multicolumn{1}{c}{$\delta_i^*$} & \multicolumn{1}{c}{$r_i^*$} & \multicolumn{1}{c}{$\overline{r}_i^*$} & \multicolumn{1}{c}{$K m_x$} & \multicolumn{1}{c}{$\mathrm{rk}(\bs{R}_i^*(\beta_i))$} & \multicolumn{1}{c}{$\delta_i^*$} \\
\midrule
EX & 33 & 57 & 48 & 47\rlap{$^\dagger$} & 10 & 35 & 59 & 48 & 48 & 11 \\
CD & 35 & 59 & 48 & 47\rlap{$^\dagger$} & 12 & 37 & 61 & 48 & 48 & 13 \\
EX+CD & 37 & 61 & 48 & 47\rlap{$^\dagger$} & 14 & 39 & 63 & 48 & 48 & 15 \\
\bottomrule
\end{tabular}
\medskip

\raggedright
\small
\textit{Note:} All rows maintain Assumptions \ref{assn_prof_lagged}, \ref{assn_KK}, and \ref{assn_scrap} together with the exclusion of $S_t$ from the entry cost. $\beta_i$ is identified in every case. $^\dagger$\,$\mathrm{rk}(\bs{R}_i^*(\beta_i))<K m_x$, so $\bs{\Pi}_i$ is not identified; in all other cases, $\bs{\Pi}_i$ is identified.
\end{table}

\begin{figure}[h!]
\centering
\includegraphics[width=1.0\linewidth]{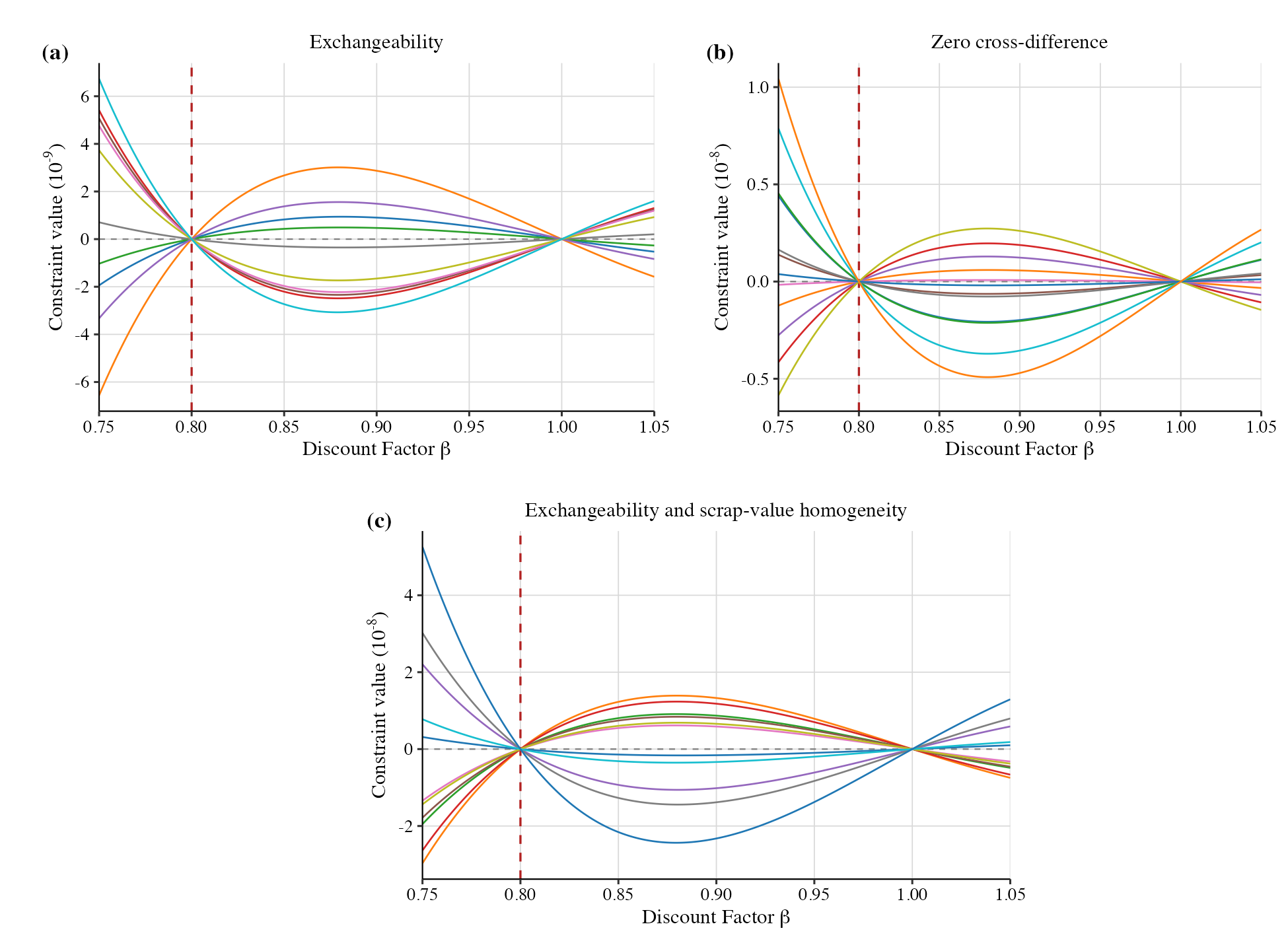}
\caption{Identifying Polynomials for Firm 1's Discount Factor with an Unknown Scrap Value}
\label{fig:dg_scrap}
\end{figure}

Figure \ref{fig:dg_scrap} plots the resulting polynomials for firm 1; for firms 2 and 3, the polynomials have no common root in $[0,1)$ other than their own discount factors, $0.9$ and $0.95$. All three panels plot \eqref{eq:prop1_new_game}, applied to the augmented system in panels (a) and (b). With the scrap value unrestricted, panel (a) plots the $10$ identifying polynomials under EX and panel (b) the $12$ polynomials under CD. Panel (c) plots the $11$ polynomials under EX with scrap-value homogeneity. In every panel, every curve intersects the horizontal axis at the true value $\beta_1 = 0.8$ and at no other value in $[0,1)$. Thus exchangeability or the zero cross-difference identifies the discount factor even when the scrap value is unknown. The payoff itself is identified only if scrap-value homogeneity is also imposed.

\section{Empirical Application: Agricultural Land Use Model}\label{sec:empirical}

We illustrate our approach with a dynamic model of agricultural land use \citep{scott13wp}. After a corn year, planting corn again yields current revenue at the expense of soil fertility. Rotating to soybeans is an investment, because soybeans fix atmospheric nitrogen that benefits the following crop. A farmer's crop choice therefore depends on the discount factor.

\subsection{Data}\label{sec:data}

The data are the Cropland Data Layer (CDL), USDA's satellite-derived annual raster classification of land cover: each $30\,\text{m}$ pixel is assigned a specific crop or land-cover category.
We use the Story County, Iowa (FIPS 19169) tiles for 2010--2023.
The choices are modeled over the 12 years 2012--2023, conditional on the previous year's crop.
The sample consists of the $583{,}390$ interior pixels classified as corn or soybean in all 14 years.
Each pixel stands in for a field, and we use the two terms interchangeably below.
Online Appendix \ref{sec:appendix_data} describes the sample selection and the data sources.

\subsection{The model with latent types}\label{sec:types}

Each year, a field owner chooses $a_t \in \{\text{corn}, \text{soy}\}$. The data pooled across fields strongly violate the first-order Markov property of $a_t$ (Table~\ref{tab:markov} in Online Appendix~\ref{sec:appendix_markov}), plausibly because soil suitability and owners' discount factors vary across fields. We therefore assume that each field belongs to one of $M$ time-invariant latent types, each with its own payoff function and discount factor $\beta_m$. Within type $m$, crop choice is first-order Markov with state $\bs{x}_t=(a_{t-1},D_t)$, where $D_t$ is a market profitability index common to all fields, and the probability of corn is $p(\text{corn}\mid a_{t-1},D_t;m)$. See, for example, \citet{kasaharashimotsu09ecma} for the identification of finite mixtures of dynamic discrete choice models and \citet{am11em} for their estimation. We apply our results type by type.

We proxy the margin of continuous corn over soybeans, net of nitrogen cost, by
\[
D_t = 174\,P_t^{\text{corn}} - 0.1159\,P_t^{N} - 53\,P_t^{\text{soy}},
\]
where $P_t^{\text{corn}}$ and $P_t^{\text{soy}}$ are USDA Risk Management Agency pre-season projected prices, $P_t^N$ is the Oct($t{-}1$)--Jun($t$) average anhydrous ammonia price per ton, $0.1159$ tons/acre is the ammonia rate for continuous corn \citep{sawyer14icm}, and $174$ and $53$ bu/acre are the yields of corn and soybeans after corn \citep{plastina23agdm}. We discretize $D_t$ into low, medium, and high regimes by tercile and keep the notation $D_t$ for the regime. The regime follows an exogenous stationary Markov process that is independent of field history and type. Table~\ref{tab:PD} reports its transition matrix $\bs{P}_D$; we estimate it over 2009--2025 because the price inputs are available beyond the crop panel. Choosing $a_t$ sets next period's crop state to $a_t$.

\begin{table}[htbp]
\centering
\caption{$D_t$ Transition Matrix $\bs{P}_D$ (2009--2025, Jeffreys-Prior Smoothed)}
\label{tab:PD}
\begin{tabular}{lccc}
\toprule
From $\backslash$ To & low & medium & high \\
\midrule
low    & 0.3846 & 0.2308 & 0.3846 \\
medium & 0.2727 & 0.4545 & 0.2727 \\
high   & 0.2941 & 0.1765 & 0.5294 \\
\bottomrule
\end{tabular}
\end{table}

We estimate the type shares and type-specific choice probabilities by maximum likelihood, with $(a_{2011},D_{2012})$ drawn from each type's stationary distribution. We choose $M$ to restore the first-order Markov property. For each $M$, we compute the within-type second-lag gap: the root mean squared difference between the corn probabilities given two lags and given one lag (Online Appendix~\ref{sec:appendix_gap}). The gap is $0.152$, $0.042$, $0.034$, and $0.036$ for $M=1,\ldots,4$ (Table~\ref{tab:M}), so we select $M=3$.

\begin{table}[htbp]
\centering
\begin{threeparttable}
\caption{Choice of the Number of Types}
\label{tab:M}
\begin{tabular}{ccc}
\toprule
Number of types $M$ & Log-likelihood & RMS second-lag gap \\
\midrule
1 & $-$2,798,250 & 0.152 \\
2 & $-$2,308,826 & 0.042 \\
3 & $-$2,277,569 & 0.034 \\
4 & $-$2,271,780 & 0.036 \\
\bottomrule
\end{tabular}
 \begin{tablenotes}[flushleft]
\footnotesize
\item \textit{Notes:} The second-lag gap is the root mean squared difference between $\Pr(a_t=\text{corn}\mid a_{t-1},a_{t-2},D_t;m)$ and $\Pr(a_t=\text{corn}\mid a_{t-1},D_t;m)$, where each cell of type, history and $D_t$ is weighted by its share of transitions under the posterior weights; see equation \eqref{eq:gap} in Online Appendix~\ref{sec:appendix_gap}.
$M=1$ is the model without heterogeneity.
The log-likelihood is that of the choices over 2012--2023 given the 2011 crop. It equals the joint log-likelihood minus the stationary initial-condition term for 2011.
\end{tablenotes}
\end{threeparttable}
\end{table}

Table~\ref{tab:types} reports the estimates. The types differ mainly in how often corn follows corn: rotators ($49.5\%$ of fields) plant corn after corn $2$--$5\%$ of the time, partial rotators ($33.1\%$) $39$--$43\%$ of the time, and corn-heavy fields ($17.5\%$) $66$--$83\%$ of the time. All three types almost always plant corn after soybeans.

\begin{table}[htbp]
\centering
\begin{threeparttable}
\caption{Estimated Types: Probability of Planting Corn}
\label{tab:types}
\begin{tabular}{lccccccc}
\toprule
 & & \multicolumn{3}{c}{Corn after corn} & \multicolumn{3}{c}{Corn after soybean} \\
\cmidrule(lr){3-5}\cmidrule(lr){6-8}
Type & Share & low & medium & high & low & medium & high \\
\midrule
Rotators & 0.495 & 0.021 & 0.019 & 0.048 & 0.997 & 0.997 & 0.995 \\
Partial rotators & 0.331 & 0.385 & 0.431 & 0.416 & 0.995 & 0.972 & 0.976 \\
Corn-heavy & 0.175 & 0.662 & 0.779 & 0.827 & 0.992 & 1.000 & 0.989 \\
\bottomrule
\end{tabular}
 \begin{tablenotes}[flushleft]
\footnotesize
\item \textit{Notes:} Types are ordered by the probability of planting corn after corn and labeled by their behavior.
Each entry is the estimated $p(\text{corn}\mid a_{t-1},D_t;m)$ for the crop planted last year and the profitability regime $D_t$ of the current year.
Shares are rounded and may not sum to one.
\end{tablenotes}
\end{threeparttable}
\end{table}

\subsection{Identification and results}\label{sec:result}

Let $u_{m,k}(a_{t-1},D_t)$ denote the per-period payoff of type $m$ from planting crop $k\in\{\text{corn},\text{soy}\}$ in state $(a_{t-1},D_t)$, so that the field owner receives $u_{m,k}(a_{t-1},D_t)+\ve_k$, and let $\bs{u}_{m,k}$ stack it over the six states. We normalize the soybean payoff to $\bs{u}_{m,\text{soy}}=\bs{0}$, assume that $(\ve_{\text{corn}},\ve_{\text{soy}})$ are i.i.d.\ type-I extreme value, and require the continuous-corn payoff $u_{m,\text{corn}}(\text{corn},D_t)$ to be weakly increasing in $D_t$. These restrictions take the form $\bs{R}_{\text{corn}}\bs{u}_{m,\text{corn}} \ge \bs{0}$, so by \eqref{ident_ineq_known} the identified set for $\beta_m$ is the set of $\beta \in [0,1)$ satisfying
\begin{equation}\label{eq:u_corn_closed_form}
\det(\bs{I}- \beta \bs{Q}_{m,\text{soy}}) \bs{R}_{\text{corn}} \bs{\psi}_{m,\text{corn}} - \bs{R}_{\text{corn}} (\bs{I}-\beta\bs{Q}_{m,\text{corn}}) \adj (\bs{I}-\beta\bs{Q}_{m,\text{soy}}) \bs{\psi}_{m,\text{soy}} \leq \bs{0},
\end{equation}
where $\bs{\psi}_{m,k}$ and $\bs{Q}_{m,k}$ are constructed from type $m$'s choice probabilities and $\bs{P}_D$. Because $D_t$ is built partly from soybean prices, the true soybean payoff likely varies with $D_t$. Online Appendix \ref{sec:appendix_norm} shows that $u_{m,\text{corn}}(\text{corn},D_t)$ nonetheless equals the current payoff margin of corn over soybean after corn, up to a constant, provided that the expected effect of this year's crop on next year's soybean payoff is the same in every regime. We do not impose monotonicity after soybeans because the probability of corn after soybeans exceeds $0.97$ in every type and does not vary systematically with $D_t$. These probabilities still enter \eqref{eq:u_corn_closed_form} through $\bs{\psi}_{m,\text{soy}}$ and $\bs{Q}_{m,\text{soy}}$ because planting soybeans today leads to the state $(\text{soy},D_{t+1})$ and thus governs the continuation value. We compute the set on a grid of $\beta$ with step $0.001$. Given the choice probabilities and $\bs{P}_D$, each $\beta$ pins down the two continuous-corn payoff increments, from the low to the medium regime and from the medium to the high regime. To measure how far the restriction is from being violated, we report the slack, the largest value over $\beta$ of the smaller of these increments. A negative slack means that the set is empty.

Table~\ref{tab:results} reports the results. For rotators, no $\beta \in [0,1)$ satisfies the monotonicity restriction. Their probability of corn after corn is below $5\%$ and hardly responds to the market regime, so we read the empty set as a failure of the restriction for this type rather than as evidence on its discount factor. For partial rotators, the identified set is $[0.681,0.848]$ with a slack of $0.01$. The exclusion of $\beta=0$ does not depend on $\bs{P}_D$. At $\beta=0$ the payoff equals the log odds of corn after corn, which must then increase with $D_t$, but for partial rotators the probability of corn after corn falls from $0.431$ in the medium regime to $0.416$ in the high regime. We report no set for corn-heavy fields, because their probability of corn after soybeans in the medium regime is within $0.001$ of $1$, where $\psi_k(\bs{p})=\gamma-\log p_k$ diverges.

The bounds for partial rotators have an intuitive interpretation. Almost every field plants corn after soybeans (Table~\ref{tab:types}), whereas partial rotators plant corn after corn only about $40\%$ of the time, so a field is worth more after soybeans than after corn. Planting corn today therefore costs continuation value, and continuous corn must pay enough today to offset this loss. The loss varies across regimes, and this variation gives both bounds. From low to medium $D_t$, the probability of continuous corn rises while the loss shrinks. If $\beta_m$ were too large, the smaller loss alone would more than explain the rise, and the implied current payoff would fall and violate monotonicity. This gives the upper bound. From medium to high $D_t$, the probability falls slightly while the loss grows. If $\beta_m$ were too small, the larger loss could not explain the fall unless the current payoff also fell. This gives the lower bound.

\begin{table}[htbp]
\centering
\begin{threeparttable}
\caption{Identified Sets for $\beta_m$ by Type}
\label{tab:results}
\begin{tabular}{lcccccccc}
\toprule
 & \multicolumn{3}{c}{Mixture} & \multicolumn{2}{c}{$k$-means (a)} & \multicolumn{2}{c}{$k$-means (b)} \\
\cmidrule(lr){2-4}\cmidrule(lr){5-6}\cmidrule(lr){7-8}
Type & Share & Identified set & Slack & Share & Identified set & Share & Identified set \\
\midrule
Rotators & 0.495 & empty & $-$0.05 & 0.610 & $[0,\,0.780]$ & 0.610 & $[0,\,0.828]$ \\
Partial rotators & 0.331 & $[0.681,\,0.848]$ & 0.01 & 0.247 & $[0,\,0.954]$ & 0.239 & $[0,\,0.771]$ \\
Corn-heavy & 0.175 & -- & -- & 0.143 & $[0,\,0.551]$ & 0.152 & $[0,\,0.111]$ \\
\bottomrule
\end{tabular}
 \begin{tablenotes}[flushleft]
\footnotesize
\item \textit{Notes:} The identified set for each type is the set of $\beta\in[0,1)$ at which the payoff from planting corn after corn is weakly increasing in $D_t$.
The slack is the largest value over $\beta$ of the smaller of the two payoff increments implied by $\beta$.
The $k$-means groups agree with the mixture's most likely type for $89.9\%$ and $90.7\%$ of pixels.
\end{tablenotes}
\end{threeparttable}
\end{table}

As a robustness check, we replace the mixture with $k$-means clustering into three groups \citep{bonhomme15em,bonhomme22em}, using (a) each field's corn share and corn-to-corn transition rate or (b) the corn-to-corn rate alone. Table~\ref{tab:kmeans} in Online Appendix~\ref{sec:appendix_kmeans} shows that both groupings recover the same three cropping patterns. For partial rotators, the identified sets in Table~\ref{tab:results} are $[0,0.954]$ and $[0,0.771]$, respectively, so discount factors in $[0.681,0.771]$ lie in the identified set under all three groupings. Unlike the mixture, $k$-means gives a nonempty set for rotators and a set for corn-heavy fields.

These results have two limitations. First, the bounds rest on small differences in choice probabilities across regimes, so they depend on how the types are classified, as the columns of Table~\ref{tab:results} show. Second, we report plug-in estimates of the identified sets. Estimation error in the choice and transition probabilities carries over to the bounds, so a confidence set would be wider. This error is likely largest for $\bs{P}_D$, which is estimated from $16$ annual transitions. Without Jeffreys-prior smoothing of $\bs{P}_D$, re-estimating the types gives $[0.459,0.629]$ for partial rotators. We leave inference for future research.

\section{Conclusion}\label{sec:conclusion}

This paper studies the identification of the discount factor and payoff function in stationary infinite-horizon dynamic discrete choice models. We show that nonparametric assumptions on per-period payoffs implied by economic theory, such as homogeneity, monotonicity, concavity, zero cross-differences, and complementarity, yield polynomial equalities and inequalities in the discount factor. These restrictions can identify the discount factor and the payoff function without normalizing the payoff of one action. They can identify the discount factor even when the payoff function is not point identified. In dynamic games, the irrelevance of other firms' lagged actions, combined with restrictions such as exchangeability, can identify firm-specific discount factors, even when the scrap value is unknown. In an application to farmers' crop choices in Story County, Iowa, we allow for latent types with type-specific payoffs and discount factors. Monotonicity of the payoff in market profitability gives the identified set $[0.681,0.848]$ for partial rotators. An important direction for future research is to develop estimation and inference procedures that implement these identification strategies.

{\titlespacing*{\section}{0pt}{2.5ex plus 1ex minus .2ex}{0.5ex plus .2ex}
\section*{Acknowledgements}
The authors thank Jaap Abbring for his detailed and insightful comments on an earlier version of the paper. The authors thank Eric Mbakop, Whitney Newey, Tang Srisuma, Quang Vuong, participants at the LMU-Todai workshop, Cowles Foundation 2025 Conference on Econometrics, and seminar attendees at the National University of Singapore, Nanyang Technological University, and the Pennsylvania State University for their valuable feedback.

\titlespacing*{\section}{0pt}{0pt plus .3ex}{0.5ex plus .2ex}
\section*{Funding}
This work was supported by the Research Grants Council of Hong Kong General Research Fund Project Number 17622225, SSHRC, and JSPS KAKENHI Grant Number JP24K04814.

\section*{Conflict of interest}
The authors declare no conflicts of interest.

}

{\singlespacing
\setlength{\bibsep}{1pt}
\bibliography{ddc}
}

\clearpage
\setcounter{page}{1}
\renewcommand{\thepage}{OA-\arabic{page}}
\begin{center}
{\Large\textbf{Online Appendix}}\\[1ex]
{\large Semiparametric Identification of the Discount Factor and Payoff Function in Dynamic Discrete Choice Models}\\[1ex]
Yu Hao, Hiroyuki Kasahara, and Katsumi Shimotsu
\end{center}

\begin{appendix}

\section{Proofs of Propositions and Lemmas}\label{sec:proofs}

Proofs are presented in order of appearance in the main text. The proof of Proposition \ref{prop_log_util_diff} is in Online Appendix \ref{sec:log_diff}.

\subsection{Proof of Proposition \ref{prop_equality_new}}

Part (a) follows immediately from \eqref{R1R2}. For part (b), by part (a), $\bs{U}=\bs{R}_1(\beta)^{-1}\bs{c}_1(\beta)$, and by \eqref{RC_partition},
\begin{equation}\label{C_decomp}
\bs{c}_1(\beta) = \begin{bmatrix}
- \bs{P}(\beta) \bs{\Psi} \\
\bs{c}_1
\end{bmatrix} 
=
-
\begin{bmatrix}
\bs{P}(\beta) \\
\bs{R}_1
\end{bmatrix}
\bs{\Psi}
+
\begin{bmatrix}
\bs{0} \\
\bs{R}_1\bs{\Psi} + \bs{c}_1
\end{bmatrix}
= 
- \bs{R}_1(\beta)\bs{\Psi}
+
\begin{bmatrix}
\bs{0} \\
\bs{R}_1\bs{\Psi} + \bs{c}_1
\end{bmatrix} .
\end{equation}
From the formula for the inverse of a partitioned matrix, we obtain
\[
\bs{R}_1(\beta)^{-1} =
\begin{bmatrix}
(\bs{I}- \beta \bs{Q}_1)^{-1} & \cdots &\bs{O} & -(\bs{I}- \beta \bs{Q}_{K})^{-1}\\
\vdots & \ddots & \vdots & \vdots \\
\bs{O} & \cdots & (\bs{I}- \beta \bs{Q}_{K-1})^{-1} & - (\bs{I}- \beta \bs{Q}_{K})^{-1}\\
\bs{R}_{11} & \cdots &\bs{R}_{1,K-1} & \bs{R}_{1K}
\end{bmatrix}^{-1}
=
\begin{bmatrix}
\bs{R}^{1}(\beta) & \bs{Q}(\beta) \bs{M}(\beta)^{-1}
\end{bmatrix} , 
\]
for a $JK \times J(K-1)$ matrix $\bs{R}^{1}(\beta)$. Consequently, multiplying \eqref{C_decomp} by $\bs{R}_1(\beta)^{-1}$ from the left gives 
\begin{equation}\label{R_inv_C}
\bs{U}=\bs{R}_1(\beta) ^{-1} \bs{c}_1(\beta) = - \bs{\Psi} + \bs{Q}(\beta) \bs{M}(\beta)^{-1} (\bs{R}_1 \bs{\Psi} + \bs{c}_1) .
\end{equation}
Substituting \eqref{R_inv_C} into $\bs{R}_2\bs{U}=\bs{c}_2$ gives $\bs{c}_2 = - \bs{R}_2\bs{\Psi} + \bs{R}_2\bs{Q}(\beta) \bs{M}(\beta)^{-1} (\bs{R}_1 \bs{\Psi} + \bs{c}_1)$. Writing $\bs{M}(\beta)^{-1} = \adj(\bs{M}(\beta))/\det(\bs{M}(\beta))$ and multiplying by $\det(\bs{M}(\beta))$ yields \eqref{eq:prop1_new}. Each entry of $\adj(\bs{M}(\beta))$ is, up to sign, the determinant of a $(J-1)\times(J-1)$ submatrix of $\bs{M}(\beta)$. Because $\bs{M}(\beta)$ has entries linear in $\beta$, $\det(\bs{M}(\beta))$ and the entries of $\adj(\bs{M}(\beta))$ are polynomials of degree at most $J$ and $J-1$, respectively. Because $\bs{Q}(\beta)$ is also linear in $\beta$, \eqref{eq:prop1_new} consists of polynomials of degree at most $J$, proving (b). \qedsymbol

\subsection{Proof of Proposition \ref{prop_beta_1}}

Because each row of $\bs{Q}_k$ is a probability distribution over $\bs{x}_{t+1}$ conditional on $\bs{x}_t$, it follows that $\bs{Q}_k \bs{1}_J=\bs{1}_J$. Hence, $(\bs{I} - \bs{Q}_k)\bs{1}_J=\bs{0}$ holds for all $k$, and $\bs{M}(1)\bs{1}_J=\bs{0}$ follows from the definition of $\bs{M}(\beta)$. Therefore, $\bs{M}(1)$ is singular, and $\det\left( \bs{M}(1)\right)=0$ follows.

We proceed to show $\bs{Q}(1) \adj(\bs{M}(1))=\bs{O}$. Recall $\mathrm{rk}(\bs{M}(1)) \leq J-1$ because $\bs{M}(1)$ is singular. From Theorem 3.2 of \citet{magnus19book}, $\mathrm{rk}(\adj( \bs{M}(1)))=1$ if $\mathrm{rk}( \bs{M}(1))= J-1$ and $\mathrm{rk}(\adj( \bs{M}(1)))=0$ if $\mathrm{rk}( \bs{M}(1)) \leq J-2$. First, suppose $\mathrm{rk}(\bs{M}(1)) \leq J-2$. Then, $\adj(\bs{M}(1))$ is the zero matrix, and $\bs{Q}(1) \adj(\bs{M}(1))=\bs{O}$ holds trivially. Next, suppose $\mathrm{rk}( \bs{M}(1)) = J-1$. Then, the null space of $\bs{M}(1)$ is one-dimensional and spanned by $\bs{1}_J$. By Theorem 3.1(b) of \citet{magnus19book}, it follows that $\adj( \bs{M}(1))= c \cdot \bs{1}_J \bs{a}\t$ for some scalar $c$ and vector $\bs{a}$. Consequently, we have $( \bs{I} - \bs{Q}_{k}) \adj(\bs{M}(1))=c \cdot ( \bs{I} - \bs{Q}_{k})\bs{1}_J \bs{a}\t=\bs{O}$ for all $k$, and $\bs{Q}(1) \adj(\bs{M}(1))=\bs{O}$ follows from the definition of $\bs{Q}(\beta)$. \qedsymbol

\subsection{Proof of Proposition \ref{prop_ineq}}
It follows from $\bs{R}_{\text{ineq}}\bs{U} \geq \bs{c}_{\text{ineq}}$ and \eqref{R_inv_C} that $\bs{c}_{\text{ineq}} \leq - \bs{R}_{\text{ineq}}\bs{\Psi} + \bs{R}_{\text{ineq}}\bs{Q}(\beta) \bs{M}(\beta)^{-1} (\bs{R}_1 \bs{\Psi} + \bs{c}_1)$. Using $\bs{M}(\beta)^{-1} = \adj (\bs{M}(\beta)) / \det (\bs{M}(\beta))$, this is equivalent to $\bs{g}_{\text{ineq}}(\beta) / \det(\bs{M}(\beta)) \leq \bs{0}$, where $\bs{g}_{\text{ineq}}(\beta)$ is defined in \eqref{eq:prop2}. Multiplying by $\det(\bs{M}(\beta))$ yields $\bs{g}_{\text{ineq}}(\beta) \leq \bs{0}$ when $\det(\bs{M}(\beta))>0$ and $\bs{g}_{\text{ineq}}(\beta) \geq \bs{0}$ when $\det(\bs{M}(\beta))<0$. \qedsymbol

\subsection{Proof of Proposition \ref{prop_ident_partial}}
Fix $\beta\in[0,1)$ and $\bs{v}$, and define
\[
\widetilde{\bs{R}}_1(\beta)\coloneqq\begin{bmatrix}\bs{P}(\beta)\\ \widetilde{\bs{R}}_1\end{bmatrix} = \begin{bmatrix} \multicolumn{2}{c}{\bs{R}_1(\beta)} \\ \bs{O}_{J-s,J(K-1)} & \bs{E}_{\mathcal{X}_R} \end{bmatrix},
\quad \widetilde{\bs{c}}_1(\beta)\coloneqq\begin{bmatrix}-\bs{P}(\beta)\bs{\Psi}\\ \widetilde{\bs{c}}_1\end{bmatrix} = \begin{bmatrix}\bs{c}_1(\beta)\\ \bs{v}\end{bmatrix}.
\]
By \eqref{eq:R_partition}, $\beta$ is in the identified set if and only if there exists $\widehat{\bs{U}}$ such that $\bs{R}_1(\beta)\widehat{\bs{U}}=\bs{c}_1(\beta)$ and $\bs{R}_2\widehat{\bs{U}}=\bs{c}_2$. We show that this holds if and only if there exists $\widetilde{\bs{U}}$ such that $\widetilde{\bs{R}}_1(\beta)\widetilde{\bs{U}}=\widetilde{\bs{c}}_1(\beta)$ and $\bs{R}_2\widetilde{\bs{U}}=\bs{c}_2$. The ``if'' part holds with $\widehat{\bs{U}}=\widetilde{\bs{U}}$ because the first $J(K-1)+s$ rows of $\widetilde{\bs{R}}_1(\beta)$ and $\widetilde{\bs{c}}_1(\beta)$ are $\bs{R}_1(\beta)$ and $\bs{c}_1(\beta)$. For the ``only if'' part, let $\widehat{\bs{u}}_K$ denote the $K$th block of $\widehat{\bs{U}}$. In the column order of \eqref{eq:Rbeta_partition}, let $\bs{d}\coloneqq(\bs{d}_L\t,\bs{d}_R\t)\t$ with $\bs{d}_R\coloneqq\bs{v}-\bs{E}_{\mathcal{X}_R}\widehat{\bs{u}}_K$ and $\bs{d}_L\coloneqq-\bs{R}_{1L}(\beta)^{-1}\bs{R}_{1R}(\beta)\bs{d}_R$, so that $\bs{R}_1(\beta)\bs{d}=\bs{0}$. Because $\mathrm{rk}(\bs{R}(\beta))=\mathrm{rk}(\bs{R}_1(\beta))$, the rows of $\bs{R}_2$ lie in the row space of $\bs{R}_1(\beta)$, so $\bs{R}_2\bs{d}=\bs{0}$. Hence $\widetilde{\bs{U}}\coloneqq\widehat{\bs{U}}+\bs{d}$ satisfies $\bs{R}_1(\beta)\widetilde{\bs{U}}=\bs{c}_1(\beta)$, $\bs{R}_2\widetilde{\bs{U}}=\bs{c}_2$, and $\bs{E}_{\mathcal{X}_R}\widetilde{\bs{u}}_K=\bs{E}_{\mathcal{X}_R}\widehat{\bs{u}}_K+\bs{E}_{\mathcal{X}_R}\bs{d}_K = \bs{E}_{\mathcal{X}_R}\widehat{\bs{u}}_K+\bs{d}_R=\bs{v}$, where $\bs{d}_K$ is the $K$th block of $\bs{d}$. Therefore, $\widetilde{\bs{R}}_1(\beta)\widetilde{\bs{U}}=\widetilde{\bs{c}}_1(\beta)$ and $\bs{R}_2\widetilde{\bs{U}}=\bs{c}_2$.

After reordering its columns, $\widetilde{\bs{R}}_1(\beta)$ equals $[\bs{R}_{1L}(\beta),\bs{R}_{1R}(\beta);\bs{O},\bs{I}_{J-s}]$, which is nonsingular. Hence the augmented system satisfies the assumptions of Proposition \ref{prop_equality_new} with $(\bs{R}_1,\bs{c}_1)$ replaced by $(\widetilde{\bs{R}}_1,\widetilde{\bs{c}}_1)$, and Proposition \ref{prop_equality_new} gives \eqref{eq:prop_partial}, whose polynomials have degree at most $J$ because $\widetilde{\bs{M}}(\beta)$ is linear in $\beta$.

It remains to show that the polynomials do not depend on $\bs{v}$. Because $\bs{v}$ enters \eqref{eq:prop_partial} only through $\widetilde{\bs{c}}_1$, it suffices to show that $\bs{R}_2\bs{\Delta}=\bs{0}$ for $\bs{\Delta}\coloneqq\bs{Q}(\beta)\adj(\widetilde{\bs{M}}(\beta))(\bs{0}\t,\bs{v}\t)\t$. First, $\bs{P}(\beta)\bs{\Delta}=\bs{0}$ because $\bs{P}(\beta)\bs{Q}(\beta)=\bs{O}$ by \eqref{eq:Pbeta} and \eqref{eq:def_MQ}. Second, $\bs{R}_1\bs{\Delta}=\bs{0}$ because 
\[
\widetilde{\bs{R}}_1\bs{\Delta}=\widetilde{\bs{M}}(\beta)\adj(\widetilde{\bs{M}}(\beta))\begin{bmatrix}\bs{0}\\ \bs{v}\end{bmatrix}=\begin{bmatrix}\bs{0}\\ \det(\widetilde{\bs{M}}(\beta)) \bs{v}\end{bmatrix},
\]
and $\bs{R}_1\bs{\Delta}$ consists of the first $s$ elements of $\widetilde{\bs{R}}_1\bs{\Delta}$. Hence $\bs{R}_1(\beta)\bs{\Delta}=\bs{0}$. Because $\mathrm{rk}(\bs{R}(\beta))=\mathrm{rk}(\bs{R}_1(\beta))$, the rows of $\bs{R}_2$ lie in the row space of $\bs{R}_1(\beta)$, so $\bs{R}_2\bs{\Delta}=\bs{0}$. \qedsymbol

\subsection{Proof of Proposition \ref{prop_beta_1_partial}}
Equation \eqref{eq:prop_partial} is \eqref{eq:prop1_new} for the augmented system, and $\widetilde{\bs{M}}(1)\bs{1}_J=\widetilde{\bs{R}}_1\bs{Q}(1)\bs{1}_J=\bs{0}$ because $(\bs{I}-\bs{Q}_k)\bs{1}_J=\bs{0}$ for all $k$. The argument in the proof of Proposition \ref{prop_beta_1} therefore applies with $\bs{M}(\beta)$ replaced by $\widetilde{\bs{M}}(\beta)$. Since each polynomial in \eqref{eq:prop_partial} has degree at most $J$ and vanishes at $\beta=1$, at most $J-1$ of its roots lie in $[0,1)$. \qedsymbol

\subsection{Proof of Lemma \ref{lem_fd_suff}}
Since $\bs{R}_1 = [\bs{O}_{J,J(K-1)},\, \bs{I}_J]$, we have $\bs{M}(\beta) = \bs{I} - \beta \bs{Q}_K$ and $\bs{M}(\beta)^{-1} = (\bs{I} - \beta\bs{Q}_K)^{-1} = \sum_{j \geq 0}\beta^j \bs{Q}_K^j$ for $\beta \in [0,1)$. Let $(\bs{I} - \beta \bs{Q}_k)(\bs{x})$ denote the row of $\bs{I}-\beta\bs{Q}_k$ corresponding to $\bs{x}$ and $\bs{Q}_k(\bs{x})$ the $\bs{x}$-row of $\bs{Q}_k$, and let $\bs{e}_{\bs{x}}$ be the unit column vector selecting state $\bs{x}$, so that $(\bs{I} - \beta \bs{Q}_k)(\bs{x}) = \bs{e}_{\bs{x}}\t - \beta \bs{Q}_k(\bs{x})$. By linearity it suffices to bound the degree of the row of $\bs{Q}(\beta) \bs{M}(\beta)^{-1}$ associated with a single utility difference $u_{k_a}(\bs{x}_a) - u_{k_b}(\bs{x}_b)$, namely
\begin{equation} \label{fd_eq_2}
\begin{aligned}
& \left[(\bs{I}-\beta\bs{Q}_{k_a})(\bs{x}_a) - (\bs{I}-\beta \bs{Q}_{k_b})(\bs{x}_b)\right](\bs{I} - \beta\bs{Q}_K)^{-1} \\
&= (\bs{e}_{\bs{x}_a}-\bs{e}_{\bs{x}_b})\t (\bs{I}-\beta\bs{Q}_K)^{-1} - \beta \left(\bs{Q}_{k_a}(\bs{x}_a) - \bs{Q}_{k_b}(\bs{x}_b)\right)(\bs{I}-\beta\bs{Q}_K)^{-1}.
\end{aligned}
\end{equation}
Using $\bs{e}_{\bs{x}}\t\bs{Q}_K = \bs{Q}_K(\bs{x})$, the coefficient of $\beta^j$ for $j \geq 1$ in \eqref{fd_eq_2} equals
\[
(\bs{e}_{\bs{x}_a}-\bs{e}_{\bs{x}_b})\t\bs{Q}_K^j - \left(\bs{Q}_{k_a}(\bs{x}_a) - \bs{Q}_{k_b}(\bs{x}_b)\right)\bs{Q}_K^{j-1}
= -\left[\bs{Q}_{k_a}(\bs{x}_a) - \bs{Q}_{k_b}(\bs{x}_b) - \bs{Q}_K(\bs{x}_a) + \bs{Q}_K(\bs{x}_b)\right]\bs{Q}_K^{j-1}.
\]
By Assumption \ref{assn_finite}, $\left[\bs{Q}_{k_a}(\bs{x}_a) - \bs{Q}_{k_b}(\bs{x}_b) - \bs{Q}_K(\bs{x}_a) + \bs{Q}_K(\bs{x}_b)\right]\bs{Q}_K^{\rho} = \bs{0}$, so this coefficient vanishes for $j-1 \geq \rho$, and \eqref{fd_eq_2} is a polynomial in $\beta$ of degree at most $\rho$. Since every restriction in $\bs{R}_2$ is a linear combination of such differences, every entry of $\bs{R}_2 \bs{Q}(\beta) \bs{M}(\beta)^{-1}$ has degree at most $\rho$. \qedsymbol

\subsection{Proof of Proposition \ref{prop_fd}}
Because \eqref{eq:prop_fd} is obtained by dividing \eqref{eq:prop1_new} by $\det(\bs{M}(\beta)) \neq 0$, the two have the same roots in $[0,1)$, and the identification claim follows from Proposition \ref{prop_equality_new}. Each entry of $\bs{R}_2 \bs{Q}(\beta) \bs{M}(\beta)^{-1}$ is a polynomial of degree at most $\rho$ by assumption, and $\bs{c}_2 + \bs{R}_2 \bs{\Psi}$ does not depend on $\beta$, so the left-hand side of \eqref{eq:prop_fd} is a system of at most $(r-J)$ polynomials each of degree at most $\rho$. \qedsymbol

\subsection{Proof of Lemma \ref{psi_lemma}}
By \eqref{v_k_x_am} and \eqref{int_Bellman}, $v_i^*$ and $V_i^*$ are the choice-specific and integrated value functions of a single-agent dynamic discrete choice problem with per-period payoff $\pi_i^*(a_i,\bs{x})$ and transition probability $Q_i^*(\bs{x}'|\bs{x},a_i)$, and $\sigma_i^*$ is its optimal decision rule. The first result therefore follows from Lemma 1 of \citet{am11em}, and the second from their Lemma 3. \qedsymbol

\subsection{Proof of Lemma \ref{lem_reduction}}
The columns of $\bs{B}$ span $\Ker(\bs{R})$, so $\bs{R}\bs{\Pi}_i=\bs{c}$ if and only if $\bs{\Pi}_i = \bs{\pi}_i^0 + \bs{B}\bs{z}$ for some $\bs{z}$. Premultiplying by $\overline{\bs{P}}_i^*$ gives $\bs{U}_i^* = \overline{\bs{P}}_i^*\bs{\pi}_i^0 + \overline{\bs{P}}_i^*\bs{B}\bs{z}$, so the expected payoffs attainable under $\bs{R}\bs{\Pi}_i=\bs{c}$ are $\overline{\bs{P}}_i^*\bs{\pi}_i^0$ plus an arbitrary vector in the column space of $\overline{\bs{P}}_i^*\bs{B}$. By the definition of $\bs{D}$, the columns of $\overline{\bs{P}}_i^*\bs{B}$ span $\Ker(\bs{D})$, because the null space of a matrix is the orthogonal complement of the column space of its transpose \citep[][B.2.1]{SeberLee03book}. Therefore, these expected payoffs are $\overline{\bs{P}}_i^*\bs{\pi}_i^0$ plus an arbitrary vector in $\Ker(\bs{D})$, which is the solution set of $\bs{D}\bs{U}_i^* = \bs{c}_i^*$ by the definition of $\bs{c}_i^*$. The final claim follows because \eqref{model_rest_star} constrains $\bs{\Pi}_i$ only through $\overline{\bs{P}}_i^*\bs{\Pi}_i$. \qedsymbol

\section{Identification of $\beta$ by Log Differences in Payoffs}\label{sec:log_diff}

In some applications, economic restrictions are linear in log-payoffs rather than in payoffs. Such restrictions also identify $\beta$.

For a vector $\bs{y}=(y_1,\ldots,y_n)\t$ with positive elements, define $\log(\bs{y})$ as $(\log(y_1),\ldots, \log(y_n))\t$. The following assumption is the counterpart of Assumption \ref{assn_eq} for log-payoffs. It also requires that $\bs{u}_K$ be known and that the elements of $\bs{r}$ sum to zero. Examples \ref{example_exp_homo} and \ref{example_homo_unknown} below satisfy both conditions.
\begin{assumption}[Log-payoff differences] \label{assn_log_diff}
$\bs{u}_K = \bs{c}_1$ for a known $\bs{c}_1$. The payoff function $\bs{U}_{-K}\coloneqq (\bs{u}_1,\ldots,\bs{u}_{K-1})$ has positive elements and satisfies $\bs{r}\t \log(\bs{U}_{-K})=c$ for a known nonzero vector $\bs{r}=(r_1,\ldots,r_{J(K-1)})\t$ and a known constant $c$. Further, the elements of $\bs{r}$ sum to zero.
\end{assumption}
Define $\bs{\Psi}_{-K}$ and $\bs{Q}(\beta)_{-K}$ as the $J(K-1)\times 1$ vector and $J(K-1)\times J$ matrix obtained by removing the last $J$ rows from $\bs{\Psi}$ and $\bs{Q}(\beta)$, respectively. Define $\bs{G}(\beta) \coloneqq -\det(\bs{I}- \beta \bs{Q}_K) \bs{\Psi}_{-K} + \bs{Q}(\beta)_{-K}\adj(\bs{I}- \beta \bs{Q}_K) (\bs{\psi}_K + \bs{c}_1)$. Let $G_k(\beta)$ denote the $k$th element of $\bs{G}(\beta)$. 
\begin{proposition} \label{prop_log_util_diff}
Suppose Assumption \ref{assn_log_diff} holds. Then the identified set of $\beta$ is the intersection of $[0,1)$, the set $\{\beta: \bs{G}(\beta)>\bs{0}\}$, and the set of solutions to 
\[
G_1(\beta)^{r_1} \cdots G_{J(K-1)}(\beta)^{r_{J(K-1)}} - \exp(c)=0.
\]
\end{proposition}
\begin{proof}[Proof of Proposition \ref{prop_log_util_diff}]
It follows from Remark \ref{rem:standard} that $\bs{u}_K = \bs{c}_1$ yields $J$ restrictions of the form $\bs{R}_1 \bs{U} = \bs{c}_1$ with $\bs{R}_1 = [ \bs{O}_{J, J(K-1)}, \bs{I}_J]$ and $\bs{M}(\beta) = \bs{I}- \beta \bs{Q}_K$. Substituting this into \eqref{R_inv_C} gives
\[
\bs{U} = - \bs{\Psi} + \bs{Q}(\beta) (\bs{I}- \beta \bs{Q}_K)^{-1} (\bs{\psi}_K + \bs{c}_1).
\]
Removing the last $J$ rows, which correspond to $\bs{u}_K$, and multiplying by $\det(\bs{I}-\beta\bs{Q}_K)$ gives
\[
\det(\bs{I}- \beta \bs{Q}_K) \bs{U}_{-K} = - \det(\bs{I}- \beta \bs{Q}_K)\bs{\Psi}_{-K} + \bs{Q}(\beta)_{-K} \adj(\bs{I}- \beta \bs{Q}_K) (\bs{\psi}_K + \bs{c}_1) = \bs{G}(\beta).
\]
Because $\det(\bs{I}- \beta \bs{Q}_K)>0$ on $[0,1)$, $\bs{U}_{-K}>\bs{0}$ holds if and only if $\bs{G}(\beta)>\bs{0}$, which is required by Assumption \ref{assn_log_diff}. Taking the logarithm elementwise gives the system
\[
\log (\bs{U}_{-K}) = - \log (\det(\bs{I}- \beta \bs{Q}_K)) \bs{1}_{J(K-1)}+ \log(\bs{G}(\beta)).
\]
Multiplying this by $\bs{r}\t$ from the left removes the $\log(\det(\bs{I}-\beta\bs{Q}_K))$ term because the elements of $\bs{r}$ sum to zero. Assumption \ref{assn_log_diff} then gives $r_1\log(G_1(\beta)) + \cdots + r_{J(K-1)}\log(G_{J(K-1)}(\beta)) =c$. Finally, taking the exponential of both sides gives the stated result.
\end{proof}

When some of $r_1,\ldots,r_{J(K-1)}$ are non-integer, the resulting identifying equation may involve non-integer powers of polynomials in $\beta$. The following examples satisfy Assumption \ref{assn_log_diff}. Let $\widetilde{\bs{w}}$ and $\lambda_2,\ldots,\lambda_L$ be as in Section \ref{subsec:equality}.
\begin{example}[Exponential of an additively separable function with a homogeneous component of known degree $\nu$]
\label{example_exp_homo}
Suppose $u_k(\bs{w},\bs{z}) = \exp(u_k^w(\bs{w},\bs{z}) + u_k^z(\bs{z}))$, where $u_k^w$ satisfies \eqref{u_w}. Then $\log u_k$ satisfies the assumptions of Example \ref{example_homo3}, which gives $m_z(L-2)$ restrictions of the form \eqref{restriction_homo3} with $\log u_k$ in place of $u_k$.
\end{example}
\begin{example}[Homogeneous function of unknown degree $\nu$]
\label{example_homo_unknown}
Suppose $u_k(\bs{w},\bs{z})$ is homogeneous of degree $\nu$ in $\bs{w}$ with possibly unknown $\nu$. Then $\log u_k$ satisfies the assumptions of Example \ref{example_homo2}, which gives $m_z(L-1)-1$ restrictions of the form \eqref{restriction_homo2} with $\log u_k$ in place of $u_k$.
\end{example}

\section{A Dynamic Entry Model with Zero Scrap Value}\label{sec:oa_zero_scrap}

In this model, the state variable is $(\bs{x}_t,\ve_t) = (w_{t},z_t,y_t,\ve_t)$, where $w_{t}$ and $z_t$ are profit shifters, $y_t=a_{t-1}$, and $\ve_t$ follows a type-I extreme value distribution. The firm's per-period payoff function (net of $\ve_t$) is

\begin{equation}\label{eq:payoff}
\begin{aligned}
u_1(w, z,y;\bs{\theta}) &= \theta_1 w + \theta_2 z + \theta_3 w z + \theta_4 z^2 - \theta_5 (y-1),\\
u_2(w, z,y;\bs{\theta}) &= 0 ,
\end{aligned}
\end{equation}
where $\theta_1 w + \theta_2 z + \theta_3 w z + \theta_4 z^2$ represents variable profit, and $-\theta_5 (y-1)$ captures the entry cost. The parameters are set as $\theta_1=1.0$, $\theta_2 = 0.005$, $\theta_3 = 0.001$, $\theta_4 = 0.002$, and $\theta_5 = 1.0$. The discount factor is set to $\beta = 0.95$.\footnote{The small magnitudes of $\theta_2,\theta_3,\theta_4$ relative to $\theta_1=1$ are chosen so that the inequality restrictions are informative. Payoffs are measured in units of the type-I extreme value shock, whose variance is $\pi^2/6\approx 1.645$. When $(\theta_2,\theta_3,\theta_4)$ are multiplied by $10$, the monotonicity, convexity, and complementarity restrictions hold at every $\beta\in[0,1)$, because the payoff differences they constrain remain positive however $\beta$ changes the implied payoffs.}

The profit shifters $(w_t,z_t)$ follow a $\text{VAR}(1)$ process
\begin{equation}\label{eq:new_transition}
\begin{aligned}
z_{t+1} &= 0.2\,\mathbbm{1}\{a_t=1\} + 0.5\, z_t + 0.04\, w_t + 0.04\,\mathbbm{1}\{y_t=1\} + \varepsilon^z_{t+1},\\
w_{t+1} &= 0.2 + 0.06\,\mathbbm{1}\{a_t=1\} + 0.5\, w_t - 0.02\, z_t + \varepsilon^w_{t+1},
\end{aligned}
\end{equation}
where $y_{t+1} = a_t$, and $\varepsilon^z_{t+1}, \varepsilon^w_{t+1}$ are mutually independent $\iid N(0, 0.01)$, independent across $t$. This model does not exhibit finite dependence because the lagged action $a_{t-1}$ affects $z_{t+1}$. We discretize $(w_t,z_t)$ using the method of \citet{tauchen1986finite}. Let $m_w$ and $m_z$ denote the number of discrete grid points for $w_t$ and $z_t$, respectively. We set $m_z=3$ and $m_w=3$, so the resulting state space $\mathcal{X}$ has cardinality $J = |\mathcal{A}| \cdot m_z m_w = 18$.\footnote{We solve the deterministic (noise-free) steady state of \eqref{eq:new_transition} under two hypothetical action sequences, ``always active'' ($a_t\equiv 1$) and ``always inactive'' ($a_t\equiv 2$). This gives $(z^{\text{act}}, w^{\text{act}}) = (0.5199,\ 0.4992)$ and $(z^{\text{inact}}, w^{\text{inact}}) = (0.0319,\ 0.3987)$. Then we center each grid at the midpoint of its two regime values and set the half-width to (regime spread)/2 $+ 2\sigma$, where $\sigma^2$ is the variance of $\varepsilon^z_{t+1}$ and $\varepsilon^w_{t+1}$. This procedure gives $\mathcal W = \{0.1987,\ 0.4490,\ 0.6992\}$ and $\mathcal Z = \{-0.1681,\ 0.2759,\ 0.7199\}$.} 

\subsection{Identifying assumptions}\label{sec:example_assns}

In contrast to the model in Section \ref{sec:scrap_value_example}, we assume that the econometrician does not know the parametric form of the profit function. The payoff function \eqref{eq:payoff} satisfies various nonparametric and semiparametric identifying assumptions discussed in Section \ref{sec:examples}. In the following, we investigate how these assumptions can identify $\beta$. Let the values of $w$ and $z$ be ordered as $w^{m_w} > w^{m_w-1} > \cdots > w^1$ and $z^{m_z} > z^{m_z-1} > \cdots > z^1$. 

\paragraph{Homogeneity in $w$}

Because $u_1(w,z,y)$ is additively separable with a component homogeneous of degree 1 in $w$, it gives $2(m_w-2) m_z = 6$ restrictions of the form
\begin{equation}\label{triplet}
\frac{u_1(w^{\ell+2},z,y) - u_1(w^{\ell+1},z,y)}{w^{\ell+2} - w^{\ell+1}} - \frac{u_1(w^{\ell+1},z,y) - u_1(w^{\ell},z,y)}{w^{\ell+1} - w^{\ell}}=0
\end{equation}
for $\ell=1,\ldots,m_w-2$ and $(z,y) \in \mathcal{Z} \times \{1,2\}$.

\paragraph{Zero cross-difference in $y$ and $(w,z)$}
The entry cost $u_1(w,z,1) - u_1(w,z,2)$ is constant across all 9 $(w,z)$ pairs. This gives $m_w m_z - 1 = 8$ restrictions of the form
\begin{equation}\label{eq:zero_cross}
u_1(w, z,y=1) - u_1(w, z,y=2) - \left[ u_1(w',z',y=1) - u_1(w',z',y=2) \right]=0
\end{equation}
for $(w, z), (w', z') \in \mathcal{W}\times \mathcal{Z}$.

\paragraph{Monotonicity in $z$}

Monotonicity in $z$ gives $2 m_w (m_z-1) = 12$ inequality restrictions of the form
\begin{equation}\label{eq:mono}
u_1(w, z^{\ell+1}, y) - u_1(w, z^{\ell}, y) \geq 0
\end{equation}
for $\ell=1,\ldots,m_z-1$ and $(w,y) \in \mathcal{W} \times \{1,2\}$.

\paragraph{Convexity in $z$}
Convexity in $z$ gives $2 m_w (m_z-2) = 6$ inequality restrictions of the form
\[
\frac{u_1(w, z^{\ell+2}, y) - u_1(w, z^{\ell+1}, y)}{z^{\ell+2} - z^{\ell+1}} - \frac{u_1(w, z^{\ell+1}, y) - u_1(w, z^{\ell}, y)}{z^{\ell+1} - z^{\ell}} \geq 0
\]
for $\ell=1,\ldots,m_z-2$ and $(w,y) \in \mathcal{W} \times \{1,2\}$.

\paragraph{Complementarity between $w$ and $z$}

Complementarity between $w$ and $z$ gives $2(m_w-1)(m_z-1) = 8$ inequality restrictions of the form
\[
u_1(w^{\ell+1}, z^{m+1},y) - u_1(w^{\ell}, z^{m+1},y) - u_1(w^{\ell+1}, z^{m},y) + u_1(w^{\ell}, z^{m},y) \geq 0
\]
for $(\ell,m)\in \{1,\ldots,m_w-1\}\times \{1,\ldots,m_z-1\}$ and $y \in \{1,2\}$.
 
\paragraph{Linearity in parameters}

The per-period payoff of action 1 can be written as a linear function of the parameters,
\[
\bs{u}_1 = \bs{H}\bs{\theta}, \quad 
\bs{H} = \begin{bmatrix}
w & z & wz & z^2 & -(y-1)
\end{bmatrix}_{(w, z, y) \in \mathcal{X} }, 
 \quad
\bs{\theta} = 
\begin{bmatrix}
\theta_1 &
\theta_2 &
\theta_3 &
\theta_4 &
\theta_5 
\end{bmatrix}\t.
\]
The matrix $\bs{H}$ is $18 \times 5$ and has rank 5. If the rows of $\bs{R}$ form a basis for $\Ker(\bs{H}\t)$, then $\bs{R}\bs{u}_1 = \bs{R}\bs{H} \bs{\theta} =\bs{0}$, which gives $18-5=13$ restrictions on $\bs{U}$.

\subsection{Identification of $\beta$}

We now investigate how these nonparametric shape restrictions identify $\beta$ without relying on the parametric form in \eqref{eq:payoff}.

\begin{figure}[h]
	\centering
	\includegraphics[width=1\linewidth]{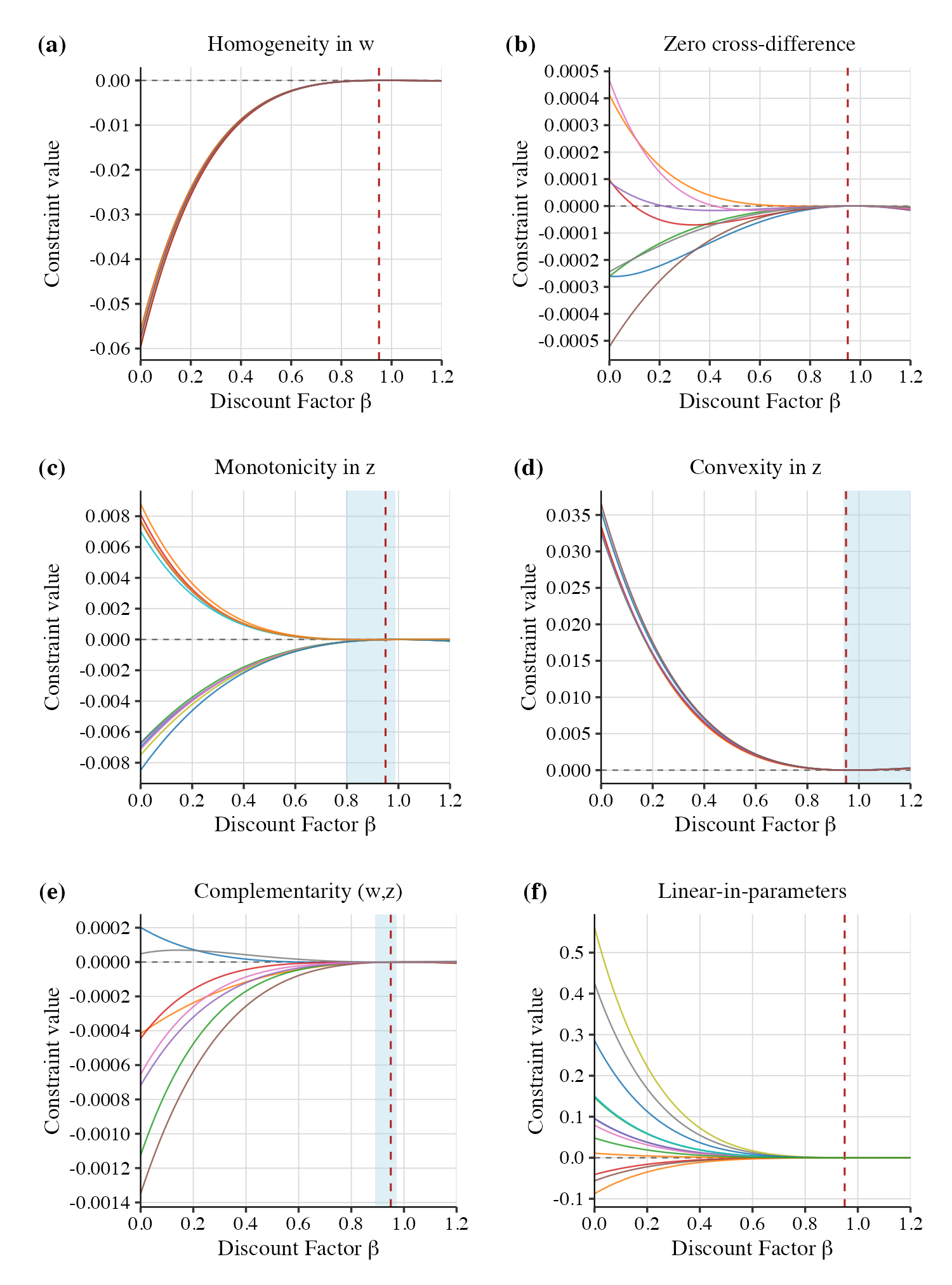}
	\caption[]{Identifying Polynomials for $\beta \in [0,1.2]$: Model with Zero Scrap Value}
	\label{fig2}
\end{figure}

\begin{figure}[h]
	\centering
	\includegraphics[width=1\linewidth]{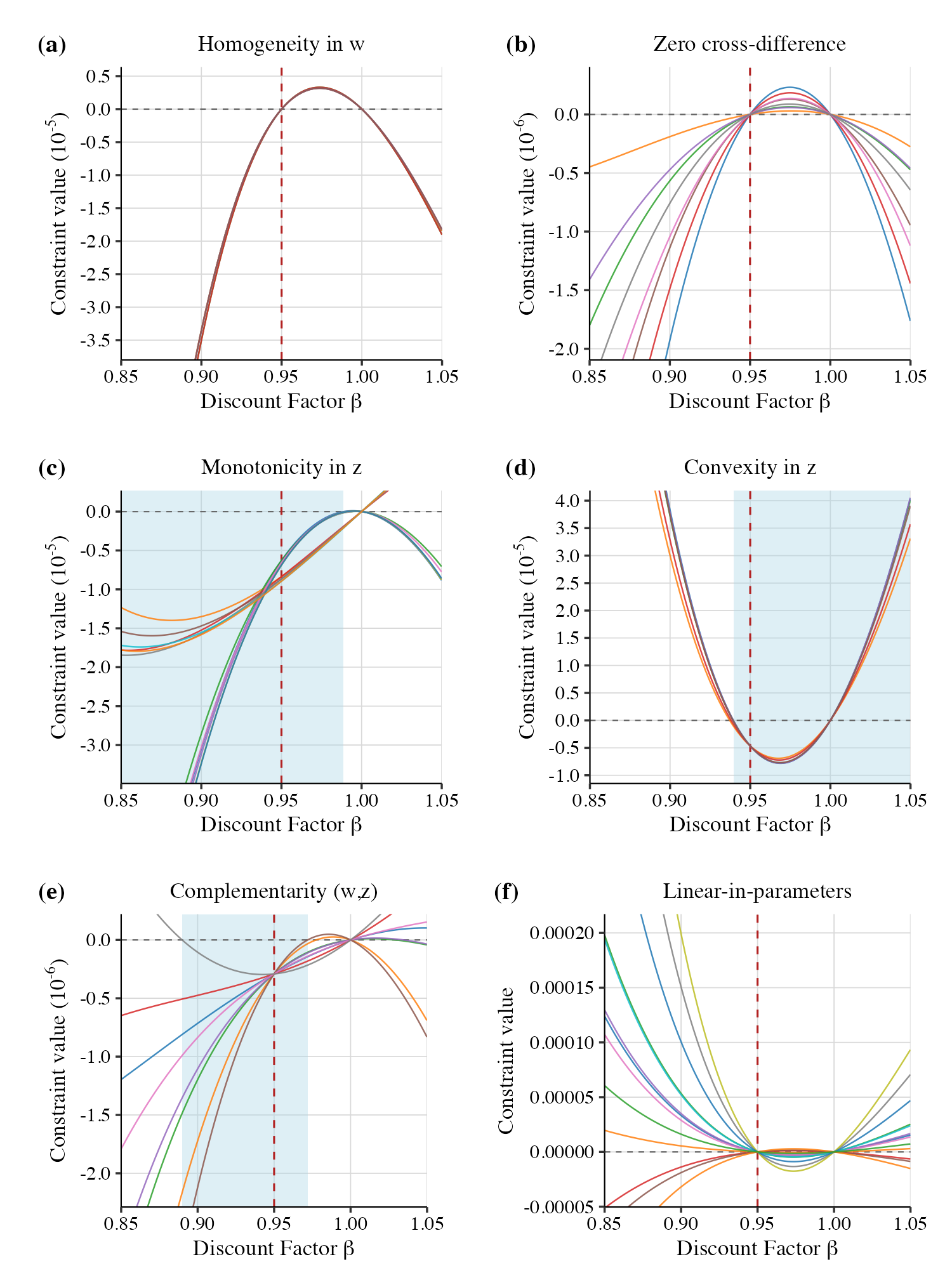}
	\caption[]{Identifying Polynomials for $\beta \in [0.85,1.05]$: Model with Zero Scrap Value}
	\label{fig3}
\end{figure}

Figures \ref{fig2} and \ref{fig3} plot the identifying polynomials in $\beta$ implied by the assumptions in Section \ref{sec:example_assns} over the ranges $\beta \in [0,1.2]$ and $\beta \in [0.85,1.05]$, respectively. In each panel, the red dashed vertical line indicates the true discount factor, $\beta=0.95$. Panels (a) and (b) plot the polynomials in \eqref{ident_eq_known} formed from the homogeneity assumption \eqref{triplet} and the zero cross-difference assumption \eqref{eq:zero_cross}, respectively. The resulting polynomials are of degree at most 18. All curves intersect the horizontal axis at $\beta=0.95$ and, consistent with Proposition \ref{prop_beta_1}, at $\beta=1$. Using Sturm's theorem, we find that every polynomial in panel (a) and four of the eight in panel (b) have exactly one root in $[0,1)$, at $\beta=0.95$, so either the homogeneity assumption or the zero cross-difference assumption alone point identifies the discount factor.\footnote{The other four polynomials in panel (b) have one additional root in $(0,0.9)$, which they do not share.}

Panels (c)--(e) depict the polynomials defined in \eqref{ident_ineq_known}, and the blue regions indicate the sets of $\beta$ values that satisfy the corresponding inequalities. A value of $\beta$ is admissible when every curve in the panel lies weakly below zero. In panel (d), the blue region extends beyond $\beta=1$ because $\det(\bs{I}- \beta \bs{Q}_K)$ changes sign there, which reverses the inequalities. Intersecting with $[0,1)$ as in Proposition \ref{prop_ineq}, the monotonicity restriction implies $\beta \in [0.797, 0.989]$, the convexity restriction implies $\beta \in [0.939, 1)$, and the complementarity assumption implies $\beta \in [0.890, 0.972]$. All three restrictions bound $\beta$ from below, and the sharpest lower bound, $0.939$, comes from convexity. Monotonicity and complementarity also bound $\beta$ from above. Each interval contains the true value $\beta=0.95$. Together they give $\beta\in[0.939,0.972]$. Inequality restrictions bound $\beta$ only when they nearly bind at the true payoffs, as they do here.

Panel (f) shows the polynomials in $\beta$ implied by the linear-in-parameters restriction $\bs{u}_1=\bs{H}\bs{\theta}$. All curves intersect the horizontal axis at the true value $\beta=0.95$. Using Sturm's theorem, we find that none has another root in $[0,1)$, which illustrates the strong identifying power of the linearity assumption. 

\clearpage

\section{Details of the Empirical Application}\label{sec:appendix_data}

The land use data are the Cropland Data Layer (CDL) described in Section \ref{sec:data}; in the raster, corn is coded $1$ and soybean coded $5$.
A pixel enters the sample if it is classified as corn or soybean in every one of the 14 years (an ``always-corn/soy'' pixel), so that a corn/soybean choice is observed every period; this gives $1{,}111{,}033$ pixels.
The CDL classifies each pixel separately, and classification errors are more frequent at field edges, so we keep only interior pixels: in every year, all eight neighboring pixels are classified as corn or soybean and all have the same crop.
This leaves $583{,}390$ pixels ($52.5\%$).
Crop prices $P_t^{\text{corn}}$ and $P_t^{\text{soy}}$ are the pre-season projected prices of the USDA Risk Management Agency (RMA), which are used to set federal crop insurance guarantees and are available from RMA's Price Discovery application (\url{https://www.rma.usda.gov/price-discovery-reporting}).
For 2009--2017, the nitrogen prices $P_t^N$ (dollars per ton of anhydrous ammonia) are taken from Table 1 of \citet{schnitkey17fdd}.
For 2018--2025, we construct the series in the same way, as the October--June average of the bi-weekly Illinois prices in USDA-AMS report GX\_GR210 (Report 3195).
Both parts of the series are Illinois prices, which we use for Story County, Iowa, following the series of \citet{schnitkey17fdd}.

$D_t$ is defined in Section \ref{sec:types}.
The yields $174$ and $53$ bu/acre are the rounded averages over 2014--2023 of the assumed yields of corn following corn and of soybean following corn in the Iowa crop budgets of \citet[p.~13]{plastina23agdm}; these assumed yields are fixed reference yields through 2018 and 30-year trend yields from 2019.
We hold these yields fixed rather than use realized county yields. Weather and soil affect both realized yields and rotation choices, so realized yields would make $D_t$ endogenous.
The nitrogen coefficient $0.1159=190/1{,}640$ converts $190$ lb N/acre into tons of anhydrous ammonia, which is $82\%$ nitrogen ($1{,}640$ lb N per ton). \citet[Table 1]{sawyer14icm} recommends this rate for corn following corn in Iowa at a nitrogen-to-corn price ratio of $0.10$, close to the 2009--2025 average of $0.096$.
The tercile cutoffs of $D_t$ are computed from 2012--2023 and applied to all years 2009--2025.
The transition matrix $\bs{P}_D$ in Table \ref{tab:PD} is estimated from the $16$ annual regime transitions over 2009--2025 with Jeffreys-prior smoothing, $\widehat{\bs{P}}_D(j,k)=(n_{jk}+1/2)/(n_j+3/2)$, where $n_{jk}$ is the number of transitions from regime $j$ to regime $k$ and $n_j=\sum_k n_{jk}$.

\subsection{The first-order Markov violation}\label{sec:appendix_markov}

The natural first specification treats $a_t$ as first-order Markov.
Table \ref{tab:markov} tests this assumption.
For each two-period history $(a_{t-1},a_{t-2})$, the table compares the conditional choice probability $\Pr(a_t=\text{corn}\mid a_{t-1},a_{t-2})$ with $\Pr(a_t=\text{corn}\mid a_{t-1})$. Both probabilities are computed within each regime $D_t$ and averaged over the three regimes using the sample shares of the regimes.
Under the first-order Markov property, the two should coincide.

\begin{table}[htbp]
\centering
\begin{threeparttable}
\caption{Probability of Planting Corn by Crop History}
\label{tab:markov}
\begin{tabular}{llcc}
\toprule
Previous Year & Two Years Ago & Conditioning on & Conditioning on \\
($a_{t-1}$)   & ($a_{t-2}$)   & 2-Year History  & 1-Year History Only \\
\midrule
Corn & Corn & 0.6035 & 0.3308 \\
Corn & Soybean & 0.1913 & 0.3308 \\
\addlinespace
Soybean & Corn & 0.9924 & 0.9912 \\
Soybean & Soybean & 0.8243 & 0.9912 \\
\bottomrule
\end{tabular}
 \begin{tablenotes}[flushleft]
\footnotesize
\item \textit{Notes:} The table reports the empirical probability of planting corn at period $t$ conditional on the row's crop history and on $D_t$, averaged over the three regimes using the sample shares of the regimes and pooled over 2012--2023 ($7.0$ million pixel-year transitions on $583{,}390$ interior pixels).
Under the first-order Markov property, the choice probability depends only on $a_{t-1}$ and should be identical across rows sharing the same $a_{t-1}$.
\end{tablenotes}
\end{threeparttable}
\end{table}

The first-order Markov property fails. Among fields planted with corn last year, those also planted with corn two years earlier are far more likely to plant corn again than those planted with soybean two years earlier.

\subsection{Computation of the second-lag gap}\label{sec:appendix_gap}

This section explains how the second-lag gap in Table \ref{tab:M} is computed.
Let $a_{it}$ denote the crop chosen in pixel $i$ in year $t$.
Given the estimated shares $\hat\pi_m$ and choice probabilities $\hat p(\cdot\mid\cdot;m)$, the posterior probability that pixel $i$ is of type $m$ is
\begin{equation*}
r_{im}=\frac{\hat\pi_m\,\hat\rho_m(a_{i,2011})\prod_{t=2012}^{2023}\hat p(a_{it}\mid a_{i,t-1},D_t;m)}{\sum_{m'=1}^{M}\hat\pi_{m'}\,\hat\rho_{m'}(a_{i,2011})\prod_{t=2012}^{2023}\hat p(a_{it}\mid a_{i,t-1},D_t;m')},
\end{equation*}
where $\hat\rho_m(a)=\Pr(a_{2011}=a\mid D_{2012};m)$ is the conditional distribution of $a_{t-1}$ given $D_t$ under type $m$'s stationary distribution of $(a_{t-1},D_t)$, which depends only on $\bs{P}_D$ and $\hat p(\cdot\mid\cdot;m)$.

For type $m$, a history $(a_1,a_2)\in\{\text{corn},\text{soy}\}^2$ and a regime $D\in\{\text{low},\text{medium},\text{high}\}$, define the weighted total and corn counts
\begin{align*}
N_m(a_1,a_2,D)&=\sum_{i}\sum_{t=2012}^{2023} r_{im}\,\mathbbm{1}\{a_{i,t-1}=a_1,\;a_{i,t-2}=a_2,\;D_t=D\},\\
C_m(a_1,a_2,D)&=\sum_{i}\sum_{t=2012}^{2023} r_{im}\,\mathbbm{1}\{a_{it}=\text{corn},\;a_{i,t-1}=a_1,\;a_{i,t-2}=a_2,\;D_t=D\},
\end{align*}
and let $N_m(a_1,D)=\sum_{a_2}N_m(a_1,a_2,D)$ and $C_m(a_1,D)=\sum_{a_2}C_m(a_1,a_2,D)$.
Define the posterior-weighted sample conditional choice probabilities for type $m$ as
\begin{equation*}
\hat P(\text{corn}\mid a_1,a_2,D;m)=\frac{C_m(a_1,a_2,D)}{N_m(a_1,a_2,D)},\qquad
\hat P(\text{corn}\mid a_1,D;m)=\frac{C_m(a_1,D)}{N_m(a_1,D)}.
\end{equation*}
We use these sample probabilities because the model's $\hat p(\cdot\mid\cdot;m)$ does not depend on $a_2$, so the gap computed from it would be zero.
The weight of a cell is its share of all weighted transitions,
\begin{equation*}
w_m(a_1,a_2,D)=\frac{N_m(a_1,a_2,D)}{\sum_{m'}\sum_{a_1',a_2',D'}N_{m'}(a_1',a_2',D')}.
\end{equation*}
The second-lag gap is
\begin{equation}\label{eq:gap}
\text{Gap}(M)=\Biggl(\sum_{m=1}^{M}\sum_{a_1,a_2,D}w_m(a_1,a_2,D)\bigl[\hat P(\text{corn}\mid a_1,a_2,D;m)-\hat P(\text{corn}\mid a_1,D;m)\bigr]^2\Biggr)^{1/2}.
\end{equation}
Cells with $N_m(a_1,a_2,D)=0$ are omitted.
For $M=1$, $r_{i1}=1$, so the posterior is trivial.
Table~\ref{tab:M} reports the gap and the log-likelihood for each $M$.

\subsection{The soybean normalization and the monotonicity restriction}\label{sec:appendix_norm}

Let $\tilde{\bs{u}}_{m,\text{corn}}$ and $\tilde{\bs{u}}_{m,\text{soy}}$ denote the payoffs of type $m$ before normalization. From \eqref{qudiff}, the corn payoff that generates the same choice probabilities under the normalization $\bs{u}_{m,\text{soy}}=\bs{0}$ is
\begin{equation}\label{eq:ucorn_norm}
\bs{u}_{m,\text{corn}} = \tilde{\bs{u}}_{m,\text{corn}} - (\bs{I}-\beta\bs{Q}_{m,\text{corn}})\bs{v}, \quad \bs{v} \coloneqq (\bs{I}-\beta\bs{Q}_{m,\text{soy}})^{-1}\tilde{\bs{u}}_{m,\text{soy}}.
\end{equation}
$v(a,D)$ is the value of planting soybean in every period starting from the state $(a,D)$. Because the transition of $D_t$ is independent of crop choice, we have
\begin{equation}\label{eq:v_soy}
v(a,D)=\tilde{u}_{m,\text{soy}}(a,D)+\beta\sum_{D'}\bs{P}_D(D,D')\,v(\text{soy},D').
\end{equation}
The last term of \eqref{eq:v_soy} does not depend on $a$, so
\begin{equation}\label{eq:vdiff_soy}
v(\text{soy},D)-v(\text{corn},D)=\tilde{u}_{m,\text{soy}}(\text{soy},D)-\tilde{u}_{m,\text{soy}}(\text{corn},D).
\end{equation}
The $(a,D)$ element of $(\bs{I}-\beta\bs{Q}_{m,\text{corn}})\bs{v}$ is $v(a,D)-\beta\sum_{D'}\bs{P}_D(D,D')\,v(\text{corn},D')$. Substituting \eqref{eq:v_soy} and \eqref{eq:vdiff_soy} into \eqref{eq:ucorn_norm} gives
\begin{equation}\label{eq:ucorn_soy_diff}
u_{m,\text{corn}}(a,D) = \tilde{u}_{m,\text{corn}}(a,D)-\tilde{u}_{m,\text{soy}}(a,D) - \beta\Delta_m(D),
\end{equation}
where
\[
\Delta_m(D)\coloneqq\sum_{D'}\bs{P}_D(D,D')\bigl[\tilde{u}_{m,\text{soy}}(\text{soy},D')-\tilde{u}_{m,\text{soy}}(\text{corn},D')\bigr].
\]
In \eqref{eq:ucorn_soy_diff}, $\tilde{u}_{m,\text{corn}}(a,D)-\tilde{u}_{m,\text{soy}}(a,D)$ is the current payoff margin of corn over soybean in state $(a,D)$, and $\beta\Delta_m(D)$ arises from the normalization. Here $\Delta_m(D)$ is the expected effect on next year's soybean payoff of planting soybean rather than corn this year. If $\Delta_m(D)$ is the same in every regime, $u_{m,\text{corn}}(\text{corn},D)$ equals the margin after corn up to a constant, so the two are monotone in $D$ together. 

\subsection{Choice probabilities of the $k$-means groups}\label{sec:appendix_kmeans}

The $k$-means groupings use two pixel-level features computed over 2010--2023: the corn share, the fraction of the 14 years planted to corn, and the corn-to-corn rate, the fraction of the 13 consecutive-year pairs in which corn follows corn. Each feature is standardized by its pixel-weighted mean and standard deviation. Because pixels with the same 14-year crop sequence have the same features, we cluster the distinct sequences, weighting each by its number of pixels. We run the algorithm from $60$ random initial centers and keep the solution with the smallest weighted within-group sum of squares.
Table \ref{tab:kmeans} reports the choice probabilities of the $k$-means groups in Table \ref{tab:results}.
The groups are ordered by the corn-to-corn transition rate, and Group $j$ corresponds to the $j$th row of Table \ref{tab:results}.

\begin{table}[htbp]
\centering
\begin{threeparttable}
\caption{Probability of Planting Corn in the $k$-Means Groups}
\label{tab:kmeans}
\begin{tabular}{lccccccc}
\toprule
 & & \multicolumn{3}{c}{Corn after corn} & \multicolumn{3}{c}{Corn after soybean} \\
\cmidrule(lr){3-5}\cmidrule(lr){6-8}
Group & Share & low & medium & high & low & medium & high \\
\midrule
\multicolumn{8}{l}{Panel A: $k$-means (a), corn share and corn-to-corn rate} \\
Group 1 & 0.610 & 0.059 & 0.060 & 0.076 & 0.996 & 0.993 & 0.991 \\
Group 2 & 0.247 & 0.462 & 0.510 & 0.523 & 0.994 & 0.973 & 0.978 \\
Group 3 & 0.143 & 0.701 & 0.835 & 0.857 & 0.997 & 0.999 & 0.997 \\
\addlinespace
\multicolumn{8}{l}{Panel B: $k$-means (b), corn-to-corn rate} \\
Group 1 & 0.610 & 0.059 & 0.060 & 0.076 & 0.996 & 0.993 & 0.991 \\
Group 2 & 0.239 & 0.452 & 0.497 & 0.519 & 0.995 & 0.972 & 0.979 \\
Group 3 & 0.152 & 0.701 & 0.834 & 0.842 & 0.994 & 0.999 & 0.991 \\
\bottomrule
\end{tabular}
 \begin{tablenotes}[flushleft]
\footnotesize
\item \textit{Notes:} Each entry is the probability of planting corn given the crop planted last year and the profitability regime $D_t$ of the current year, estimated within the group.
Groups are formed by weighted $k$-means with $M=3$ on (a) each pixel's corn share and corn-to-corn transition rate and (b) the corn-to-corn rate alone.
\end{tablenotes}
\end{threeparttable}
\end{table}

In both groupings, the corn probability after corn weakly rises with the regime in every group.
This differs from the mixture, where it falls from the low to the medium regime for rotators and from the medium to the high regime for partial rotators (Table \ref{tab:types}).

\end{appendix}

\end{document}